\documentclass[twocolumn]{aastex6}
\usepackage{times}
\usepackage{amsmath}
\usepackage{textcomp}
\usepackage{amssymb}
\usepackage{color}

\newcommand{\lum}{erg\,s$^{-1}$}
\newcommand{\fermi}{{\it Fermi}}

\newcommand{\swift}{{\it Swift}}
\newcommand{\phflux}{\mbox{${\rm \, ph \,\, cm^{-2} \, s^{-1}}$}}

\newcommand{\gm}{$\gamma$}

\slugcomment{ApJ accepted}

\shorttitle{2FHL FSRQs}
\shortauthors{Paliya et al.}

\begin{document}

\title{Leptonic and Hadronic Modeling of {\it Fermi}-LAT Hard Spectrum Quasars and 
Predictions for High-Energy Polarization}

\author{Vaidehi S. Paliya$^{1}$, Haocheng Zhang$^{2}$, Markus B{\"o}ttcher$^{3}$, M. Ajello$^{1}$, A. Dom{\'{\i}}nguez$^{4}$, M. Joshi$^{5}$, D. Hartmann$^{1}$, and C. S. Stalin$^{6}$} 
\affil{$^1$Department of Physics and Astronomy, Clemson University, Kinard Lab of Physics, Clemson, SC 29634-0978, USA}
\affil{$^2$Department of Physics and Astronomy, Purdue University, West Lafayette, IN 47907, USA}
\affil{$^3$Centre for Space Research, North-West University, Potchefstroom, 2520, South Africa}
\affil{$^4$3Grupo de Altas Energ{\'{\i}}as, Universidad Complutense, E-28040 Madrid, Spain}
\affil{$^5$Institute for Astrophysical Research, Boston University, 725 Commonwealth Avenue, Boston, MA 02215, USA}
\affil{$^6$Indian Institute of Astrophysics, Block II, Koramangala, Bangalore-560034, India}
\email{vpaliya@g.clemson.edu}

\begin{abstract}
We present the results of a study of the time-averaged spectral energy distributions (SEDs) of eight flat spectrum radio quasars (FSRQs) present in the second catalog of high energy sources detected beyond 50 GeV by the {\it Fermi}-Large Area Telescope (2FHL). Both leptonic and hadronic scenarios are adopted to explain the multi-wavelength SEDs and we find them to be marginally consistent with the 2FHL spectra above 50 GeV. We derive the expected degree of X-ray and \gm-ray polarizations both for the average and elevated activity states and note that: (i) a hadronic radiative model consistently predicts a higher degree of high energy polarization compared to leptonic ones, and (ii) the X-ray polarization degree is higher than the \gm-ray polarization in the leptonic scenario, but similar to the \gm-ray polarization if the observed radiation is powered by hadronic processes. From the leptonic modeling, the location of the \gm-ray emitting region is found to be at the outer edge of the broad line region (BLR) and is consistent with the \gm\gm~opacity estimates for the \gm-ray absorption by the BLR. We conclude that a majority of the FSRQs could be detected by the upcoming Cherenkov Telescope Array, though future high energy polarimeters will be able to detect them only during elevated activity states, which could provide supportive evidence for the hadronic origin of the X-ray and \gm-ray emission.

\end{abstract}


\section{Introduction}{\label{sec:Intro}}

Blazars are radio-loud active galactic nuclei (AGN) with their relativistic jets pointed towards the observer
 \citep[][]{1995PASP..107..803U}. Due to the peculiar orientation of the jet, the flux across the electromagnetic 
spectrum, from radio waves to very high energy \gm-rays, is strongly enhanced by relativistic Doppler boosting. 
Blazars are sub-divided into two categories: flat spectrum radio quasars (FSRQs) and BL Lac objects,
with FSRQs exhibiting broad emission lines (equivalent width $>$5\AA). The observation of strong optical-UV emission lines
from FSRQs indicates the presence of a luminous broad line region (BLR), which, in turn, suggests
an efficient accretion process illuminating it \citep[e.g.,][]{2012MNRAS.421.1764S}. 

The spectral energy distribution (SED) of blazars consists of two broad non-thermal components. The low energy (radio to UV/X-ray) component in the blazar SED is understood to originate from synchrotron
emission by electrons in the relativistic jet. However, the origin of the high-energy SED component from X-rays to $\gamma$-rays 
is less understood. In the leptonic emission scenario, the high energy radiation is produced via inverse Compton 
(IC) scattering of low energy photon fields that can be either synchrotron emission \citep[synchrotron self 
Compton or SSC; e.g.,][]{1985ApJ...298..114M} or can originate outside the jet \citep[External Compton or EC; 
e.g.,][]{1987ApJ...322..650B}. Accordingly, in the canonical picture of the powerful FSRQs, it is assumed that 
the primary site of the $\gamma$-ray emission lies inside the BLR or outside the BLR but inside the torus, where intense radiation fields from the BLR and torus provides seed 
photons for the IC mechanism \citep[e.g.,][]{2009MNRAS.397..985G}. Alternatively, the hadronic scenario suggests 
that both primary electrons and protons are accelerated to ultrarelativistic energies. Here, the high-energy 
emission is dominated by synchrotron emission of primary protons and secondary particles in electromagnetic
cascades initiated by photon-pion and photo-pair production \citep{Mannheim92,Aharonian00,Mucke01,Boettcher13}. 

Both leptonic and hadronic scenarios have been successful in reproducing the steady-state spectra of blazars 
\citep{Boettcher13}. However, they require very different jet energetics and particle dynamics. Interestingly, 
\cite{Zhang13} have shown that the high-energy polarization signatures can be dramatically different in these two 
scenarios. This is because, in general, the relativistic Compton scattering that dominates the high-energy emission 
in the leptonic model generally produces lower polarization degrees than the proton and cascade synchrotron emission 
in the hadronic model. Thus, X-ray and $\gamma$-ray polarization signatures can be used to distinguish the origin 
of the high-energy emission from FSRQs, making them promising targets for the future X-ray polarimeters 
\citep[e.g., IXPE;][]{2016SPIE.9905E..17W}.

\citet[][]{2010ApJ...716...30A} have classified blazars based on the location of the synchrotron peak 
in their SEDs. A source is defined as low synchrotron peaked (LSP) if the rest-frame synchrotron peak 
frequency ($\nu^{\rm peak}_{\rm syn}$) is less than 10$^{14}$ Hz. On the other hand, in the case of
$10^{14}$ Hz$<\nu^{\rm peak}_{\rm syn}<10^{15}$ Hz and $\nu^{\rm peak}_{\rm syn}>10^{15}$ Hz, 
blazars are classified as intermediate synchrotron peaked (ISP) and high synchrotron peaked (HSP) objects, respectively. 
In general, FSRQs are LSP sources and this indicates that their IC peak is typically located at relatively low ($\sim$~MeV) energies. 
Accordingly, FSRQs exhibit a soft \gm-ray spectrum. In addition to that, absorption of \gm-ray photons via pair-production 
with the BLR radiation field can lead to an additional softening of the \gm-ray spectrum of powerful FSRQs 
\citep[e.g.,][]{2010ApJ...717L.118P}, provided the \gm-ray emission region is located inside the BLR. In most
cases, the BLR (with a spherical geometry) can be considered as opaque for \gm-rays with energies greater than 
20 GeV/($1+z$). The detection of very high energy (VHE, $E>50$ GeV) \gm-rays from a few FSRQs 
\citep[e.g.,][]{2011ApJ...730L...8A} therefore, indicates that the \gm-ray emitting region must be located close 
to or outside the outer boundary of the BLR \citep[e.g.,][]{2011A&A...534A..86T}. Overall, a detailed quantitative 
study of the \gm-ray absorption by the BLR radiation field can provide clues about the location of the \gm-ray emitting region in FSRQs. 

In this work, we present a study of the time-averaged spectra of 8 FSRQs that are included in the second catalog of 
hard \fermi-Large Area Telescope (LAT) sources \citep[2FHL;][]{2016ApJS..222....5A} with the primary motivation 
to understand the radiation processes (leptonic or hadronic) dominating the high-energy emission of these objects. 
We derive the \gm\gm~opacity self-consistently with the location of the \gm-ray emission region and predict the 
degree of X-ray and $\gamma$-ray polarization in both leptonic and hadronic emission scenarios. We also briefly 
discuss the role of these peculiar objects in probing the extragalactic background light \citep[EBL,][]{2001ARA&A..39..249H}. Our aim is to 
study the overall time-averaged broadband behavior of these FSRQs rather than any of their specific activity states. In 
Section \ref{sec2}, we describe the sample. The details of the data reduction methodologies are given in 
Section \ref{sec3}. We outline the adopted leptonic and lepto-hadronic/hadronic emission models in Section 
\ref{sec4} and briefly discuss the methods adopted to derive the central black hole mass and the accretion disk 
luminosity in Section \ref{sec5}. Results are presented and discussed in Section \ref{sec6} and we conclude in Section \ref{sec8}. 
Throughout this paper, we assume a flat cosmology with 
$H_0=67.8$ km s$^{-1}$ Mpc$^{-1}$ and $\Omega_{\rm M}=0.308$ \citep[][]{2016A&A...594A..13P}.
 
\section{Sample}{\label{sec2}} 
The 2FHL is a catalog of sources detected in the 50~GeV$-$2~TeV energy range by the \fermi~Large Area Telescope
(LAT) in its first 80 months of the operation using the latest Pass 8 data set \citep[][]{2013arXiv1303.3514A}. 
There are only 10 FSRQs present in the 2FHL. For comparison, the recently released 3FHL contains more than 150 
FSRQs detected above 10 GeV \citep[][]{2017ApJS..232...18A}. This indicates a flux cut-off in the range 10$-$50~GeV, 
possibly due to the steep \gm-ray spectra of FSRQs. Moreover, EBL and possibly BLR absorption of the high energy \gm-ray photons
can also lead to the decreased number of significant \gm-ray detections. Among 
the 10 FSRQs, 2FHL J0043.9+3424 ($z=0.97$) does not have any existing multi-wavelength observations, particularly 
X-rays, and 2FHL J0221.1+3556 is a gravitationally lensed quasar\footnote{Recently, \citet[][]{2017MNRAS.470.2814F} 
found the redshift of the lensing galaxy doubtful.} \citep[$z=0.94$; e.g.,][]{2016ApJ...821...58B}. 
Since there is no clear consensus about the lensing magnification factor, we do not consider this object as 
well as J0043.9+3424 and rather focus on remaining 8 sources with the well-characterized broadband SEDs. In Table 
\ref{tab:basic_info}, we present the basic 2FHL information on these FSRQs, and in Figure \ref{fig_gamma_sed} the 
observed 2FHL spectra of all the sources are shown, along with their 3FGL and 1FHL \gm-ray SEDs.
 
\section{Data Compilation and Reduction}{\label{sec3}}
We use 2FHL \gm-ray SEDs of all the objects as reported in 3 energy bands by \citet[][]{2016ApJS..222....5A} and 
also consider their publicly available 1FHL (3 energy intervals) and 3FGL \gm-ray spectra 
\citep[5 bands;][]{2013ApJS..209...34A,2015ApJ...810...14A}. The \gm-ray SEDs, from 0.1 GeV$-$2~TeV, are corrected 
for EBL absorption following \citet[][]{2011MNRAS.410.2556D}. One can argue about the possible variability between these 
catalogs, since they cover different time periods. However, as noted before, our primary motivation is to study the average properties
of these objects, hence considering time-averaged \gm-ray SEDs is appropriate. Moreover, as can be seen in Figure \ref{fig_gamma_sed},
\gm-ray spectra from the three catalogs join smoothly (except for J2000.9$-$1749), thus supporting our assumption.

We use all of the \swift-X-ray Telescope \citep[XRT;][]{2005SSRv..120..165B} and UltraViolet and Optical Telescope 
\citep[UVOT;][]{2005SSRv..120...95R} observations covering the period of data collection for the 2FHL. In particular, 
the XRT data are analyzed using the online tool ``\swift-XRT data product 
generator\footnote{http://www.swift.ac.uk/user\_objects/}" \citep[][]{2009MNRAS.397.1177E}. This tool
automatically corrects for pile-up, if any, and suitably selects source and background regions \citep[see][for details]{2009MNRAS.397.1177E}.

The downloaded 
source spectra are rebinned to have atleast 20 counts per bin and we perform the spectral fitting in XSPEC 
\citep[][]{1996ASPC..101...17A}. We take the Galactic neutral hydrogen column density ($N_{\rm H}$) from \citet[][]{2005A&A...440..775K}
and use two models: namely power-law and log-parabola, to perform the spectral fitting. The $N_{\rm H}$ value is kept
frozen during the fitting. The best-fit model is determined by comparing
the $\chi^2$ values derived for the power-law (null hypothesis) and log-parabola models and computing the {\tt f-test} 
probability\footnote{https://heasarc.gsfc.nasa.gov/xanadu/xspec/manual/node83.html}. 
We retain the log-parabola model if the null-hypothesis probability is $<$10$^{-4}$, thus indicating the presence of a significant ($>$5$\sigma$) curvature in the XRT spectrum. 
The results of the X-ray spectral analysis are provided in Table \ref{tab:x-ray} where we also give the total number
of XRT observations for each source.

Note that we have combined all of the XRT measurements taken during the period covered in the 2FHL catalog. Since blazars
are known to exhibit large amplitude flux variations, one has to consider possible impact it can have on the results also keeping in mind the fact the pointed
mode operation of \swift~compared to all-sky scanning mode operation of the \fermi-LAT. We have partially
taken variability into account by excluding Windowed Timing (WT) mode XRT data. It is well-known that the WT mode is used
to observe a bright object \citep[e.g., Mrk 421;][]{2011ApJ...736..131A} or when a source is in an elevated activity state \citep[e.g.,][]{2017ApJ...848..103K}.
By excluding the WT mode data, we have rejected very high flux states, such as 2010 November flare of J2254.0+1613 (3C 454.3). Moreover, in Section \ref{subsec:pol}, we briefly
discuss the possible implications of blazar variability to predict the high-energy polarization. In addition to that, since our primary
objective is to study the overall average behavior of 2FHL FSRQs, rather than any of their specific activity state, we believe that the results derived under this
assumption are robust.

UVOT snapshots are first summed using the tool {\tt uvotimsum} 
and then source magnitudes are extracted from a circular region of 5$^{\prime\prime}$ centered at the target quasar, 
using the task {\tt uvotsource}. The background is estimated from a nearby circular region of 30$^{\prime\prime}$ 
free from source contamination. We correct source magnitudes for Galactic extinction following 
\citet[][]{2011ApJ...737..103S} and convert to flux units using the calibrations of \citet[][]{2011AIPC.1358..373B}.

\section{Radiative Models}{\label{sec4}}
The basic assumption, when averaging years of the multi-frequency data, is that the emitting region is 
considered as a quasi-stationary acceleration zone, where a continuous flow of particles (leptons or hadrons) are 
injected and then radiate, moving along with a certain bulk Lorentz factor $\Gamma$ \citep[e.g.,][]{2016A&A...592A..22H}. This is because a single emitting region moving along the jet at 
relativistic speed, in several years it moves by several light-years, expanding accordingly (and seeing varying external photon fields), while 
we keep the location of the emitting region fixed. Therefore, the SED parameters derived from such an averaged analysis provide information
about an overall average behavior of the blazar rather than any of its specific activity state.

\subsection{Leptonic Emission Model}
We use a simple one-zone leptonic emission model to describe the broadband SEDs of 2FHL FSRQs, following 
\citet[][]{2009MNRAS.397..985G}. In particular, we assume a spherical emission region (or blob) located at a distance 
$R_{\rm diss}$  from the central black hole of mass $M_{\rm BH}$. 
The blob is filled with highly energetic electrons. Here, we do not consider a self-consistent cooling model. 
Instead, we assume that the electron population follows a smooth broken power law distribution of the following form:

\begin{equation}
N(\gamma)  \, = \, N_0\, { (\gamma_{\rm b})^{-s1} \over
(\gamma/\gamma_{\rm b})^{s1} + (\gamma/\gamma_{\rm b})^{s2}}
\end{equation}
where $N_0$ is the normalization constant (cm$^{-3}$) and $s1$, $s2$ are the spectral indices below and above the break 
energy $\gamma_{\rm b}$, respectively. We consider both SSC 
and EC for the high-energy emission. The accretion disk is considered as \citet{1973A&A....24..337S} type and its 
spectrum is given by a multi-temperature annular blackbody with a temperature distrubiton 
\citep[][]{2002apa..book.....F}

\begin{equation}
T =\, \left[ {  3 R_{\rm S}\, L_{\rm disk }  \over 16 \pi\, \eta_{\rm acc}\, \sigma_{\rm SB}\, R^3 }  
\left\{ 1- \left( {3 R_{\rm S} \over  R}\right)^{1/2} \right\} \right]^{1/4}  
\end{equation}
where $L_{\rm disk}$ is the disk luminosity, $R_{\rm S}$ is the Schwarzschild radius, $\sigma_{\rm SB}$ is Stefan-Boltzmann constant, and $\eta_{\rm acc}$ is the 
accretion efficiency taken as 10\%. The BLR and the dusty torus are assumed to reprocess a fraction 
(10\% and 30\%, respectively) of $L_{\rm disk}$. Their spectral profiles are characterized by a spherical 
blackbody located at a distance $R_{\rm BLR} = 10^{17} \, L^{1/2}_{\rm disk,45}$ cm and $R_{\rm torus} = 
2.5 \times 10^{18}L^{1/2}_{\rm disk,45}$ cm, respectively, where $L_{\rm disk,45}$ is the accretion disk luminosity 
in units of $10^{45}$ \lum. We also consider the presence of the X-ray emitting corona recycling 30\% of 
$L_{\rm disk}$ and its spectral shape is assumed as a flat power law with an exponential cutoff. 
In the comoving frame, the radiative energy densities of these external AGN components are calculated 
as a function of $R_{\rm diss}$ \citep[][]{2009MNRAS.397..985G}. We calculate the powers that the jet
carries in the form of electrons ($P_{\rm e}$), magnetic field ($P_{\rm m}$), 
and cold protons ($P_{\rm p}$). In particular, the kinetic jet power, is derived by assuming equal number 
densities of emitting electrons and cold protons \citep[e.g.,][]{2008MNRAS.385..283C}.

\subsection{Lepto-hadronic Emission Model}
We use the code of \cite{Boettcher13} for the one-zone hadronic model. Unlike the leptonic model, our hadronic model 
involves a semi-analytical evolution of particle injection, cooling, and escape, to a quasi-equilibrium 
state by solving steady state Fokker-Planck equations. For the electrons and positron pairs, our model 
includes synchrotron cooling, which is the dominating energy loss process in the strong magnetic field
required for these models and energy-independent electron escape. We also consider injection terms representing primary 
electron injection, pion/muon decay, and $\gamma\gamma$ pair production. Here, electron SSC can be important 
in the high-energy SED component, but in general the SSC cooling rate is much lower than that of the 
synchrotron process. For the protons, since the radiative cooling time scale is much longer than that 
of the electrons, we also include adiabatic and photon-pion production losses. In this way, particle cooling is self-consistently evaluated. The quasi-equilibrium proton 
distribution is then derived based on the energy loss terms, primary proton injection, and energy-independent 
proton escape. Our choice of parameters are generally consistent with physical conditions of small muon and pion contributions described in \cite{Boettcher13}, thus we ignore their contributions to the spectrum. We choose 
the primary proton injection spectra as straight power-law distributions with a turnover at the high energy $\gamma_{b,p}$,

\begin{equation}
\begin{aligned}
Q_{p}(\gamma) & =Q_{0,p}(\gamma/\gamma_{b,p})^{−s_{1,p}}, & 1\leq\gamma<\gamma_{b,p} \\
& =Q_{0,p}(\gamma/\gamma_{b,p})^{−s_{2,p}}, & \gamma_{b,p}\leq \gamma<\gamma_{2,p}
\end{aligned}
\end{equation}
where $s_{1,p}$ is the power-law index and $s_{2,p}$ is the power-law index of the high-energy turnover. This turnover is a natural result of the Fermi acceleration based on numerical simulations \citep{Sironi11,Guo16}, which can be approximated by a short power-law that covers less than one decade of particle Lorentz factor. For the primary electron spectra, because of the very strong cooling, we expect that the high-energy electrons are sufficiently cooled, thus the high-energy turnover is not observable. Therefore, we choose a straight power-law spectra,

\begin{equation}
\begin{aligned}
Q_{e}(\gamma) & =Q_{0,e}\gamma^{−s_e}, & \gamma_{1,e}\leq\gamma<\gamma_{2,e}
\end{aligned}
\end{equation}
where $\gamma_{1,e}$ is the low-energy cutoff of the electron spectra, which corresponds to a background thermal temperature, $\gamma_{2,e}$ is the high-energy cutoff, and $s_e$ is the electron spectral index.

\subsection{High-Energy Polarization Model}{\label{xpol}}

We predict the high-energy (X-ray and $\gamma$-ray) polarization degree for both the leptonic and hadronic models
following \cite{Zhang13}. Their calculations considered a perfectly ordered magnetic field,
thus representing upper limits to the expected degree of polarization. Therefore, a generalization with the correction 
for a partially ordered magnetic field is necessary, as described below.

The general formalism for the observed high-energy polarization degree is

\begin{equation}
\Pi(\nu)=Z_{\rm m}\frac{P_{\rm pol}(\nu)}{P_{\rm tot}(\nu)}
\end{equation}
where $Z_{\rm m}$ is a correction factor due to the partially ordered magnetic field, and $P_{\rm pol}$, $P_{\rm tot}$ 
are the polarized and total radiation power in a perfectly ordered magnetic field. Given that the low-energy synchrotron component and high-energy 
SED component are co-spatial for both leptonic and hadronic models, the optical polarization degree and high-energy polarization degree should be corrected by the same factor for a partially ordered magnetic field. The same assumption has been used in 
\cite{Bonometto73} and \cite{Zhang13}. More recently, this conjecture of equal depolarization factors has been 
confirmed for a 3D multi-zone hadronic model developed by \cite{Zhang16}, where the magnetic field is partially 
ordered.

To evaluate the correction factor $Z_{\rm m}$, we first collect the average optical polarization degree 
($\Pi_{\rm o}$) for each source (see Table \ref{tab:pol}). 
We collect this information from Steward and RoboPol observatories \citep[][]{2009arXiv0912.3621S,2014MNRAS.442.1693P}.
The theoretical upper limit for the optical polarization degree is around $70-75 \, \%$ for an electron power-law distribution of index $2-3$ \citep{Rybicki85}. Given that the average optical polarization degrees are not obtained simultaneously with our {\it Fermi} observations, here
we choose a conservative value at $70\%$. Clearly the average observed optical polarization degree $\Pi_{\rm o}$ 
is lower than the theoretical value of $70\%$. This is mainly because of two depolarization effects, namely the partially ordered magnetic field and 
unpolarized thermal contributions to the optical emission. The correction introduced by the partially ordered magnetic field is $Z_m$, while the correction by the unpolarized thermal contribution is $P_{\rm syn}/(P_{\rm syn}+P_{\rm th})$, where $P_{\rm syn}$ and $P_{\rm th}$ are the synchrotron and thermal radiation power derived from spectral 
fitting. Therefore, we have 

\begin{equation}
\Pi_{\rm o}=Z_{\rm m} \frac{P_{\rm syn}}{P_{\rm syn}+P_{\rm th}}\times 70\%
\end{equation}
Given $\Pi_{\rm o}$ for each source and the fact that 
leptonic and hadronic models predict very similar low-energy SED component, so that the ratio between the primary electron synchrotron and thermal emission are identical for both models, 
$Z_{\rm m}$ is then identical for the two models. We list the resulting values of $Z_{\rm m}$ in Table 
\ref{tab:pol}.

For the leptonic model, the high-energy emission is a combination of SSC ($P_{\rm SSC}$) and EC ($P_{\rm EC}$). 
EC radiation is essentially unpolarized. Therefore, the frequency-dependent polarization degree is predicted to be

\begin{equation}
\Pi_{\rm lep}(\nu)=Z_{\rm m}\frac{P_{\rm SSC,pol}(\nu)}{P_{\rm SSC,tot}(\nu)+P_{\rm EC,tot}(\nu)}
\end{equation}
where $P_{\rm SSC,pol}(\nu)$ is the polarized SSC power in a perfectly ordered magnetic field, evaluated following 
\cite{Bonometto73} and \cite{Zhang13}, and $P_{\rm SSC,tot}(\nu)$ and $P_{\rm EC,tot}(\nu)$ are the SSC and EC powers 
derived from the modeling. On the other hand, the X-ray and $\gamma$-ray emission in the hadronic model are 
from proton synchrotron ($P_{\rm p}$), pair synchrotron ($P_{\rm pair}$), and an SSC contribution from the primary 
electrons ($P_{\rm SSC}$). Similarly, the polarization degree is

\begin{equation}
\Pi_{\rm had}(\nu)=Z_{\rm m}\frac{P_{\rm p,pol}(\nu)+P_{\rm pair,pol}(\nu)+P_{\rm SSC,pol}(\nu)}{P_{\rm p,tot}(\nu)+P_{\rm pair,tot}(\nu)+P_{\rm SSC,tot}(\nu)}
\end{equation}

\section{Black Hole Mass and the Accretion Disk Luminosity}{\label{sec5}}
The black-hole mass, $M_{\rm BH}$, and accretion-disk luminosity, $L_{\rm disk}$, are the two crucial ingredients 
to model the accretion-disk contribution to the SED of an FSRQ. With the knowledge of the disk luminosity, the 
external photon fields can be fully parameterized in terms of distance of the emission region from the central 
engine \citep[][]{2009MNRAS.397..985G}. This is also crucial to determine the \gm\gm~pair production optical 
depth as a function of the dissipation distance which depends on the energy density of the interacting radiation 
fields \citep[e.g.,][and references therein]{2016ApJ...821..102B}. 

Two widely accepted methods to calculate 
$M_{\rm BH}$ and $L_{\rm disk}$ are (a) to use single-epoch optical spectroscopy with the assumption that the 
BLR is virialized \citep[e.g.,][]{2012ApJ...748...49S} and (b) the modeling of the optical-UV SED with a \citet[][]{1973A&A....24..337S} disk if this
part of the SED is accretion-disk dominated. It has been found in recent studies that both the methods reasonably 
agree \citep[see, e.g.,][]{2017ApJ...851...33P}. 

We have collected the optical spectroscopic emission line parameters from the literature to derive both $M_{\rm BH}$ and $L_{\rm disk}$.
We take the
Mg {\sc ii} line luminosity of J0456.9$-$2323 and J1427.3$-$4204 from \citet[][]{1993A&AS..100..395S} and \citet[][]{1989A&AS...80..103S} respectively. 
By following the empirical relation and line coefficients of \citet[][]{2012ApJ...748...49S}, we derive both $L_{\rm disk}$ and $M_{\rm BH}$.
In particular, $L_{\rm disk}$ is computed following the scaling relations of \citet[][]{1991ApJ...373..465F} and 
\citet[][]{1997MNRAS.286..415C} and assuming that the BLR reprocesses 10\% of $L_{\rm disk}$.
\citet[][]{2012ApJ...748...49S} reported $M_{\rm BH}$ and Mg {\sc ii} line luminosity for J0957.6+5523 and J1224.7+2124, which we use
to calculate $L_{\rm disk}$. For J1256.2$-$0548, J1512.7$-$0906, and J2254.0+1613, we take their $M_{\rm BH}$ and $L_{\rm disk}$ 
from 
the literature \citep[][]{1999ApJ...521..112P,2002ApJ...579..530W,2007AJ....133.2187D,2011MNRAS.410..368B,2015ApJ...803...15P}. 
For J2000.9$-$1749, Mg {\sc ii} line information are available in \citet[][]{1984ApJ...277...64O} which we use to derive $M_{\rm BH}$ and $L_{\rm disk}$.

\section{Results and Discussion}{\label{sec6}}

\begin{figure*}
\hbox{
\includegraphics[width=6.2cm]{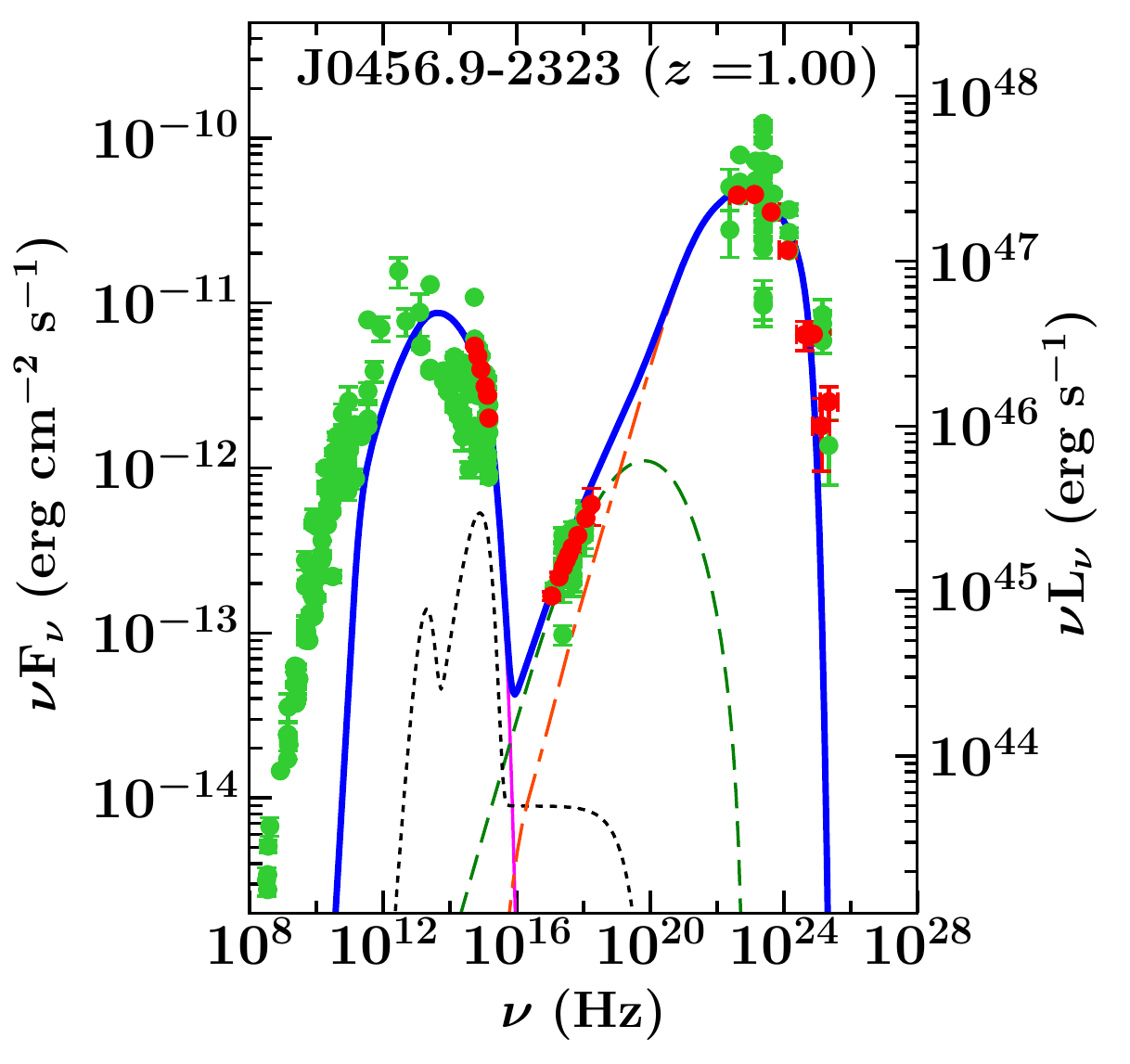}
\includegraphics[width=6.2cm]{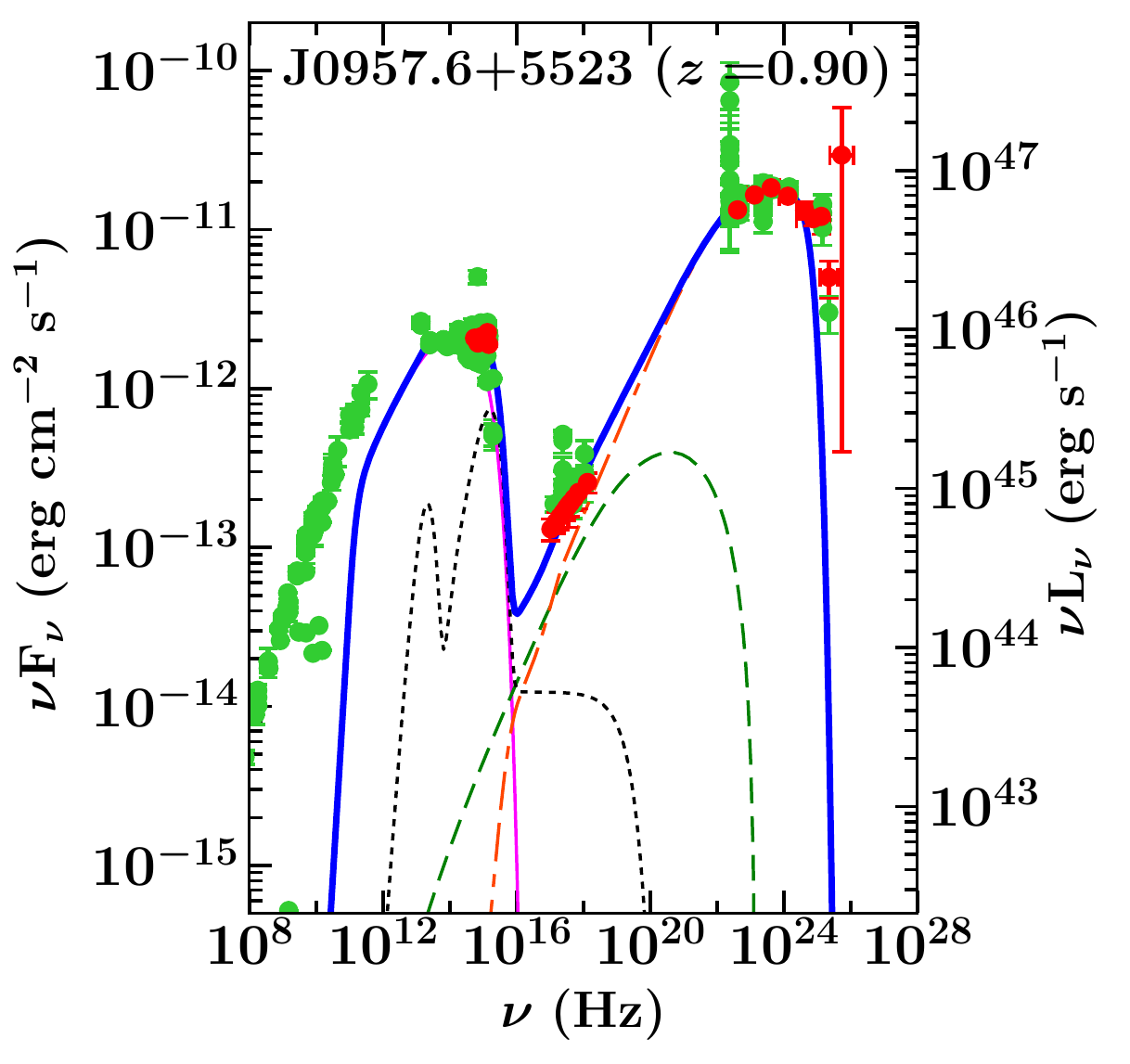}
\includegraphics[width=6.2cm]{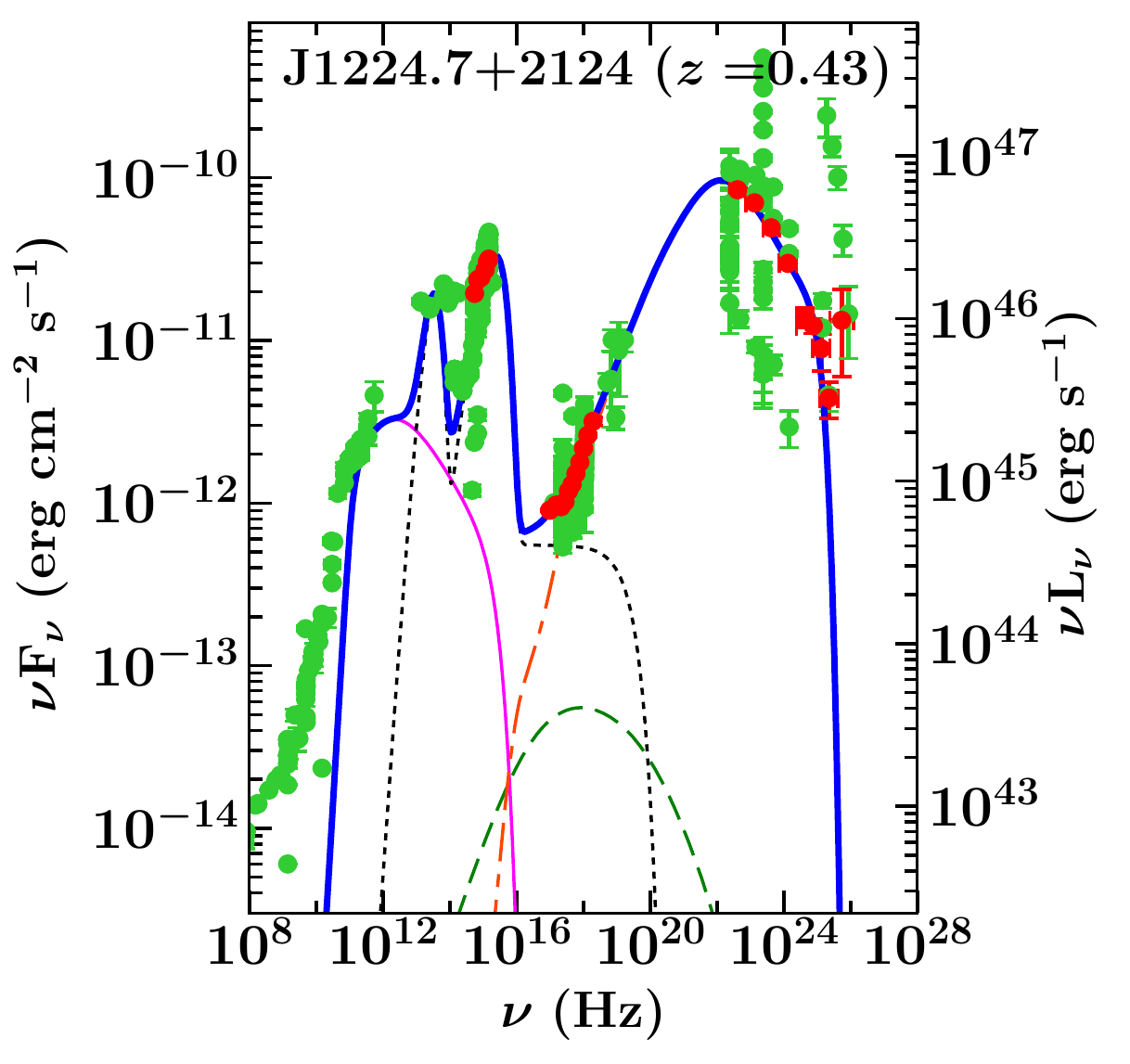}
}
\hbox{
\includegraphics[width=6.2cm]{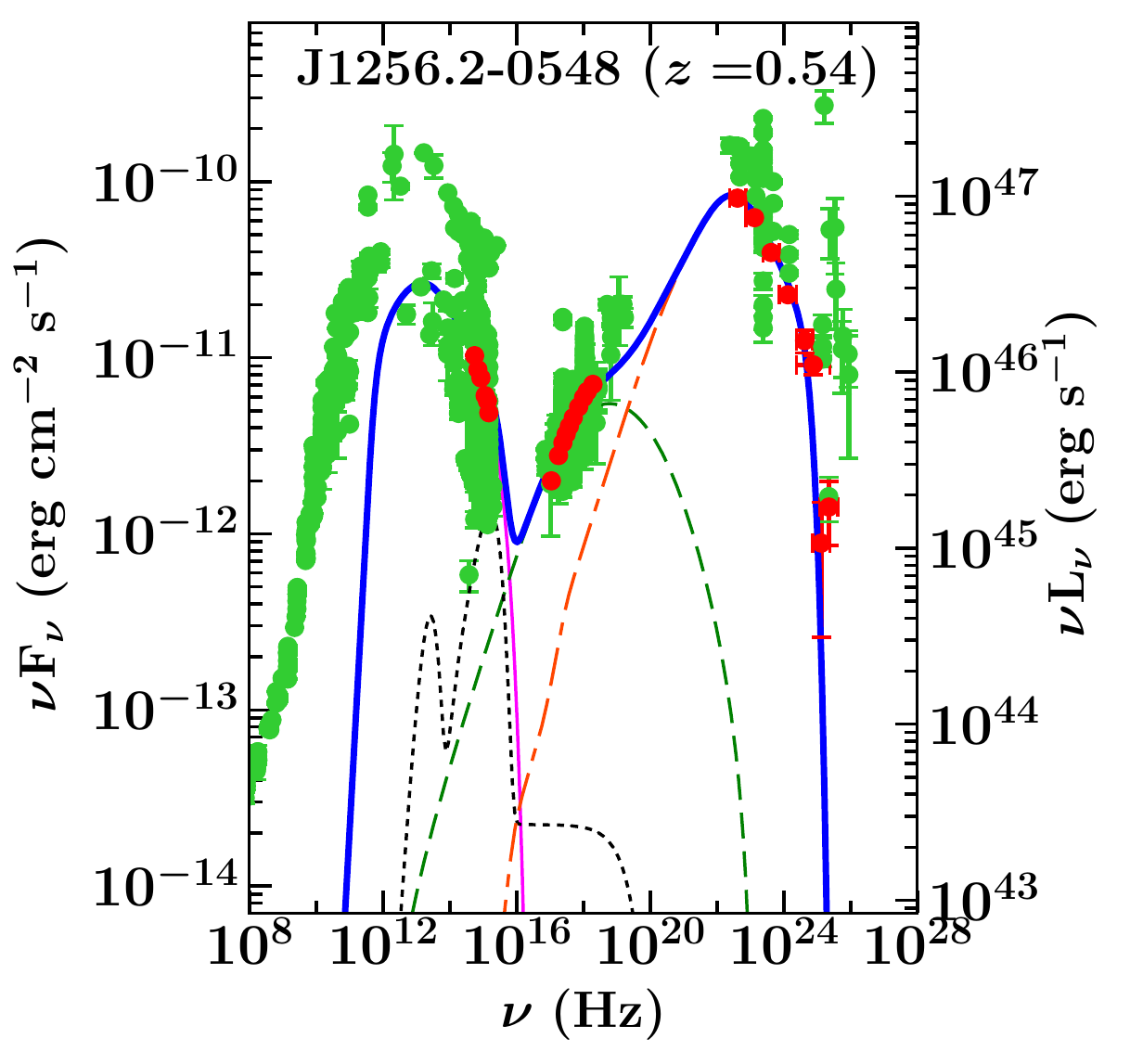}
\includegraphics[width=6.2cm]{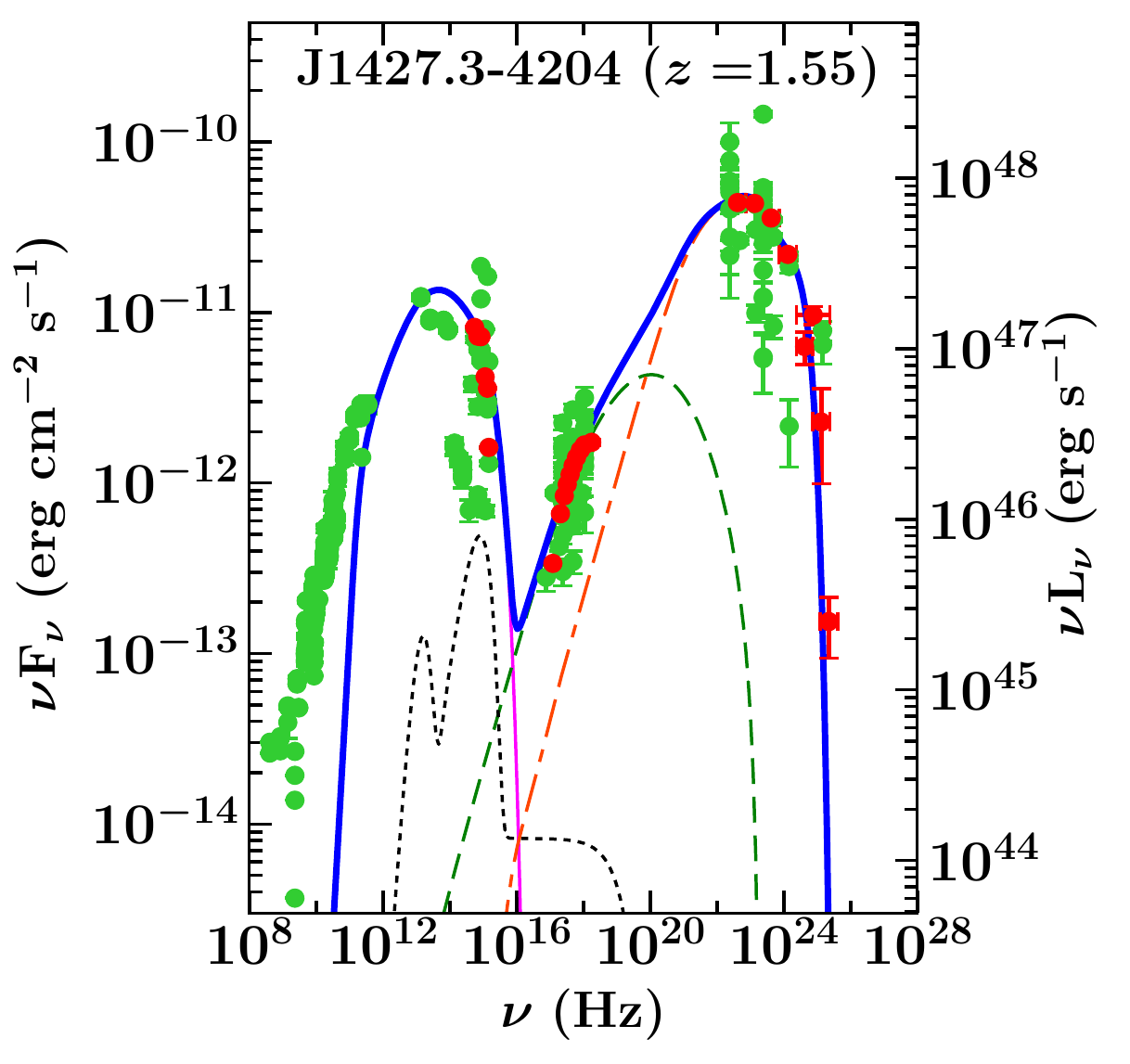}
\includegraphics[width=6.2cm]{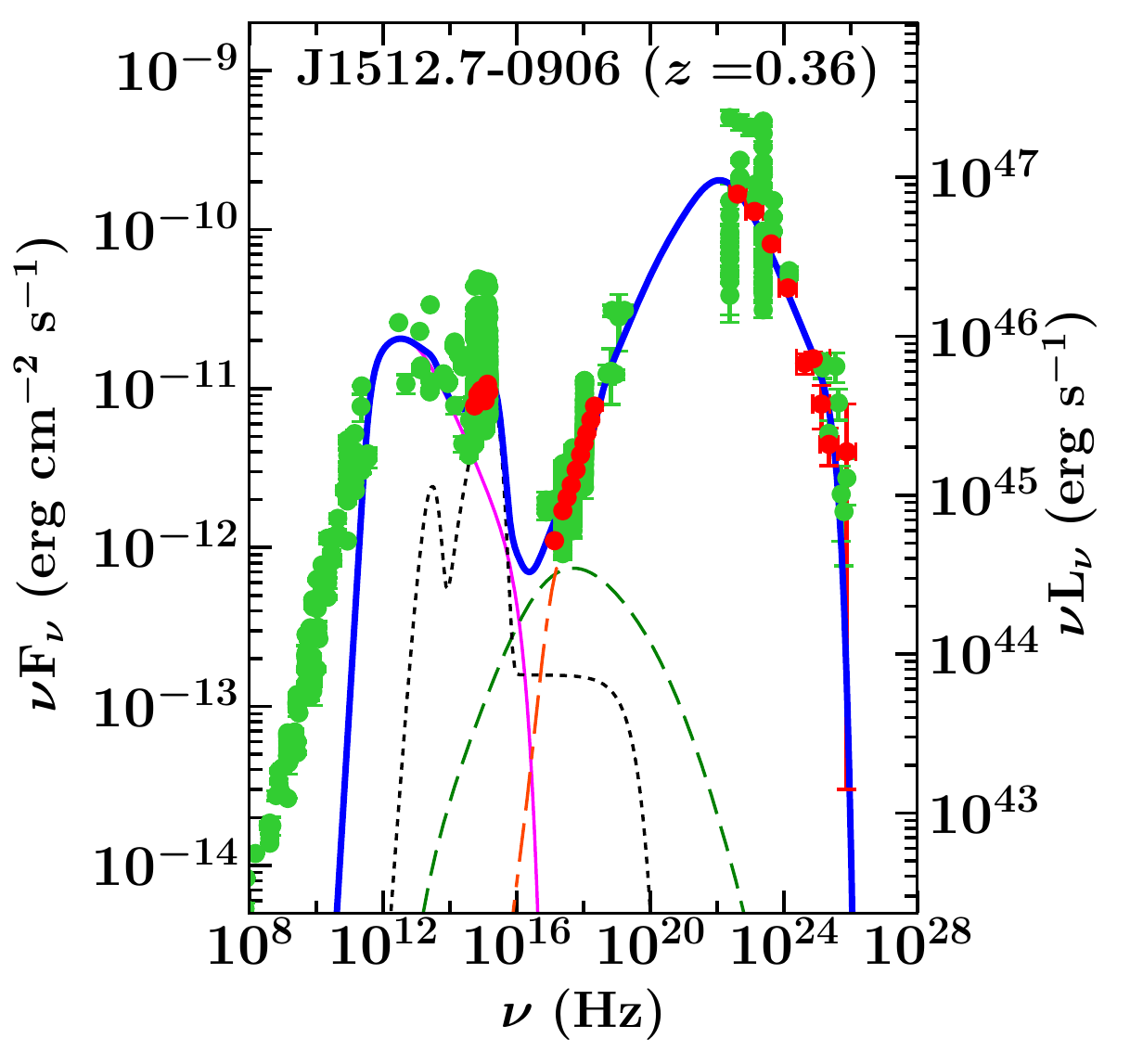}
}
\hbox{\hspace{3.0cm}
\includegraphics[width=6.0cm]{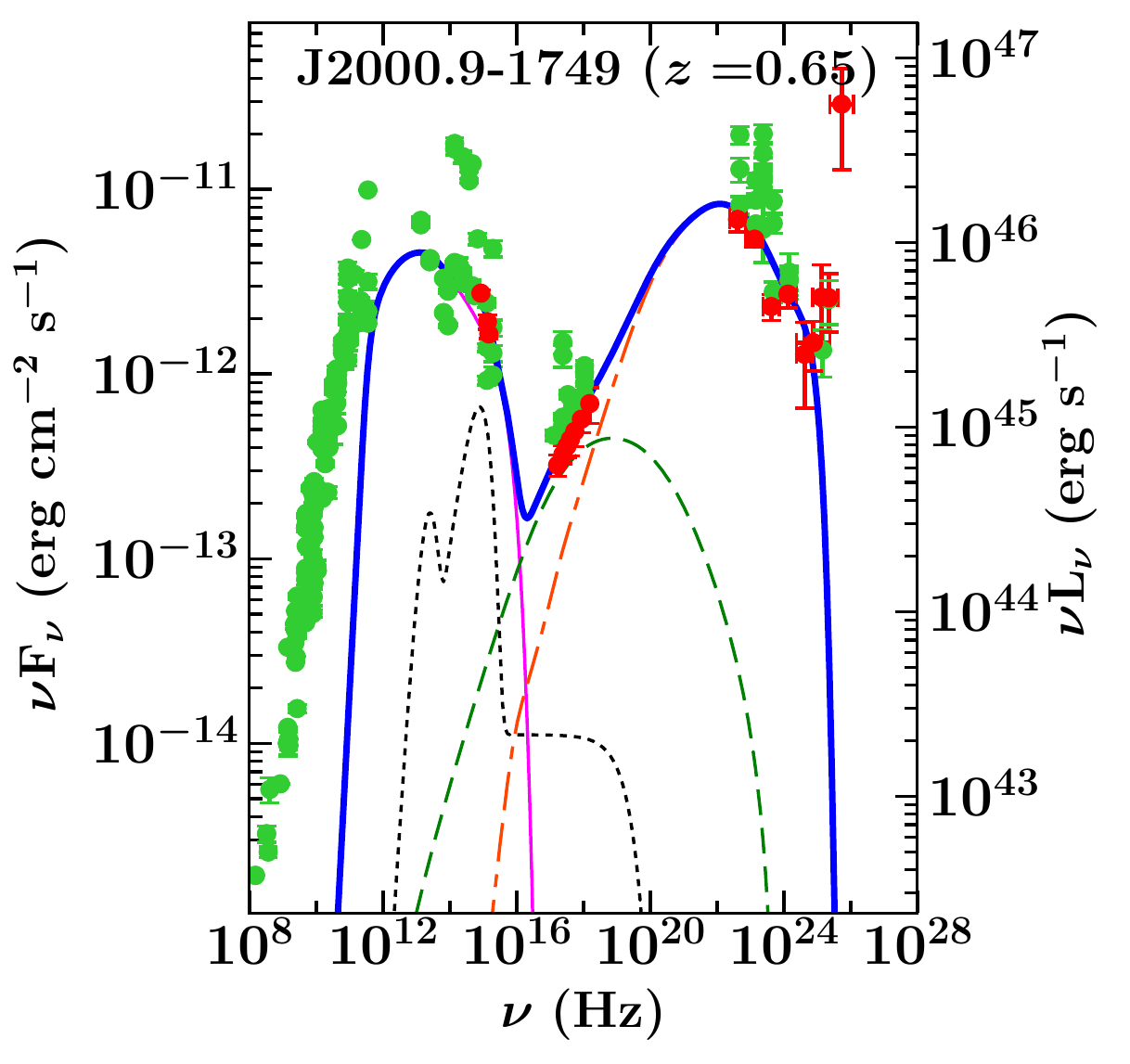}
\includegraphics[width=6.0cm]{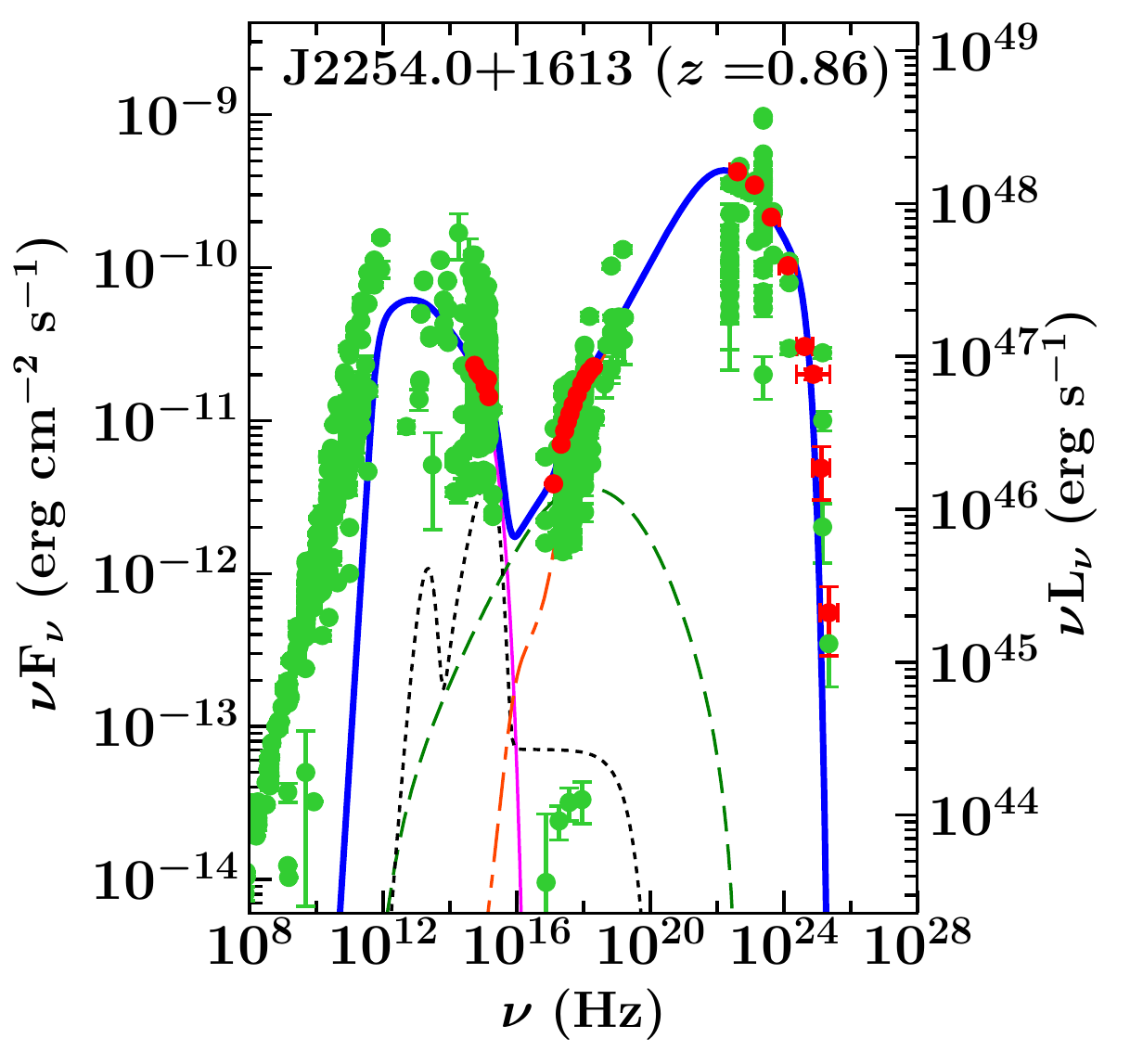}
}
\caption{Leptonic modeling of the broadband SEDs of 2FHL FSRQs studied here. The data used for the modeling is shown with red filled 
circles, whereas, green cricles represent the archival information. Pink thin solid, green dahsed and orange 
dash-dash-dot lines correspond to the synchrotron, SSC, and EC mechanisms, respectively. The thermal emissions 
from the dusty torus, the accretion disk, and the X-ray corona are shown with black dotted line. The overall radiative 
output is represented by blue thick solid line. Note that the \fermi-LAT data points are corrected for EBL absorption 
following \citet[][]{2011MNRAS.410.2556D}.\label{fig_leptonic_sed}} 
\end{figure*}

\begin{figure*}
\hbox{
\includegraphics[width=6.2cm]{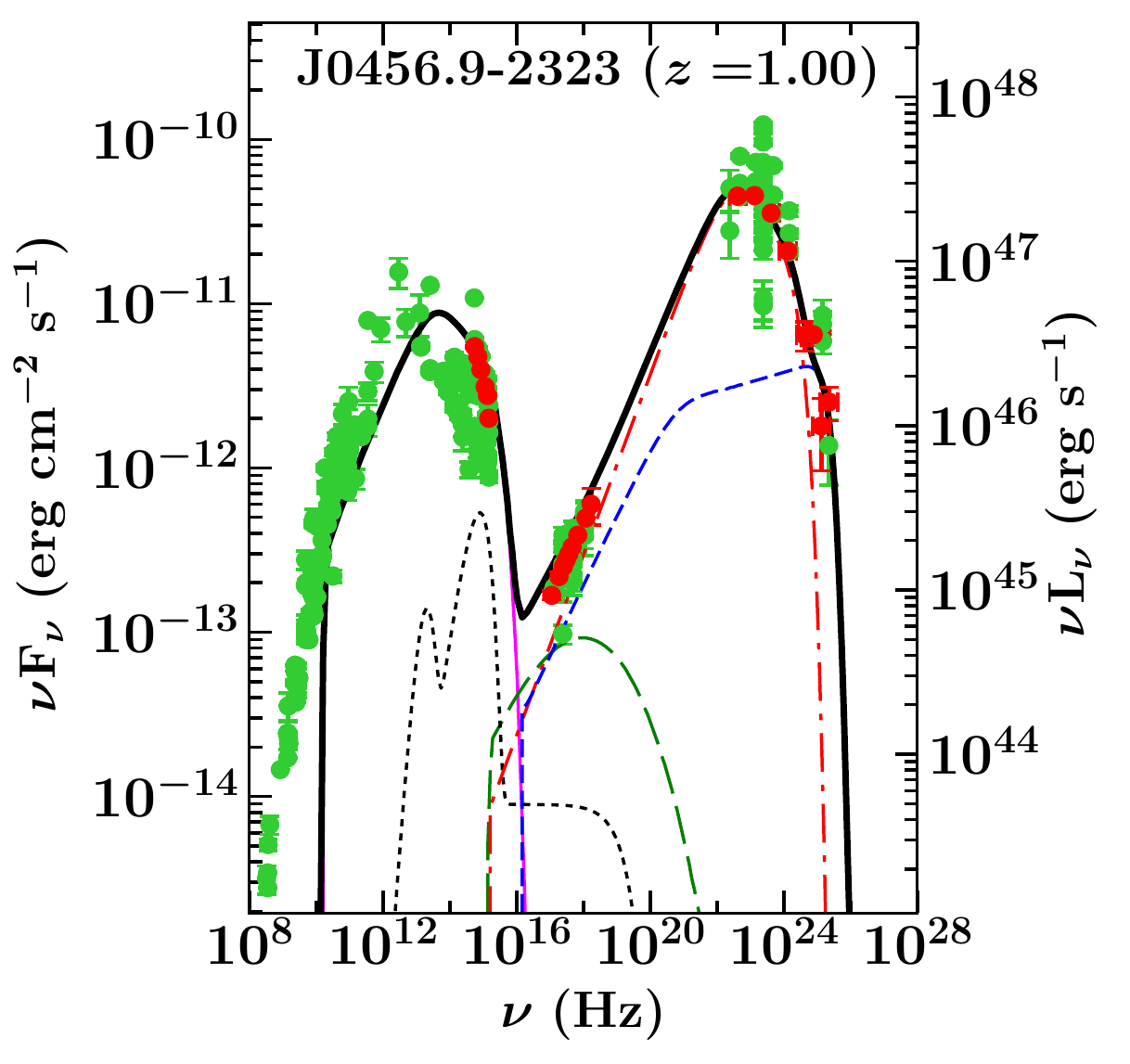}
\includegraphics[width=6.2cm]{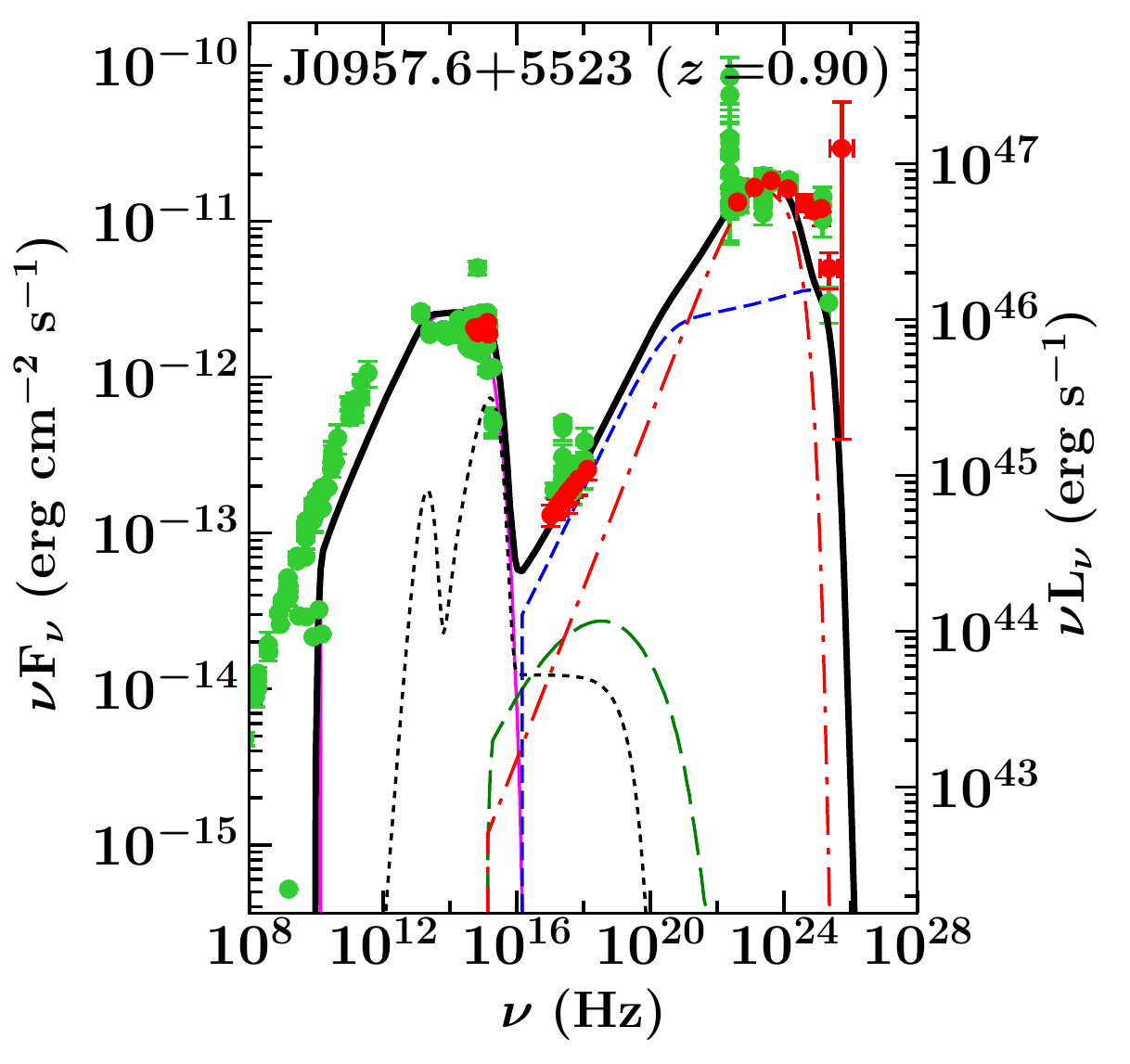}
\includegraphics[width=6.2cm]{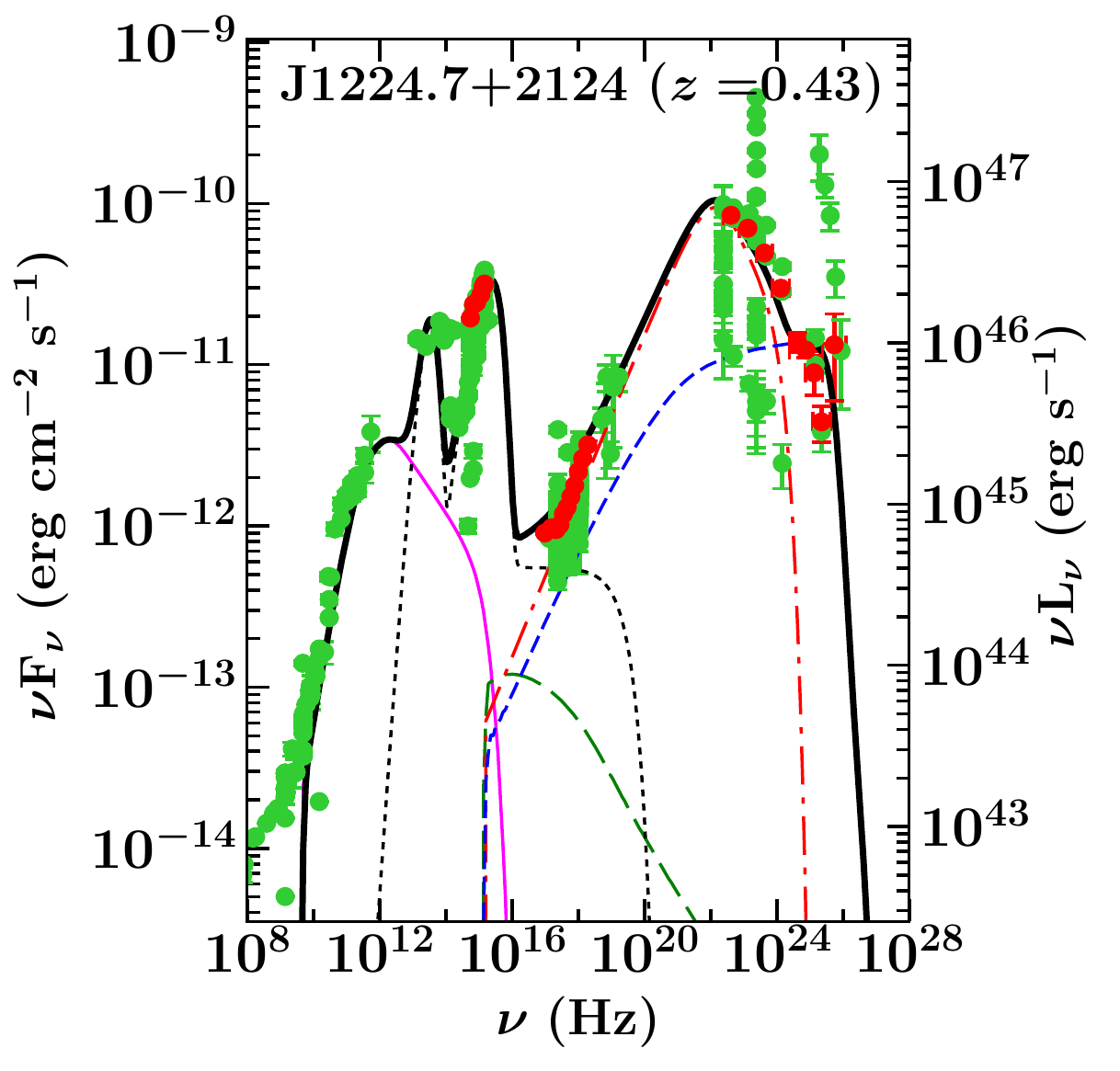}
}
\hbox{
\includegraphics[width=6.2cm]{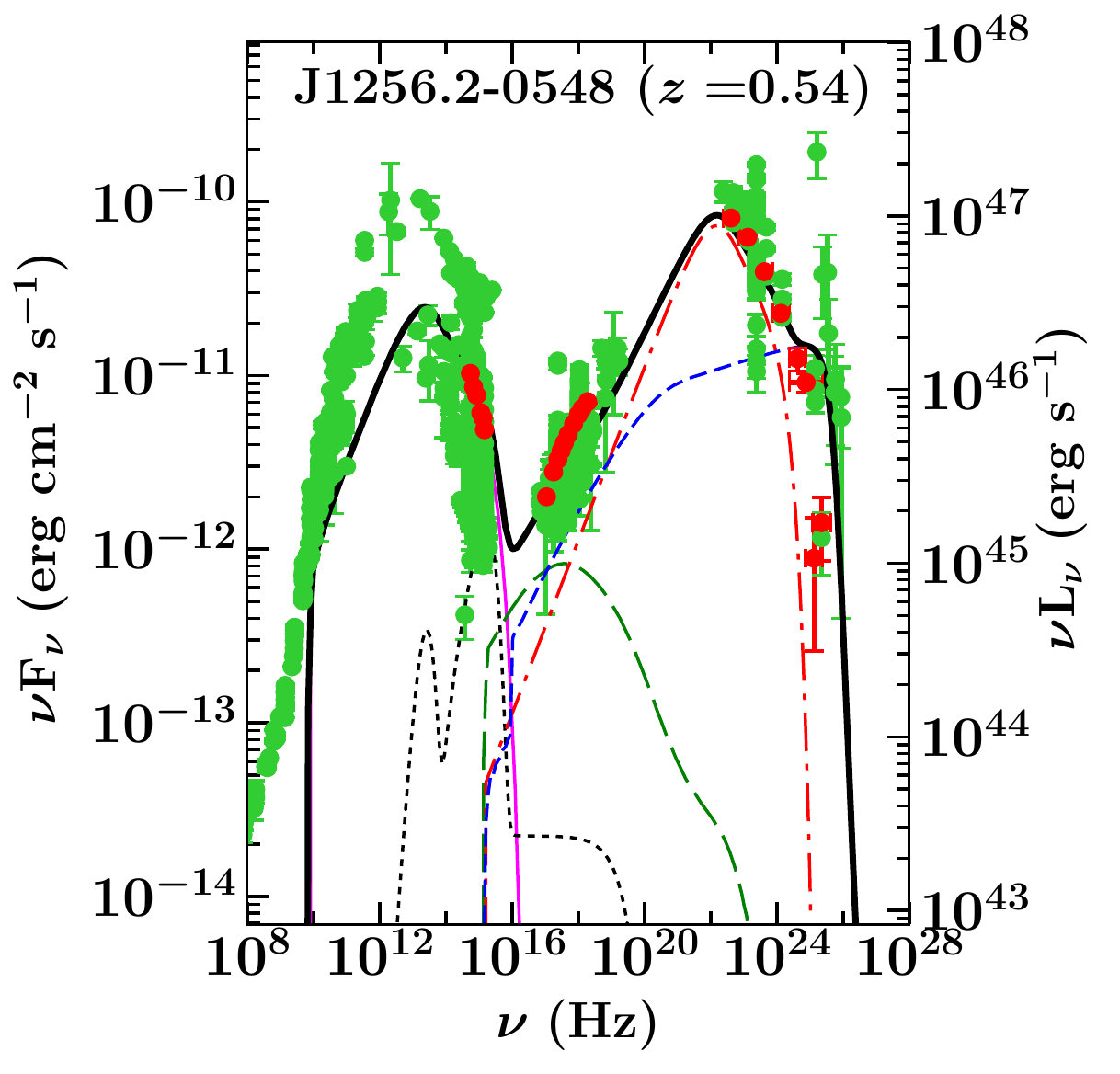}
\includegraphics[width=6.2cm]{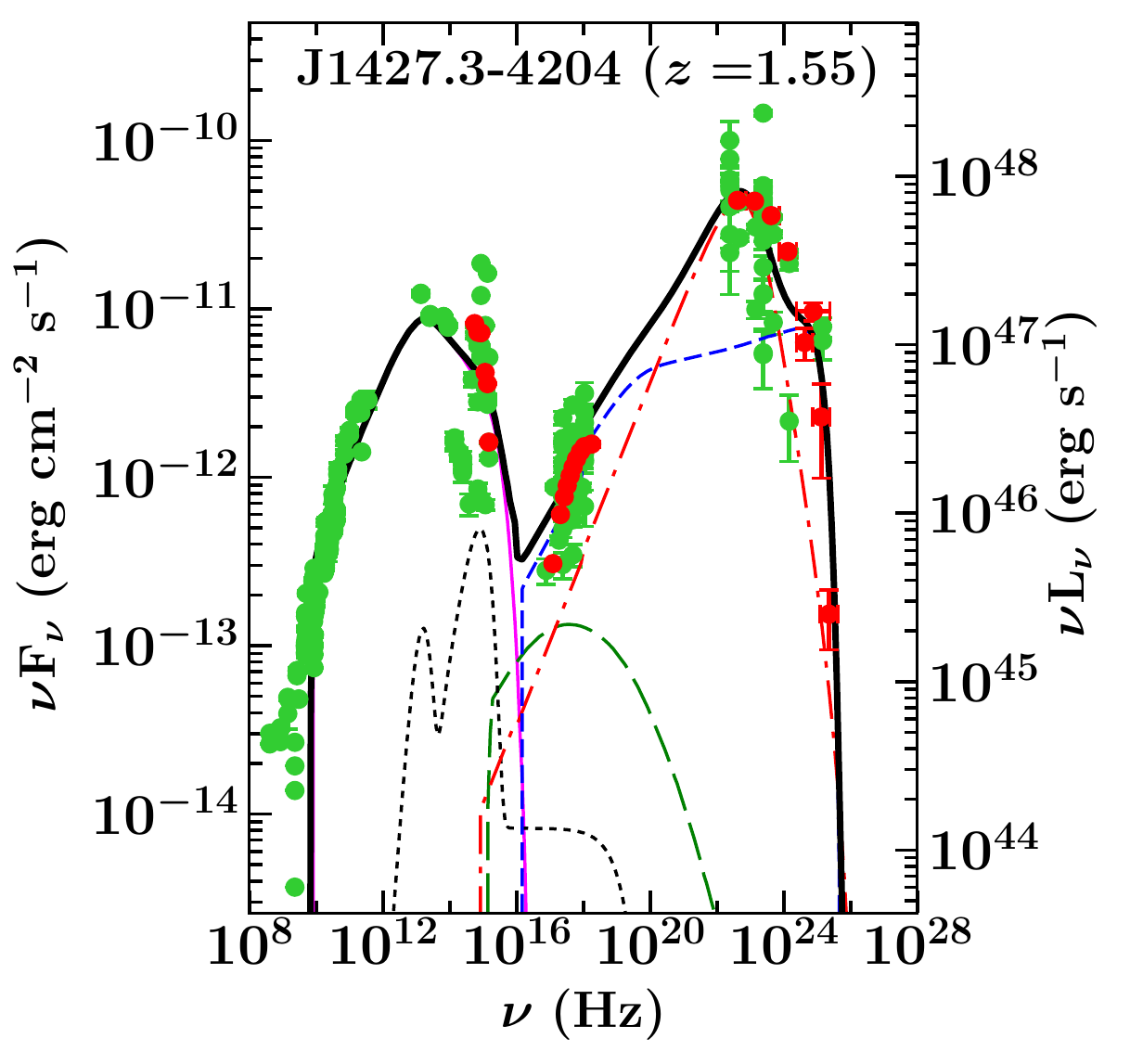}
\includegraphics[width=6.2cm]{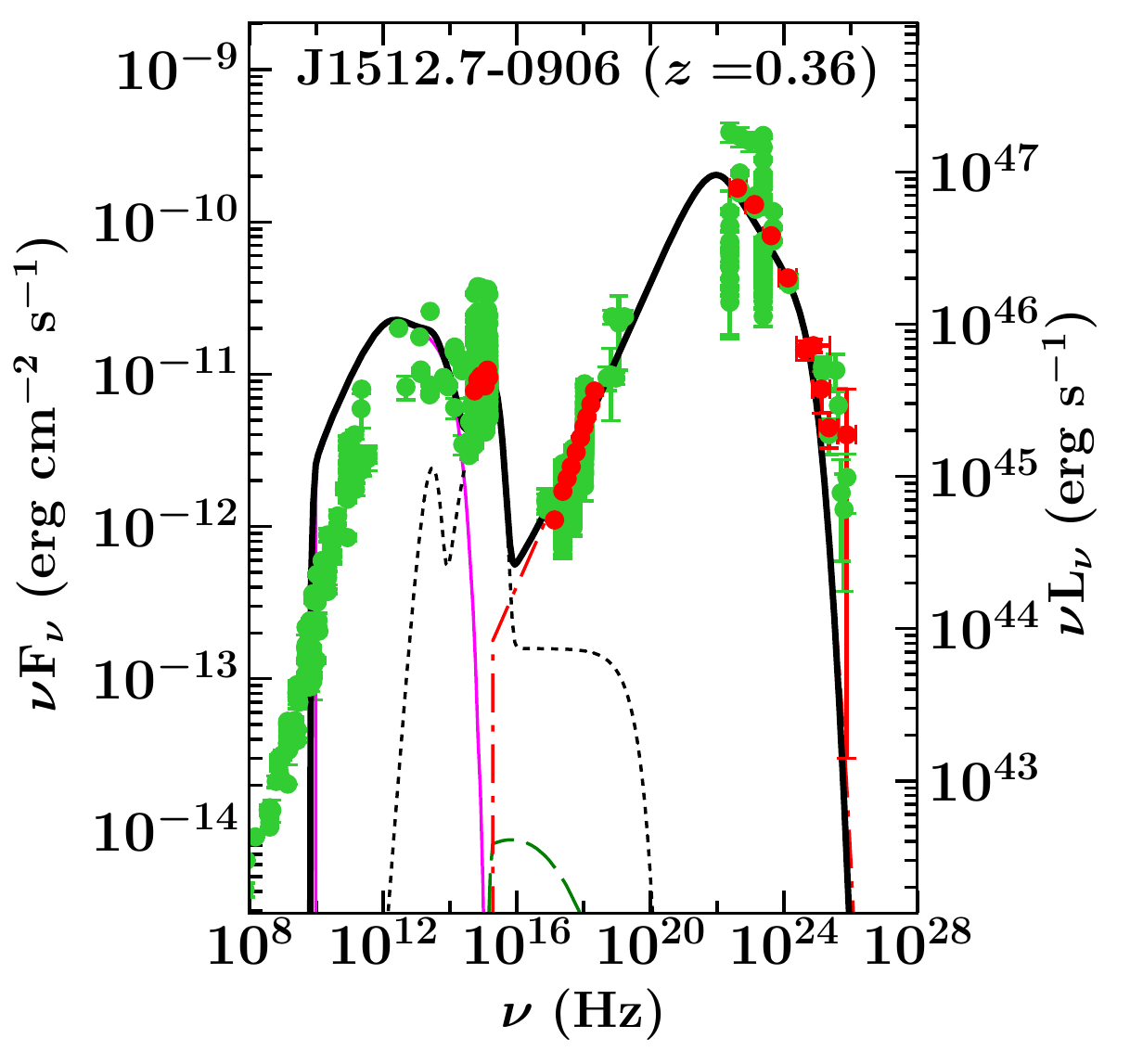}
}
\hbox{\hspace{3.0cm}
\includegraphics[width=6.0cm]{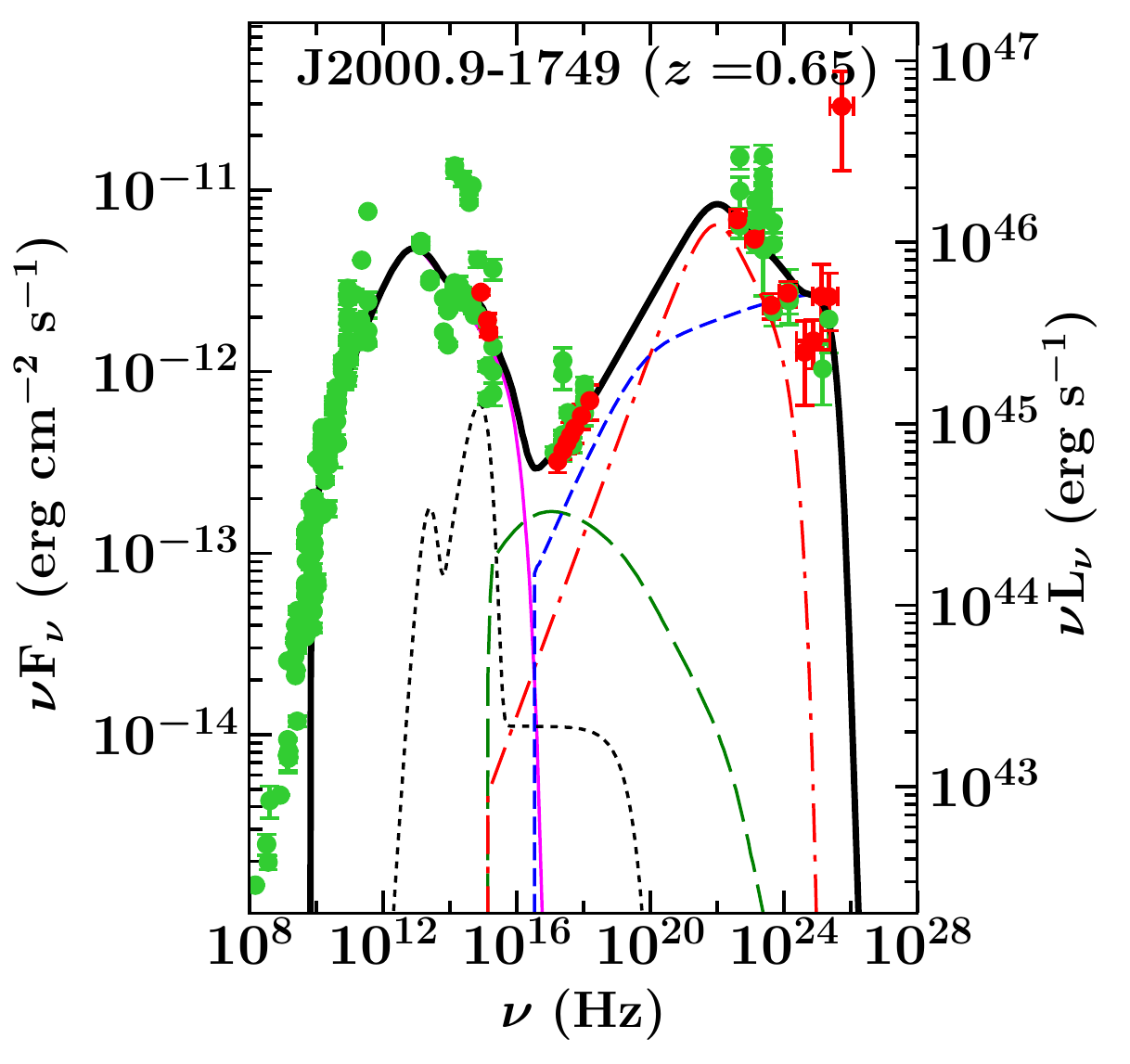}
\includegraphics[width=6.0cm]{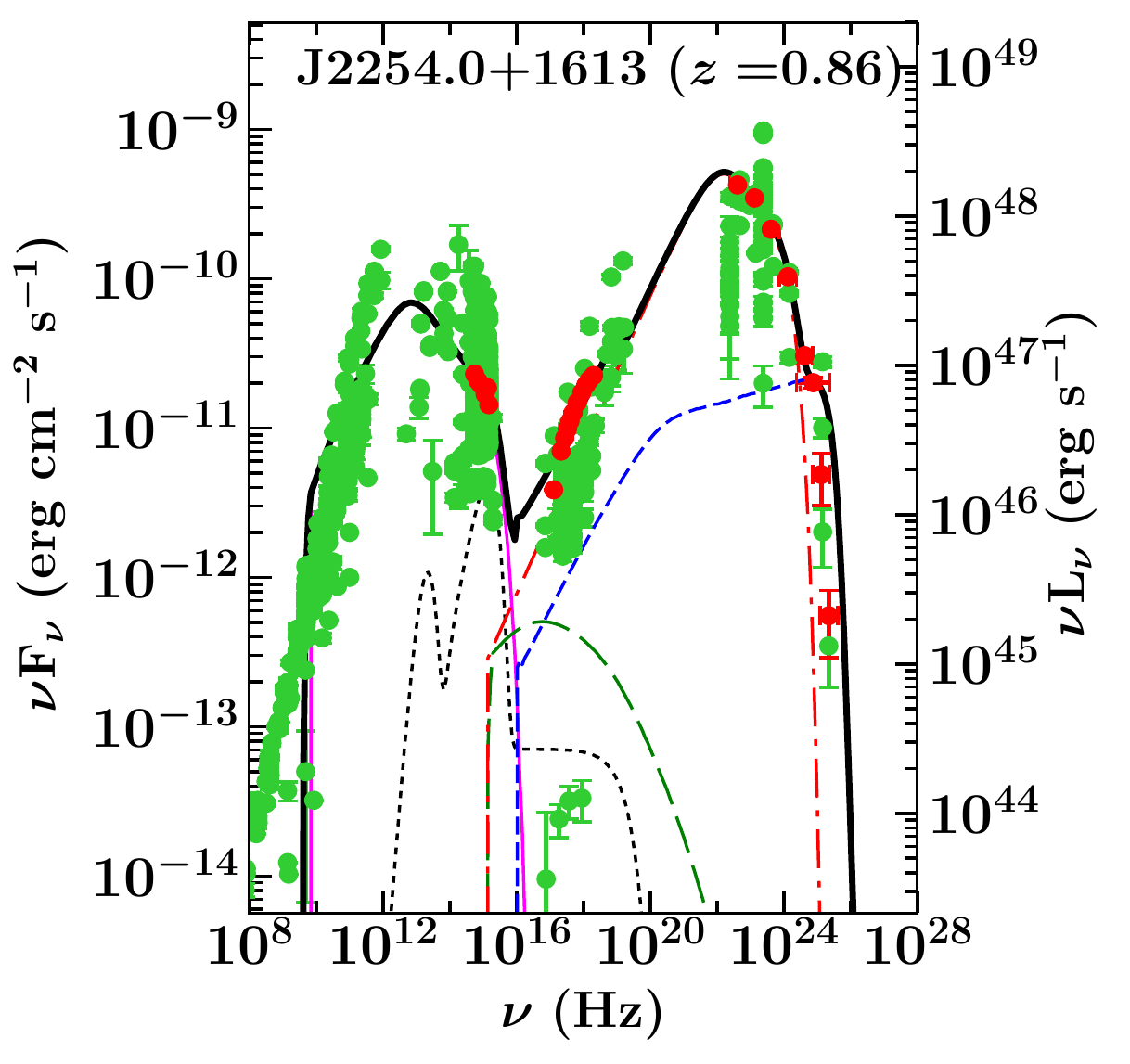}
}
\caption{Broadband SEDs of 2FHL FSRQs modeled with lepto-hadronic radiative model. Electron synchrotron and SSC models are shown 
with pink thin solid and green long dashed lines, respectively. On the other hand, proton synchrotron and SSC emissions are represented by 
red dash-dot and blue small dashed lines, respectively. Black thick solid line corresponds to sum of all the radiative components. 
Other information are same as in Figure \ref{fig_leptonic_sed}.\label{fig_hadronic_sed}} 
\end{figure*}

\begin{figure*}
\hbox{
\includegraphics[width=6.2cm]{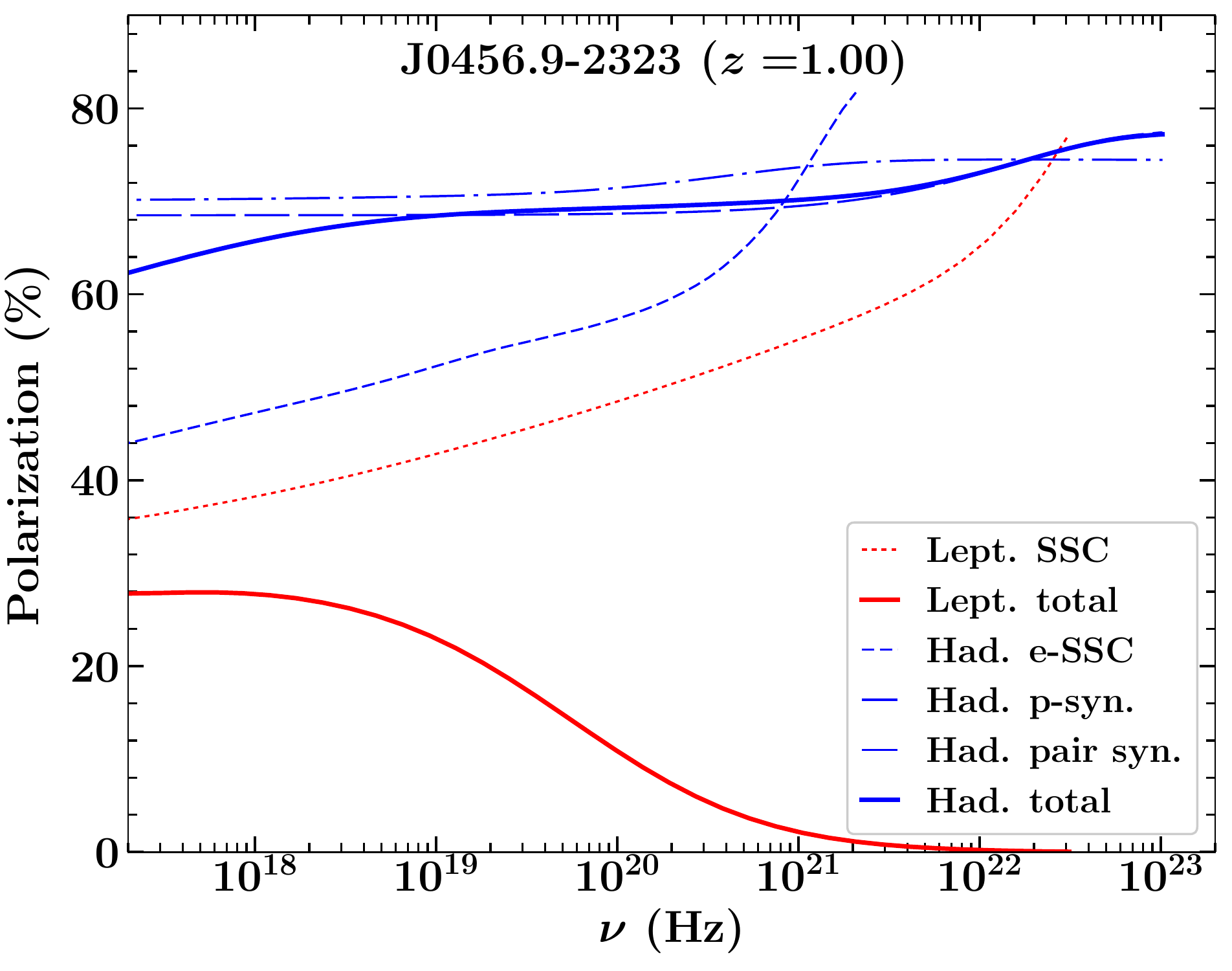}
\includegraphics[width=6.2cm]{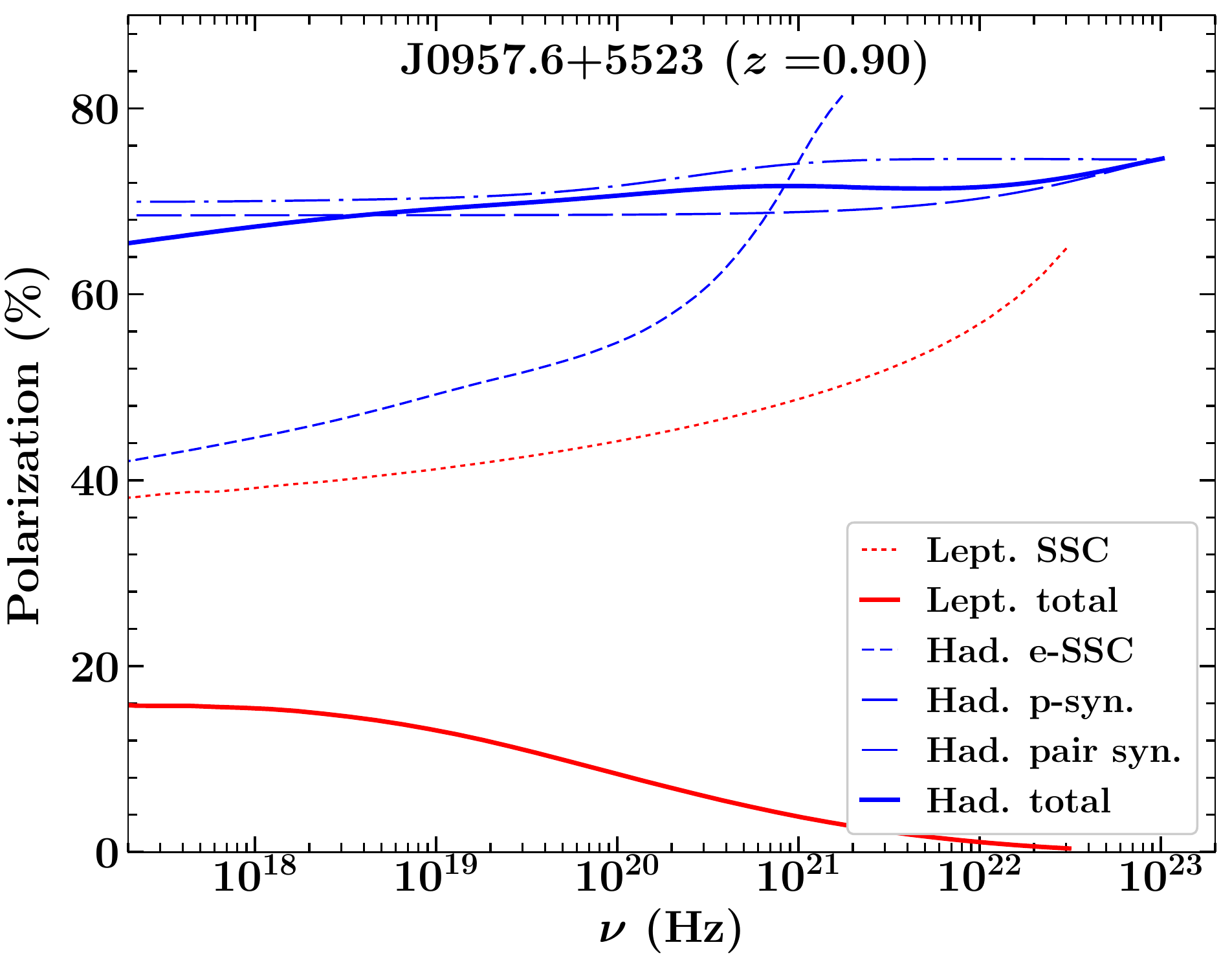}
\includegraphics[width=6.2cm]{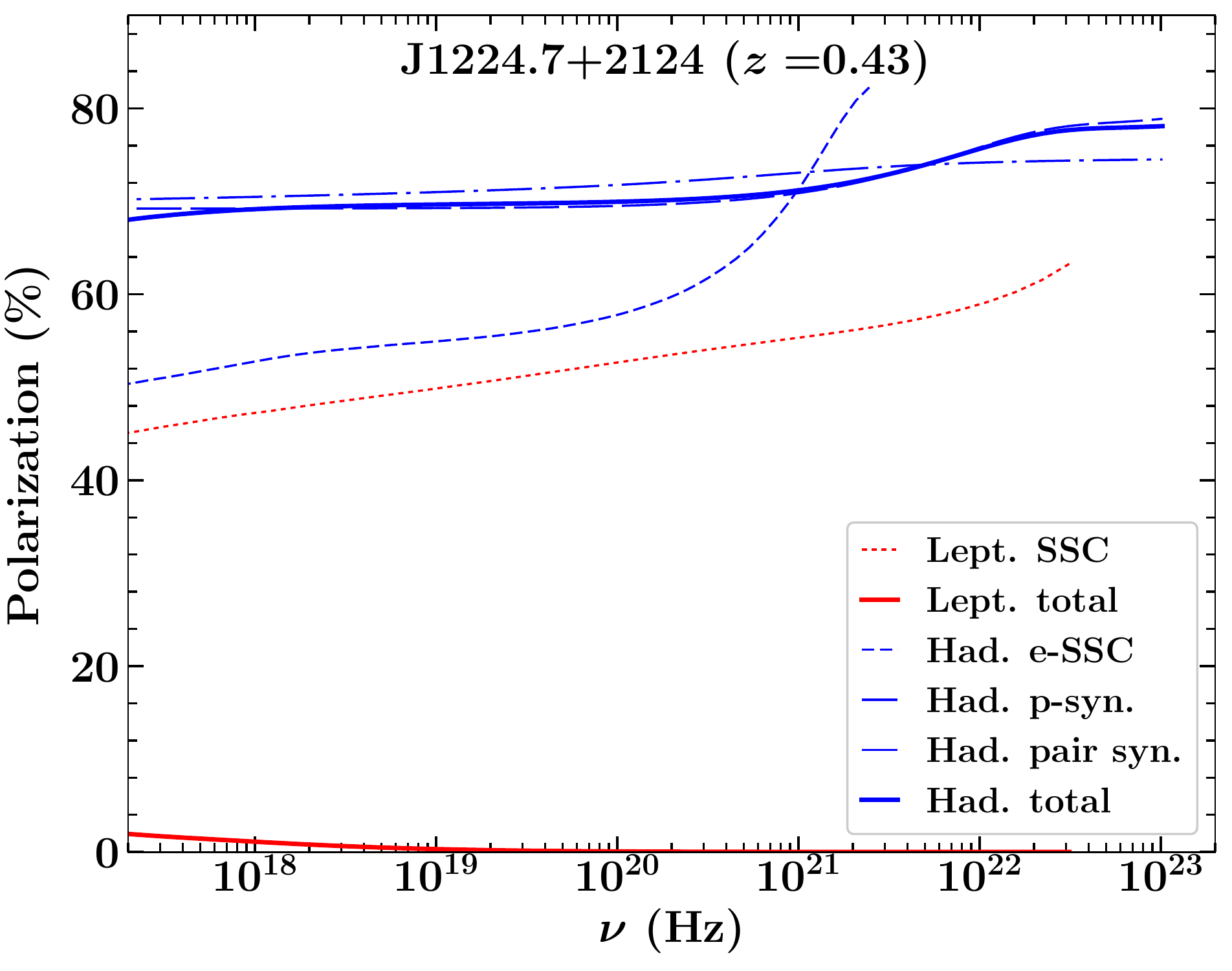}
}
\hbox{
\includegraphics[width=6.2cm]{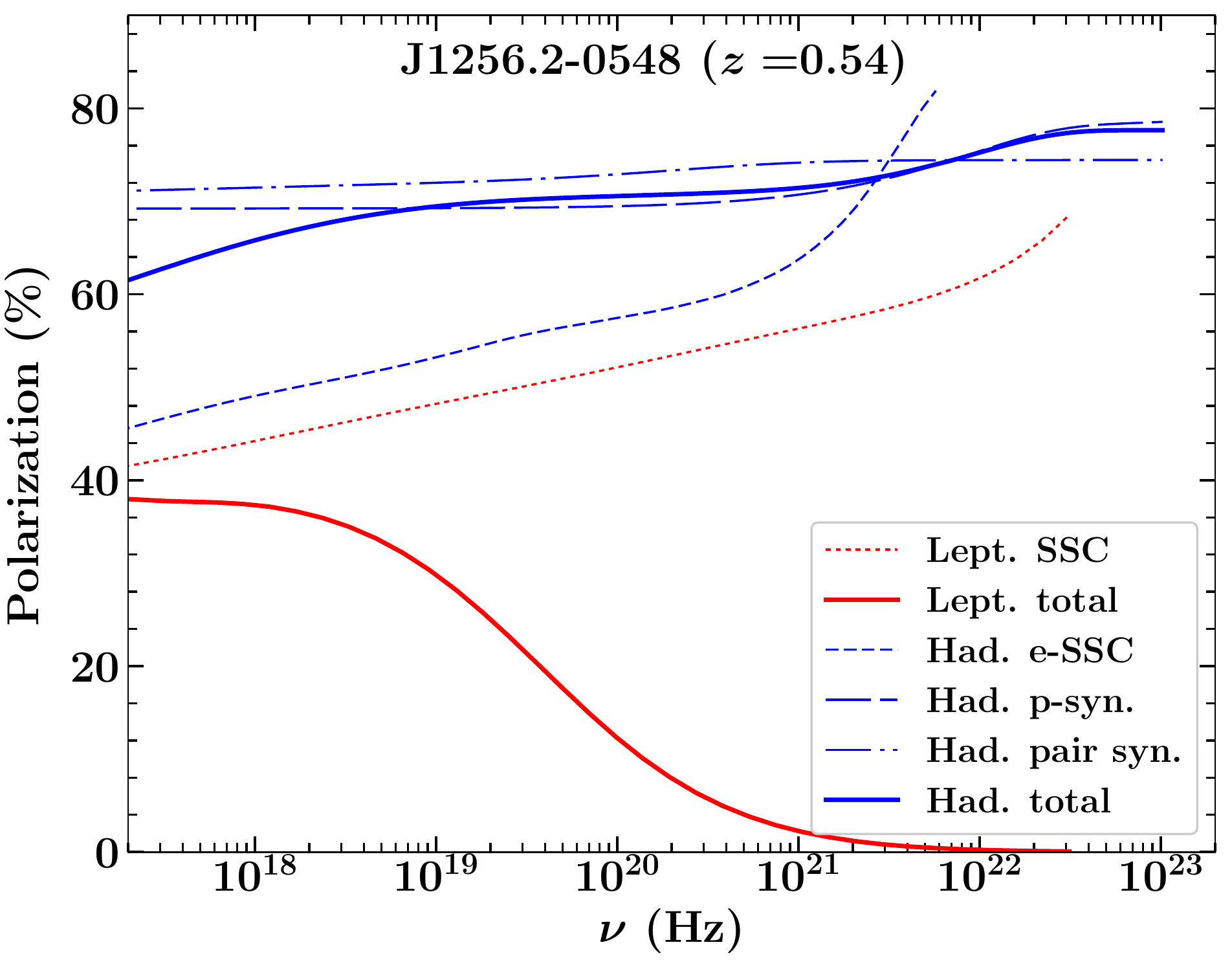}
\includegraphics[width=6.2cm]{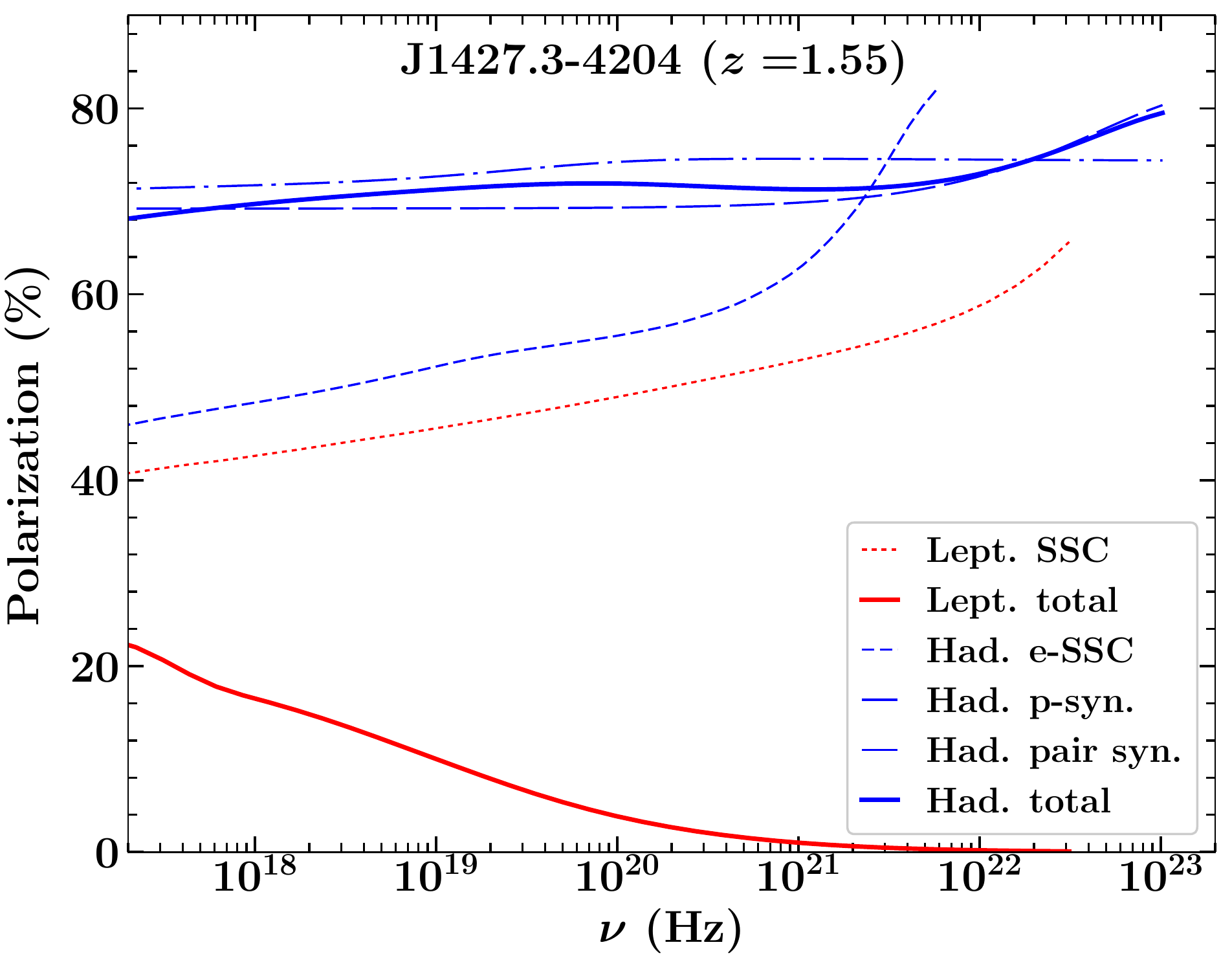}
\includegraphics[width=6.2cm]{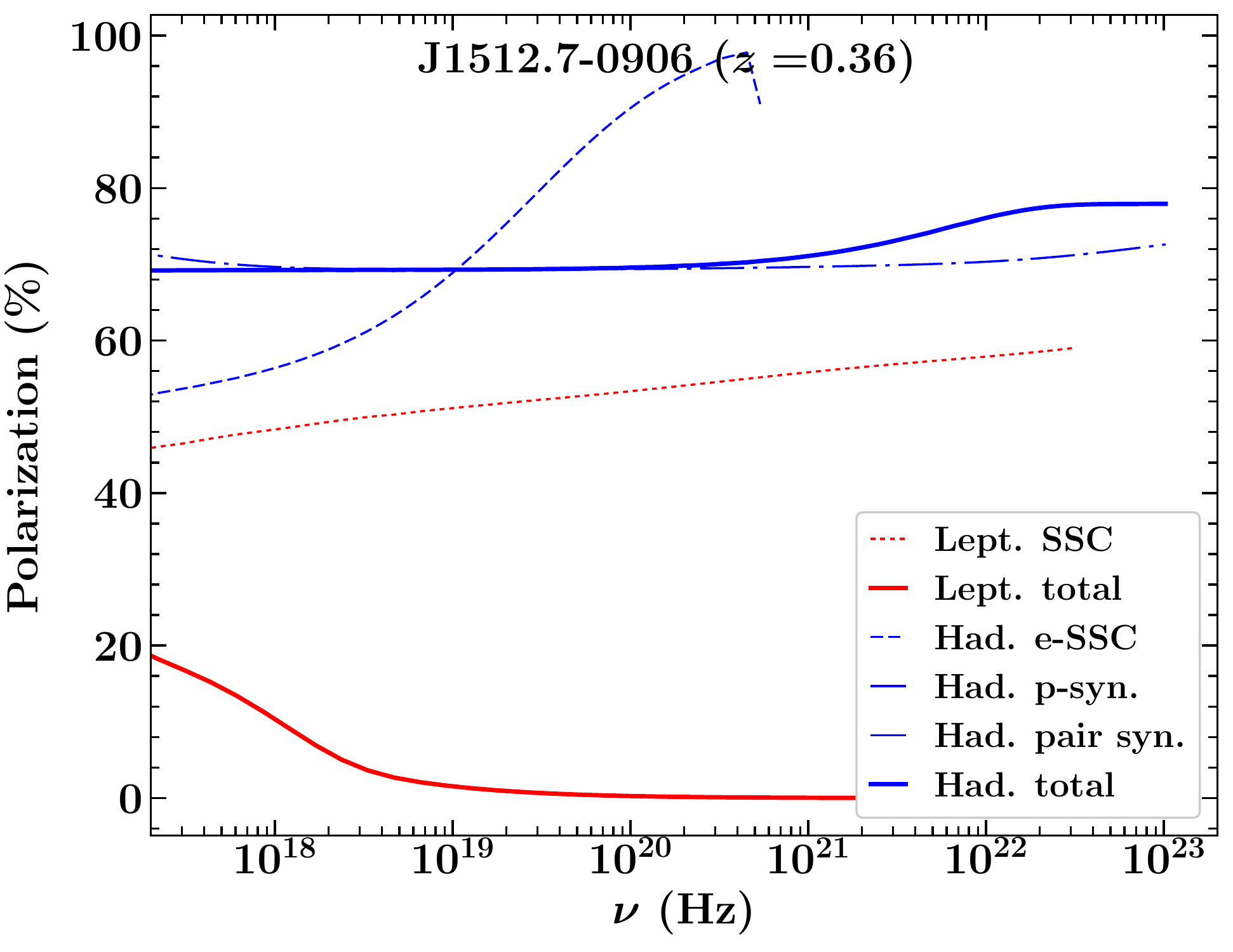}
}
\hbox{\hspace{3.0cm}
\includegraphics[width=6.0cm]{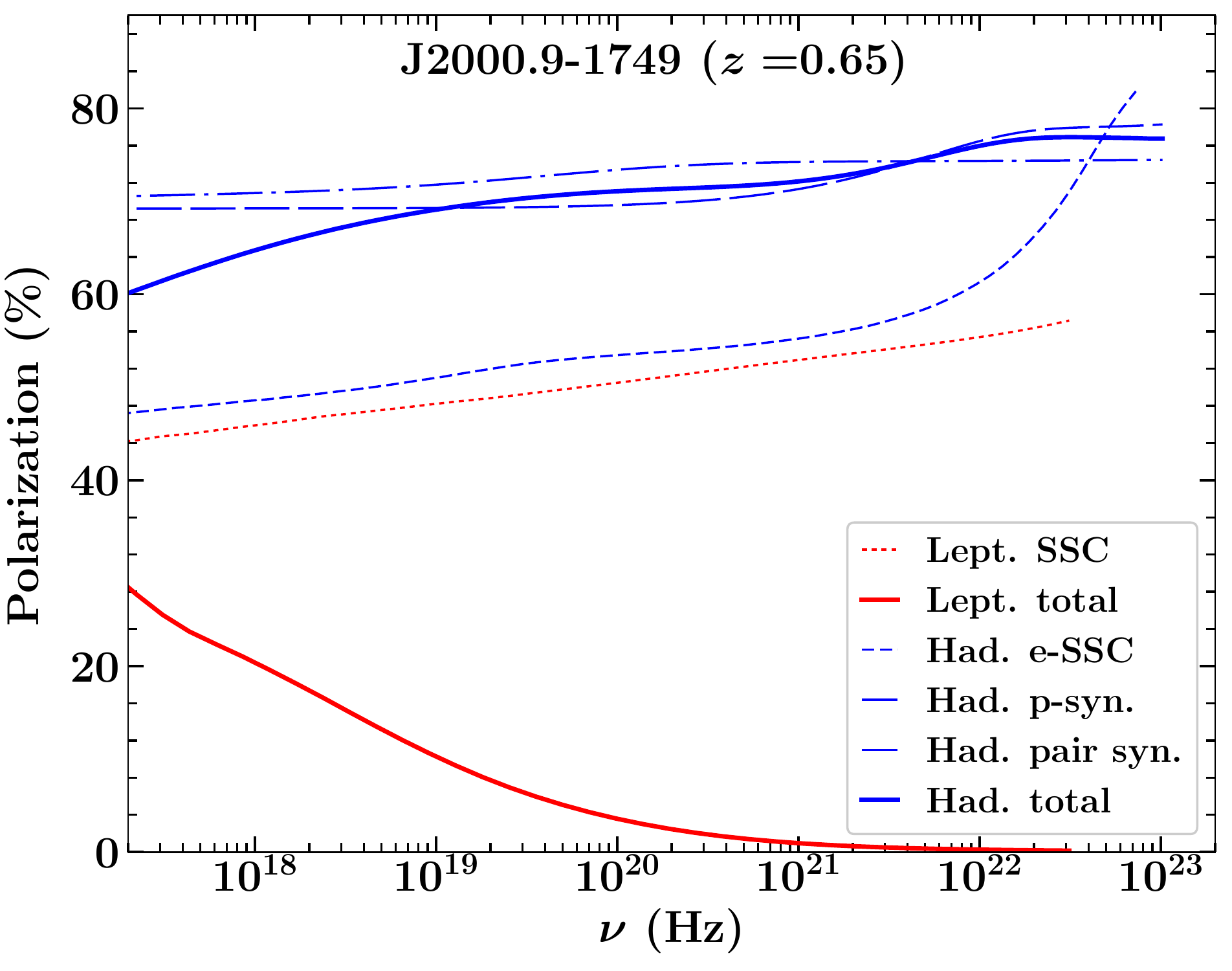}
\includegraphics[width=6.0cm]{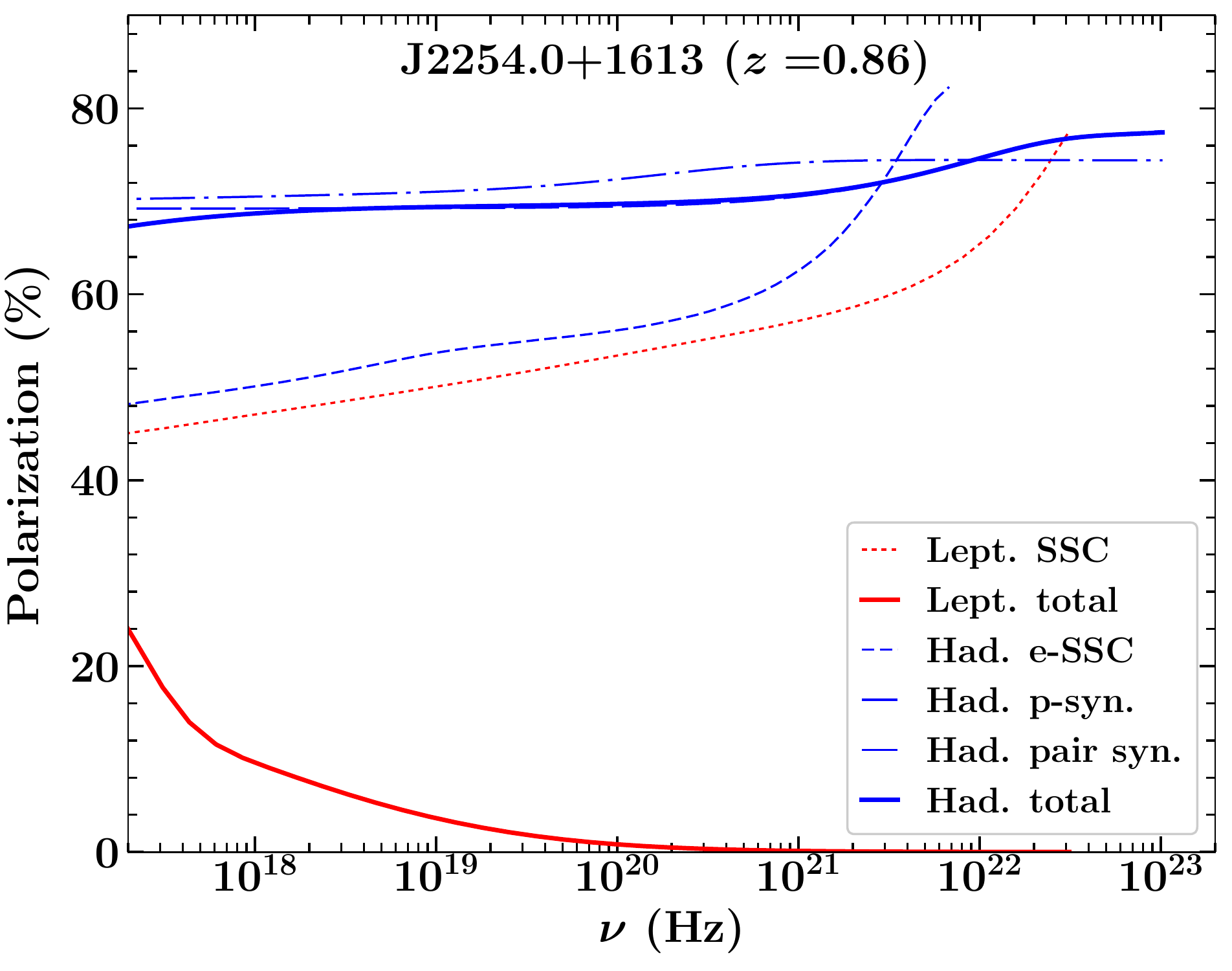}
}
\caption{The degree of X-ray polarization computed by considering both leptonic and hadronic emission scenarios. 
The components are appropriately labeled. See the text for details.\label{fig_pol}} 
\end{figure*}

We generate steady-state, i.e. time-averaged, broadband SEDs of all 8 FSRQs following the details outlined in Section \ref{sec3} and 
reproduce them using both leptonic and lepto-hadronic emission scenarios. The leptonic model SEDs are presented 
in Figure \ref{fig_leptonic_sed} and the associated SED parameters are given in Table \ref{tab:leptonic_sed_param}. 
The results of the lepto-hadronic SED modeling are shown in Figure \ref{fig_hadronic_sed} and we provide the physical 
parameters derived from the modeling in Table \ref{tab:hadronic_sed_param}. Using the results of the SED modeling, we compute the degree of 
X-ray and \gm-ray polarization predicted by the leptonic and hadronic models. The results are shown in 
Figure \ref{fig_pol}. Table \ref{tab:pol} provides our prediction about the degree of polarization that would be detected 
from sources under study at 1 keV and at 1 MeV energies, correcting for partially ordered magnetic field as described in Section \ref{xpol}.
\subsection{Leptonic Modeling}

Our SED modeling procedure does not involve any statistical fitting method and hence there 
could be possible degeneracy in the SED parameters. However, depending on the quality
of the observations, the SED parameters are reasonably constrained. Before the modeling, 
we fix to the following parameters either due to a priori knowledge
or based on physical considerations: $M_{\rm BH}$ and $L_{\rm disk}$ (Section \ref{sec5}), 
$\gamma_{\rm min}$, $\theta_{\rm view}$ and the fraction of 
$L_{\rm disk}$ reprocessed by the BLR, torus and the X-ray corona 
\citep[e.g.,][]{2005AJ....130.1418J,2009MNRAS.397..985G,2013AJ....146..120L}. 
This fixes the radiative energy densities used for EC calculation.
 Among eight free parameters: $N_0, s1, s2, B, R_{\rm diss}, \Gamma, \gamma_{\rm b},$ 
 and $\gamma_{\rm max}$, the slopes of the electron energy distribution, $s1$ and $s2$, 
 can be constrained from the shapes of the X-ray and \gm-ray SEDs \citep[see also,][]{2017ApJ...851...33P}. 
Whenever the optical spectrum is found to be synchrotron dominated, it provides further 
constrains to the high-energy slope $s2$. We determine the size of the 
emission region by assuming it to cover the entire jet cross-section whose semi-opening 
angle is assumed as 0.1 rad. The Compton dominance, which
is the ratio of the high-to-low-energy humps, enables us to determine the ratio of the radiation 
to magnetic energy density, $U_{\rm rad}/U_{\rm mag}$, and hence 
constrains the location of the emission region. This is because, in our model, these quantities 
are a function of $R_{\rm diss}$. Furthermore, for a major fraction of sources, 
we find the optical-UV emission to be synchrotron dominated, which suggests a 
high level of SSC. Once the synchrotron spectrum is determined 
from the optical-UV SED, a high level of SSC emission demands a relatively low magnetic 
field $B$. This is because, with smaller $B$, higher number of electrons are
needed to achieve the same synchrotron flux level. As the electrons number density goes 
up, so the SSC flux level. Similarly, both SSC and EC fluxes (constrained from the 
observed X-ray and \gm-ray SEDs) also provide a tight constraint to $\Gamma$. An increase 
in $\Gamma$ (or equivalently Doppler factor $\delta$) 
decreases the electron number density since fewer electrons are needed to maintain a given 
synchrotron flux level, thus decreasing both SSC and EC. However, overall EC 
flux increases since the enhancement in $\delta$ also lead to increase in the comoving-frame 
external photon densities \citep[][]{1995ApJ...446L..63D}.

\begin{figure*}
\hbox{
\includegraphics[width=6.2cm]{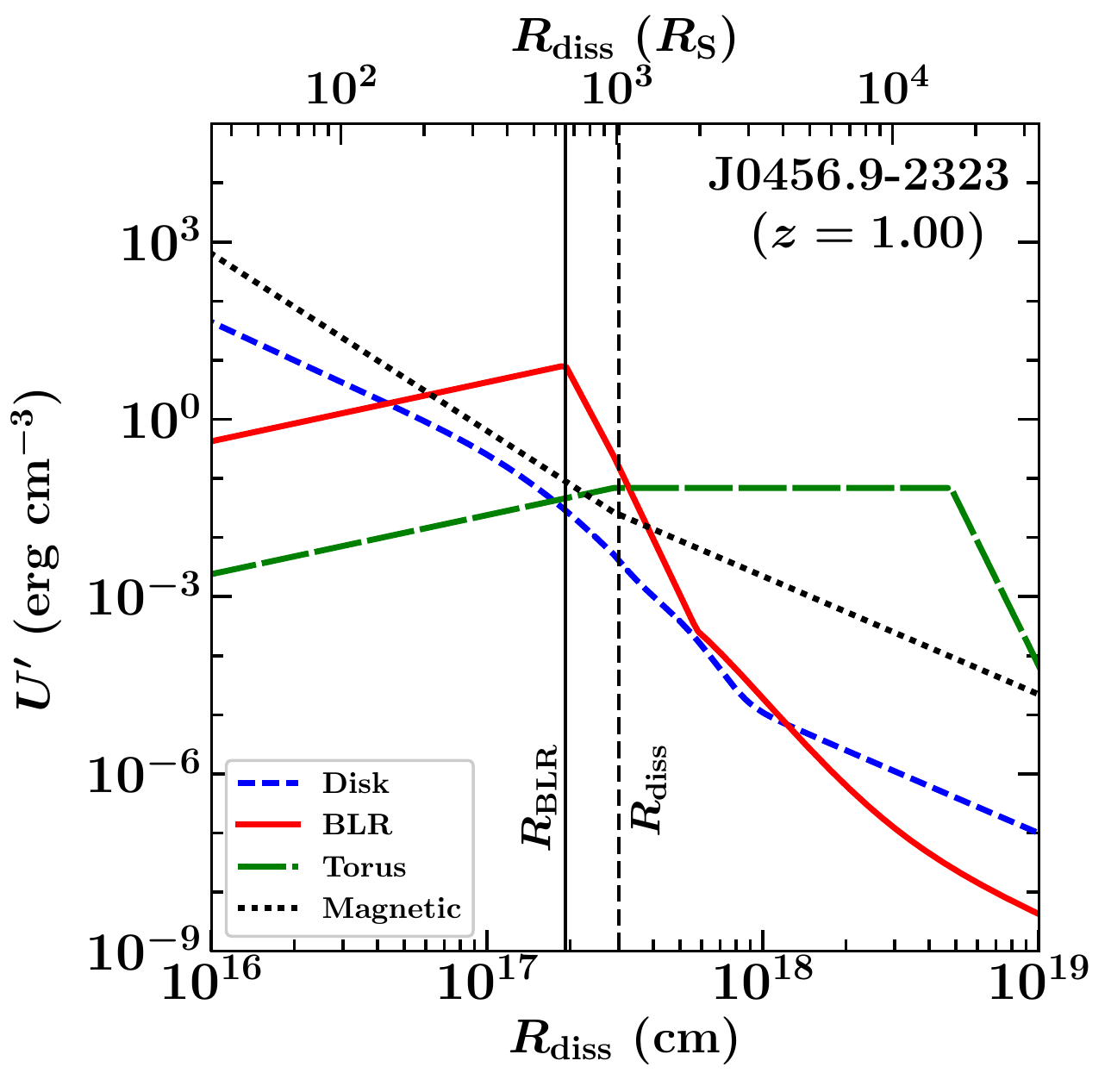}
\includegraphics[width=6.2cm]{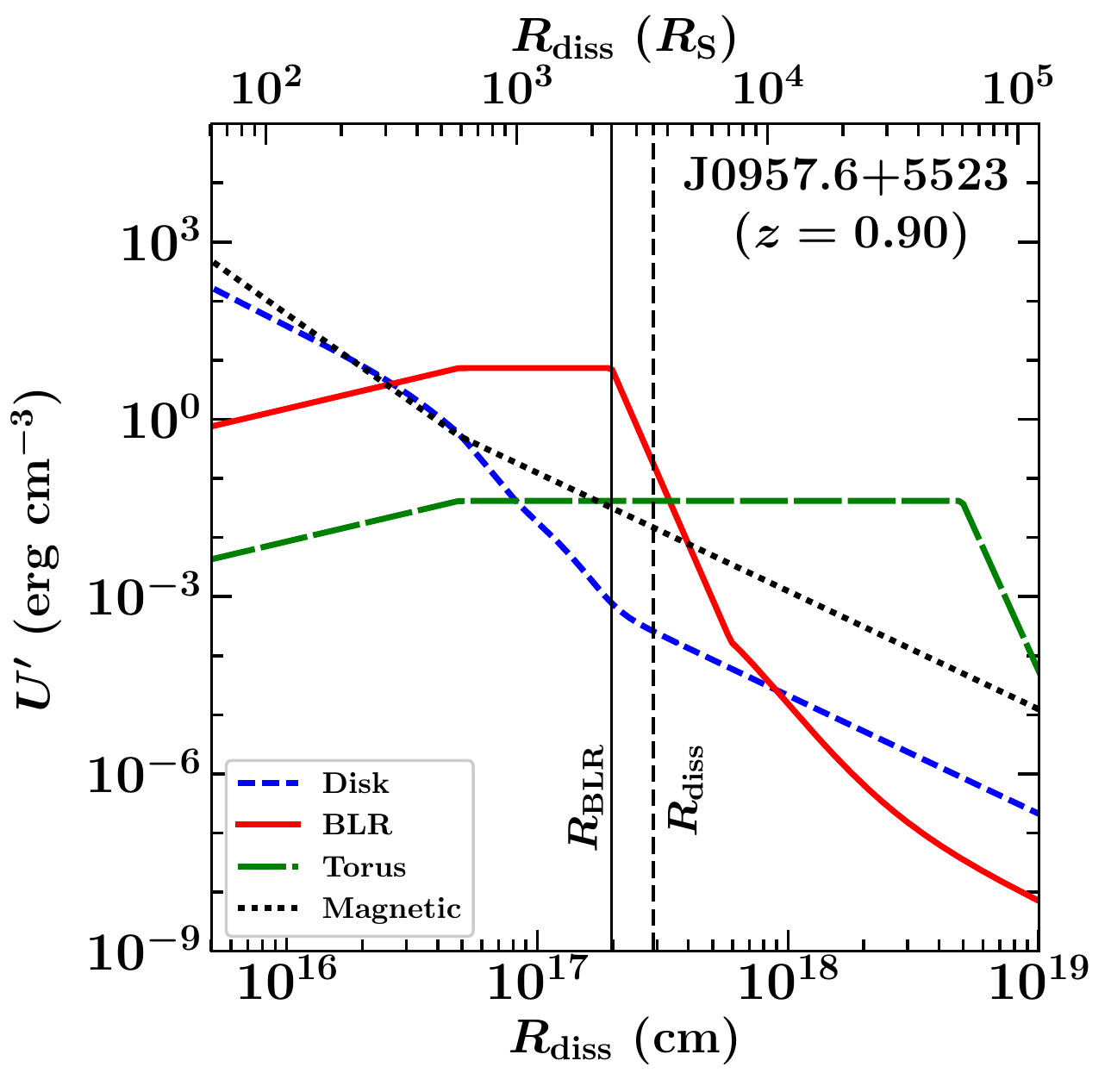}
\includegraphics[width=6.2cm]{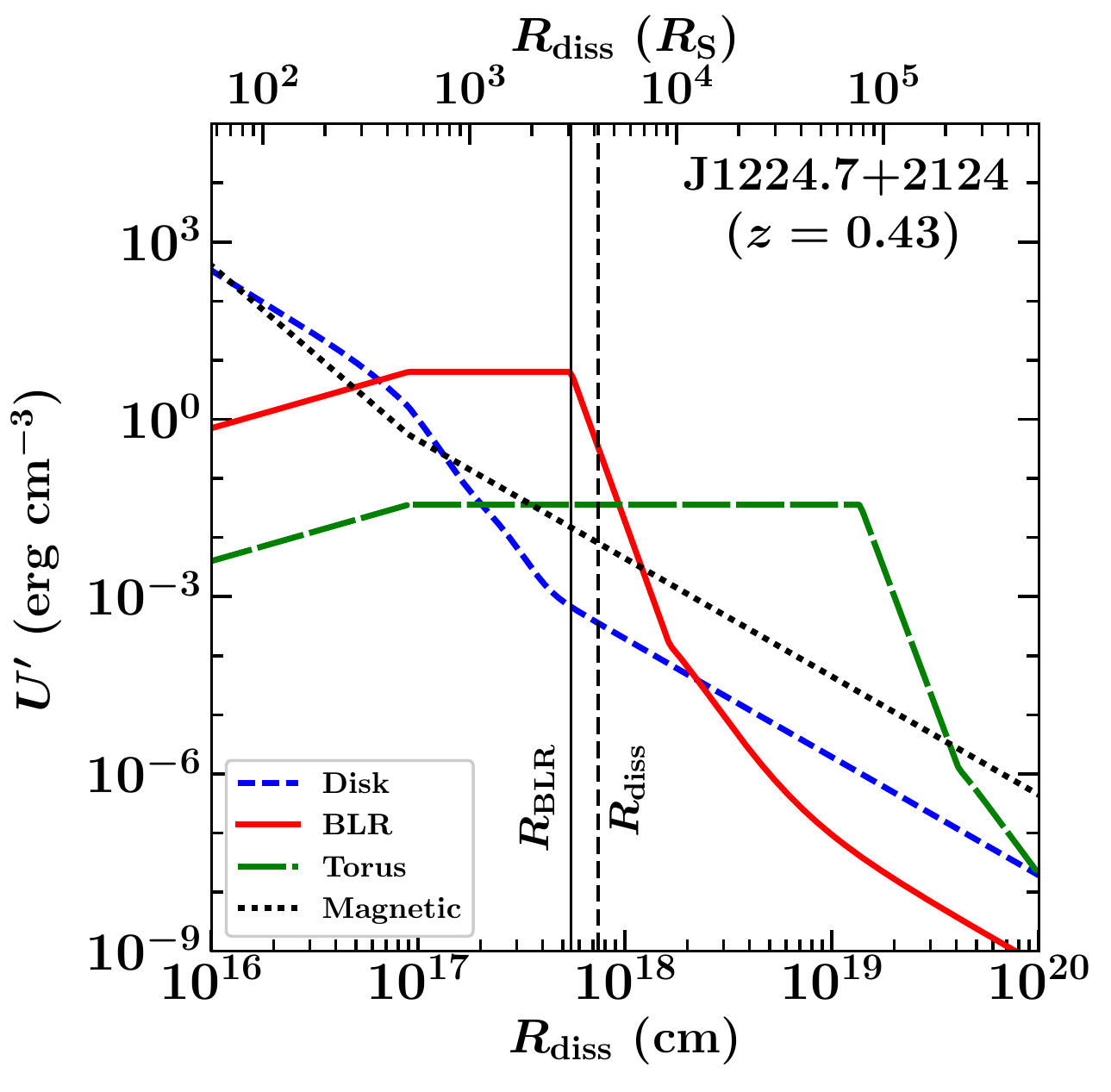}
}
\hbox{
\includegraphics[width=6.2cm]{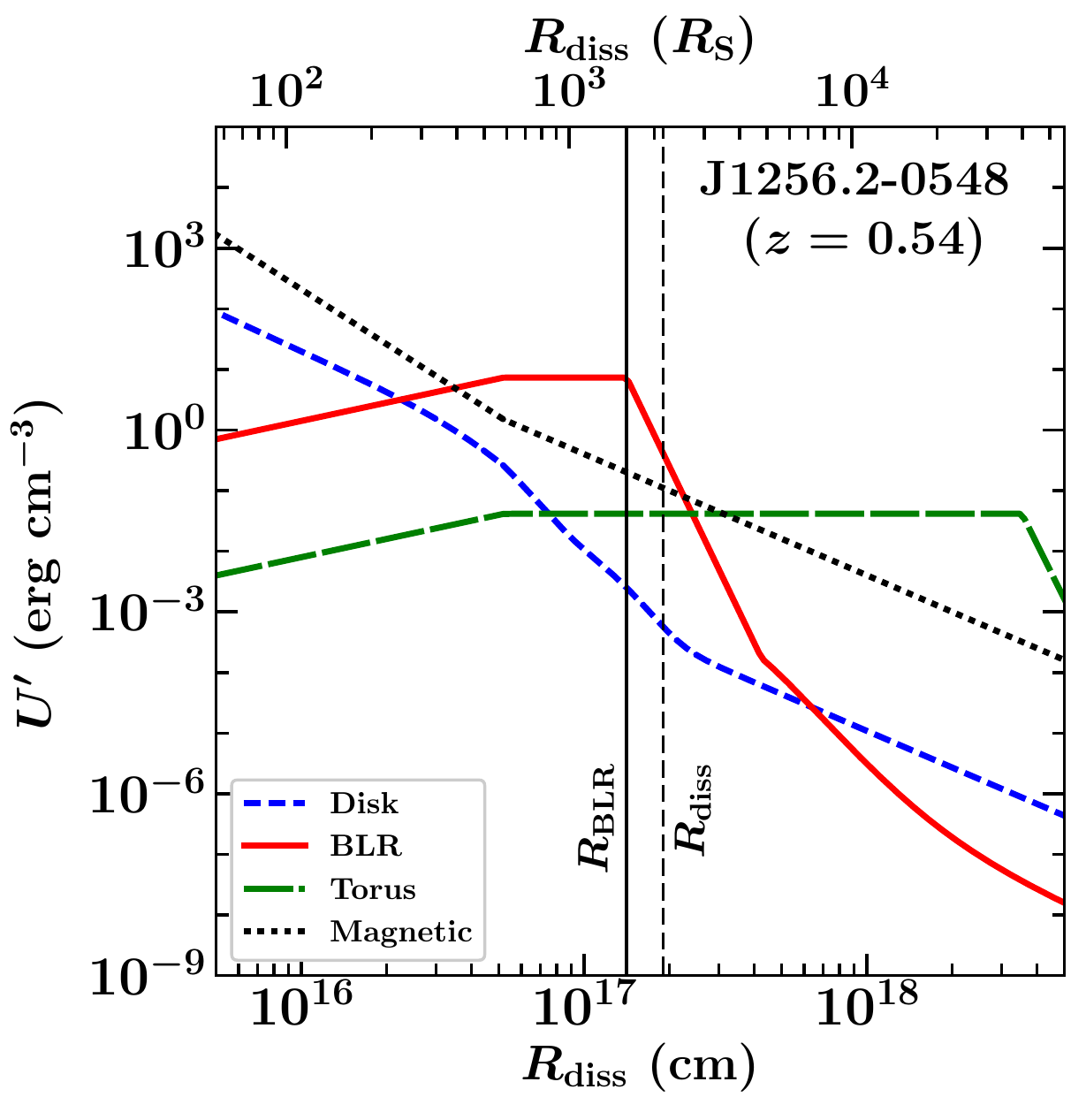}
\includegraphics[width=6.2cm]{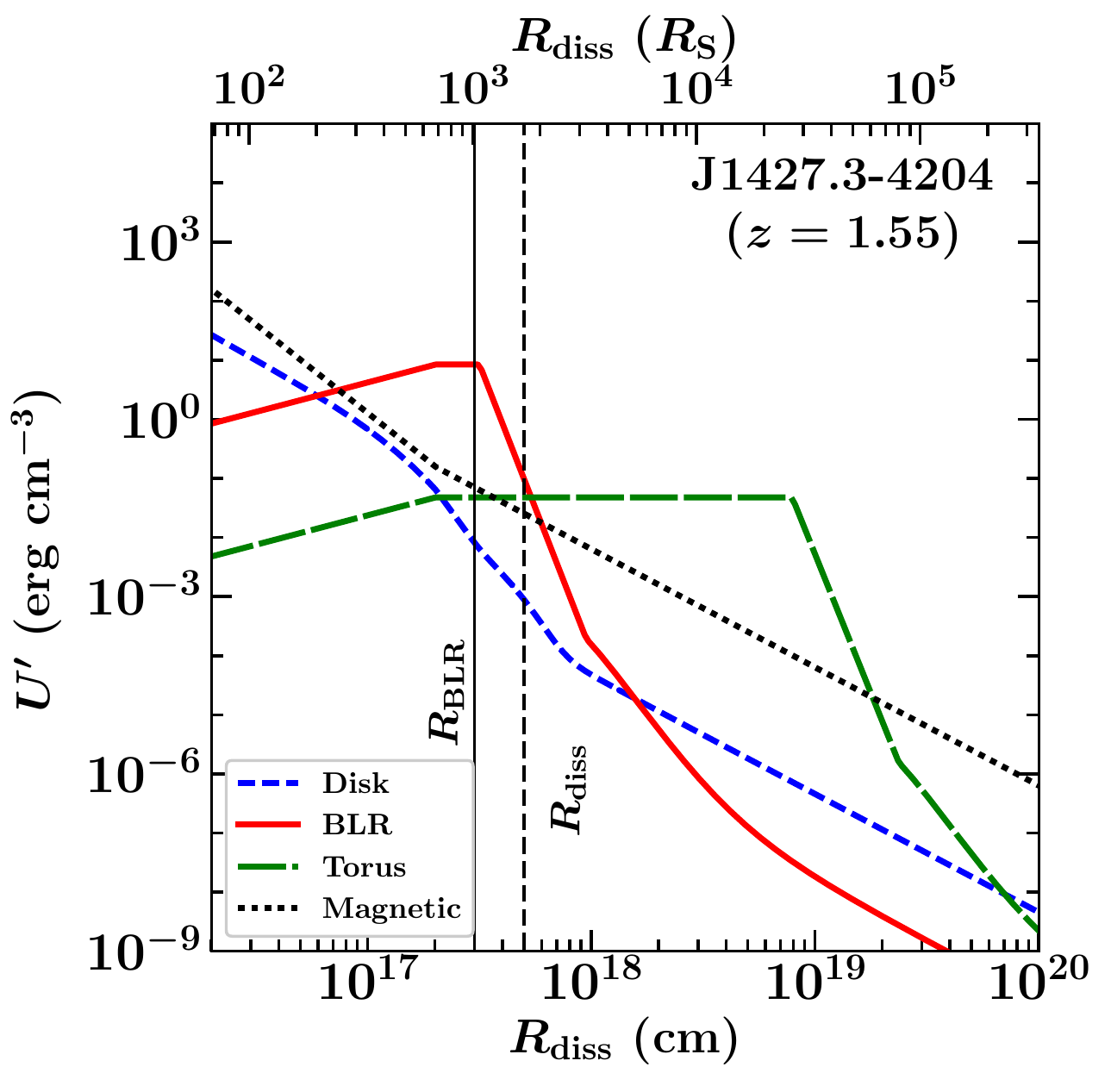}
\includegraphics[width=6.2cm]{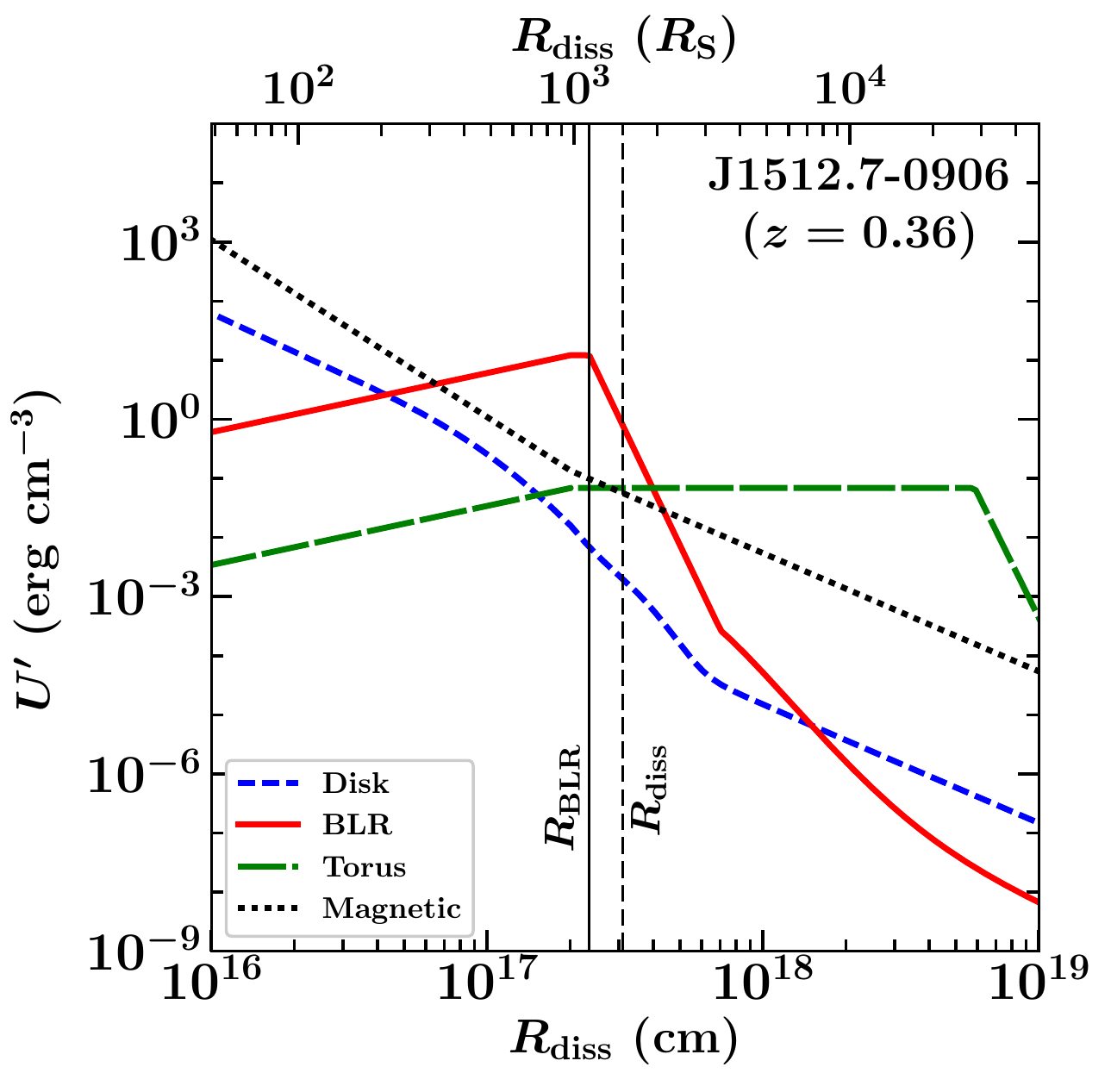}
}
\hbox{\hspace{3.0cm}
\includegraphics[width=6.0cm]{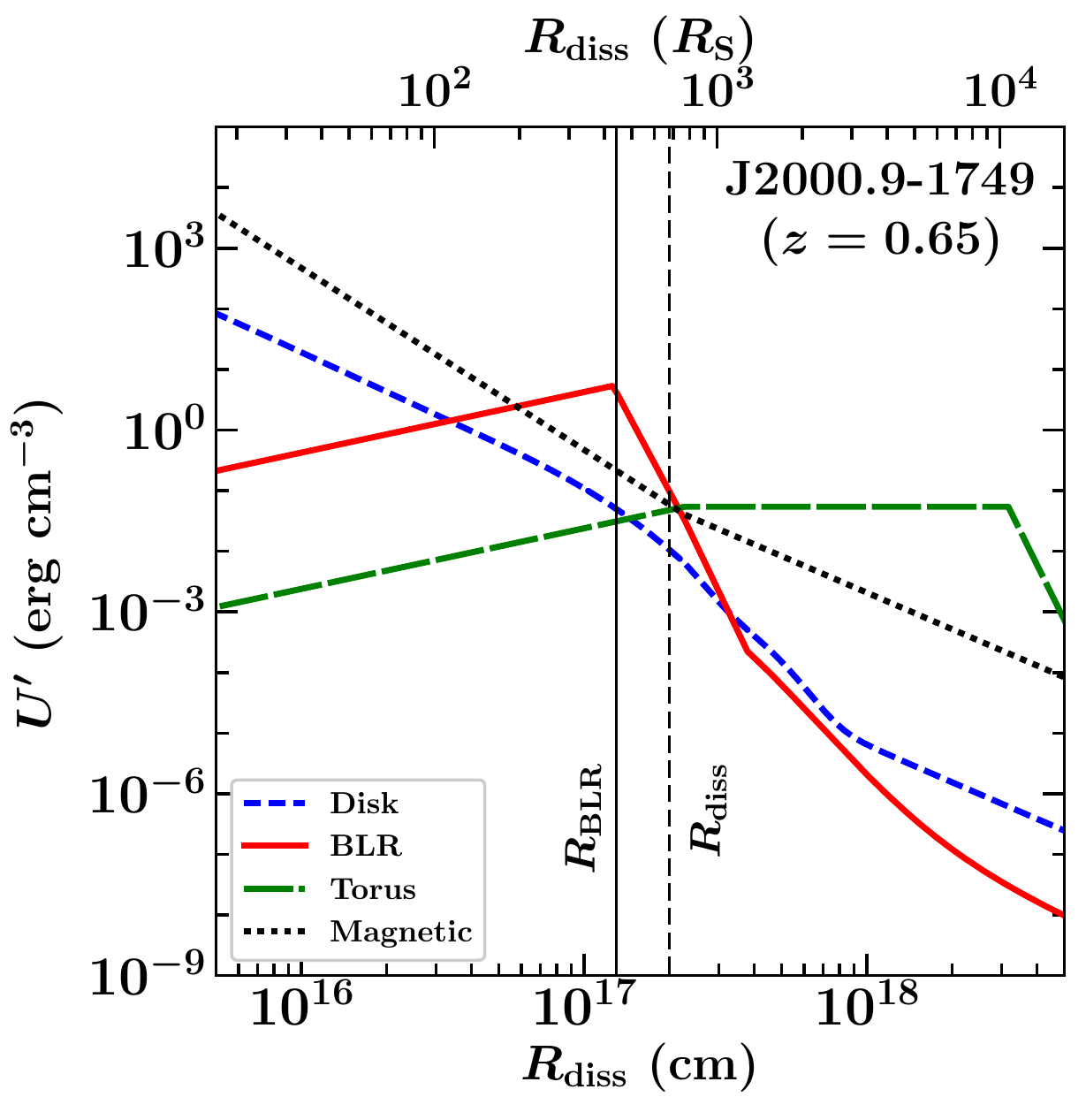}
\includegraphics[width=6.0cm]{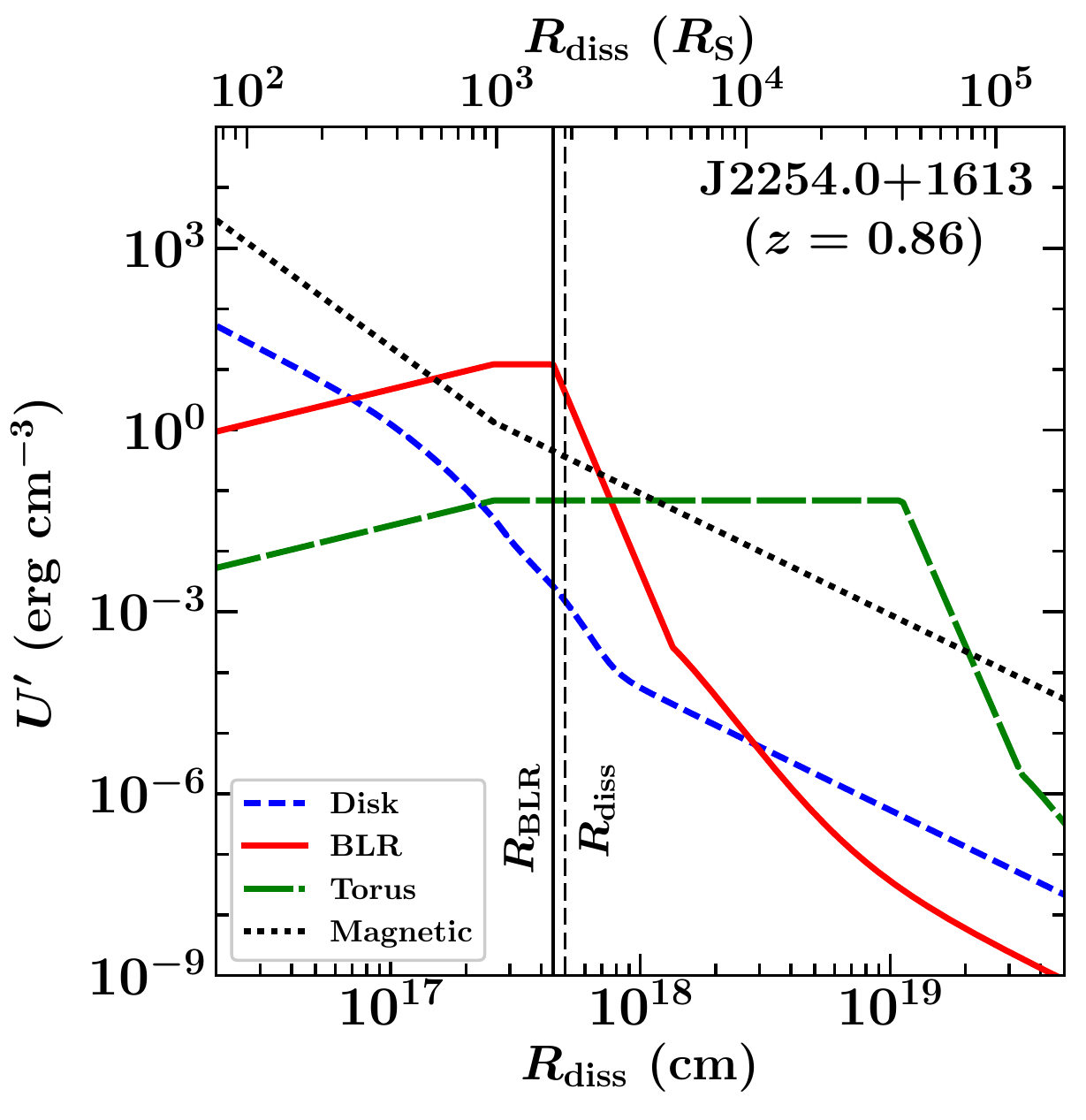}
}
\caption{Dissipation distance dependence of the comoving frame radiative energy densities. Vertical black solid and dashed lines represent the inner boundary of the BLR and the location of the emission region as inferred from the leptonic SED modeling, respectively. Note that the bulk Lorentz factor $\Gamma$ varies as min[(R$_{\rm diss}$/3R$_{\rm S})^{1/2}, \Gamma_{\rm max}$], i.e. an accelerating jet followed by a constant moving phase \citep[see,][]{2009MNRAS.397..985G}.}\label{fig_leptonic_ene_den}
\end{figure*}

The leptonic modeling reasonably reproduces most of the observed SEDs of 2FHL FSRQs. The accretion disk emission 
is observed at optical-UV frequencies in J1224.7+2124, J1427.3$-$4204, and J1512.7$-$0906. On the 
other hand, the optical-UV emission in the remaining 5 sources is dominated by the synchrotron radiation. 
The high-energy spectra of these objects can be reproduced by a combination of SSC and EC processes. In our 
model, the radiative energy densities are a function of the dissipation distance from the central black hole. 
Our modeling parameters suggest that the emission region is outside of the inner boundary of the BLR 
(Table \ref{tab:leptonic_sed_param}). This is evident from the fact that all of the sources are detected 
above 50 GeV, which requires the effect of \gm-ray absorption by BLR photons to be negligible. However, the emission 
region is likely close to the BLR, because all the objects have a Compton dominance significantly larger than unity, implying 
that the external photon energy density dominates over the magnetic energy density. In 
Figure \ref{fig_leptonic_ene_den}, we present the variation of the comoving-frame radiative energy densities 
with distance from the central black hole and also show the location of the emission region as inferred from 
the modeling. We can see that both BLR and torus energy densities contribute to the observed \gm-ray emission, 
but the dominant fraction comes from the BLR \citep[see also,][for similar arguments]{2014ApJ...785..132J}. 
The EC peak frequency ($\nu^{\rm IC}_{\rm peak}$) is a further
diagnostic of the primary \gm-ray emission mechanism. In the Thomson regime, the peak of the IC component is

\begin{equation}
\nu^{\rm EC}_{\rm peak} \simeq \frac{\nu_{\rm seed}\Gamma^2\gamma_{\rm b}^{2}}{(1+z)}
\end{equation}
where $\nu_{\rm seed}$ is the characteristic frequency of seed photons for EC mechanism ($\sim10^{15}$~Hz or 
$\sim10^{13}$~Hz for BLR and torus photons, respectively).
The derived $\gamma_{\rm b}$ has a rather low value and therefore a higher energy seed photons
can explain the observed IC peak located around MeV$-$GeV energies, which is consistent with
BLR photons as the seed photons for IC. Moreover, above 50 GeV or so, interaction of BLR photons with jet electrons occurs
at Klein-Nishina energies, whereas, IC scattering of torus photons still remains within the 
Thomson regime \citep[e.g.,][]{2013ApJ...771L...4C,2014ApJ...782...82D}.
To summarize, the emission region is probably located at the outer edge 
of the BLR.

A Compton dominated SED, as observed for all of the sources, indicates a considerably smaller magnetic jet power compared to the 
kinetic luminosity. This, in turn, hints a low magnetization of the emission region \citep[e.g.,][]{2015MNRAS.449..431J}. Furthermore, a comparison of the the kinetic jet 
power with the accretion disk luminosity (Table \ref{tab:leptonic_sed_param}) suggests the jet power to exceed 
the accretion luminosity, which is now a well-known fact \citep[][]{2014Natur.515..376G}. However, note that the jet power computation is a strong
function of assumed number of protons per electron. If a few pairs are present in the emission region, thus reducing the number density of protons, the budget of
the jet power will decrease \citep[see, e.g.,][]{2016ApJ...831..142M,2017MNRAS.465.3506P}. Moreover, \citet[][]{2016MNRAS.457.1352S} 
proposed a spine-sheath structured jet which predicts a lower jet power with respect to that computed by assuming a uniform single-zone emission.

\subsection{Hadronic Modeling}

The hadronic model produces similarly good fits compared to the leptonic model. In particular, it predicts higher flux beyond $50~\rm{GeV}$ than the leptonic model. Here, we fix the viewing angle and the Lorentz factors the same as the leptonic fits to reduce the number of free parameters. The self-consistent treatment of cooling effects employed in the hadronic fits promises less degeneracy in the model parameters than the leptonic fits. The cascading secondaries typically have softer spectra than the primary protons \citep{Boettcher13}. Given the very hard spectra from X-ray to $\gamma$-ray, all sources require a dominating proton synchrotron contribution for the high-energy spectral component. Therefore, the underlying proton spectrum is well constrained by the observed X-ray to $\gamma$-ray SED, which suggests a single power-law shape with a softer turnover near the high energy end. This turnover, however, is not consistent with the cooling break that are self-consistently treated in the hadronic code. Instead, we suggest that it is due to the particle acceleration. In practice, shock and magnetic reconnection can produce a power-law shaped spectrum with a turnover at the high energy end. Numerical simulations have shown that this turnover is not an exponential cutoff, but rather like a softer power-law that extends to about one decade in the particle Lorentz factor \citep[e.g.,][]{Sironi11,Guo16}. Our proton spectral parameters are generally consistent with the numerical particle acceleration simulations. Therefore, if the $\gamma$-ray of these hard spectrum {\it Fermi} sources is produced through proton synchrotron, then the highest protons should not be efficiently cooled within one light crossing time. Based on our fitting, this suggests an upper limit on the magnetic field $B\lesssim 100~\rm{G}$.

It is clear that the emission beyond $\sim 50~\rm{GeV}$ is not perfectly consistent with an exponential cutoff. We suggest that this feature is due to the synchrotron of cascading pairs. We notice that this contribution is generally small compared to the proton synchrotron. Since the cascading pair flux is proportional to the low-energy photon density, then the low-energy photon density should be small. Therefore, the emission region is likely beyond the BLR, so that the low-energy photons are from the primary electron synchrotron. Given the observed luminosity, the low photon density indicates a large emission blob. Our fitting results suggest that the size is on the order of $\sim 10^{15}~{\rm cm}$.

We find that the X-ray spectra are well fit by the synchrotron of primary protons and cascading pairs. Therefore, the SSC from the primary electrons must be very low. Then the ratio between the optical emission, which is dominated by the synchrotron of the primary electrons, and the X-ray emission, which is the upper limit of the primary electron SSC, gives a lower limit on the magnetic field strength. Our fitting suggests a lower limit on the order of $\sim 10~\rm{G}$. This magnetic field range suggests that the low-energy spectral component must originate from synchrotron of primary electrons. Thus the underlying electron spectral shape can be constrained by the observed optical spectra.

Our fitting results generally suggest that the jet energy composition is $P_e<P_m \lesssim P_p$. Our results are generally consistent with previous hadronic model fits \citep[see, e.g.,][]{Boettcher13}. In the hadronic model, the proton power $P_p$ is generally larger than the accretion luminosity, which often requires a super Eddington accretion \citep[][]{Zdziarski15}. We notice that, although all the models parameters are constrained by the multi-wavelength spectra, the parameter space is rather large. For example, the magnetic field strength can range from $\sim 10~\rm{G}$ to $\sim 100~\rm{G}$. To further constrain the model parameters, additional observational constraints, such as the time-dependent signatures, are necessary. \cite{Zhang16} has demonstrated that the time-dependent multi-wavelength light curves and polarization signatures can stringently constrain the hadronic model parameters. However, a detailed time-dependent, variability focused study is beyond the scope of this paper.

\subsection{X-ray Polarization and Anticipation for Future X-ray Polarimetric Satellites}\label{subsec:pol}

It is obvious that the high-energy polarization signatures are drastically different between the leptonic and hadronic models. 
Since all eight sources are FSRQs, they exhibit strong thermal components, which lead to high EC contribution in the high-energy 
component. Generally speaking, EC can be considered unpolarized. Moreover, SSC reduces the seed synchrotron polarization, 
thus its polarization degree is mostly $\lesssim 40\%$. As we can see in Figure \ref{fig_pol}, the maximal leptonic polarization 
degree is only $20-40\%$ at X-rays, where SSC generally dominates, and then quickly drops to zero towards higher energies, 
where EC becomes dominant. On the other hand, the synchrotron emission of protons and cascading pairs dominates the high-energy 
emission in the hadronic model. Here the polarization degree is $70-80\%$. This makes the maximal hadronic polarization degree 
much higher than that of the leptonic model.

An interesting feature of the hadronic modeling is that all these hard spectrum FSRQs have a straight and dominating proton 
synchrotron SED component. The SSC of primary electrons and the synchrotron of cascading pairs only mildly lower the polarization 
degree at X-rays. Therefore, the X-ray and $\gamma$-ray polarization of the FSRQs are nearly identical. For the leptonic model, 
however, the X-ray polarization is clearly higher than $\gamma$-ray. This feature can be examined by future X-ray and 
$\gamma$-ray polarimeters as a further diagnostic of the two models for FSRQs.

To give a better prediction of what we expect from future high-energy polarimetry, we estimate the corrected high-energy 
polarization degree at 1 keV and 1 MeV. The depolarization factor due to partially ordered magnetic field is taken into account, 
as detailed in Section \ref{xpol}. We use both the average and high activity state polarization degrees, obtained by Steward 
Observatory and RoboPol \citep[][]{2009arXiv0912.3621S,2014MNRAS.442.1693P}, to estimate the potential range of high-energy 
polarization degree. The results are listed in Table \ref{tab:pol}. During the average state, the optical polarization degree is relatively 
low, $\sim 10\%$. We estimate that the corrected leptonic polarization degree is $\sim 5\%$, and the hadronic is $\gtrsim 10\%$ at 1 keV. 
Since the flux level and the polarization degree are rather low during the average state, it is hard for next generation polarimeters 
to detect high-energy polarization signatures. In addition, it is well known that the optical polarization signatures are highly variable, 
which indicate changes in the magnetic field \citep{Marscher14,Zhang15}. The same can happen to the high-energy polarization, so 
that the averaged polarization degree may be even lower than what we estimate here. Therefore, we argue that the X-ray and $\gamma$-ray 
polarization signatures may not be detectable for either leptonic or hadronic model by averaging a long period of high-energy polarization monitoring.

On the other hand, during the elevated activity state, not only the flux can be higher, but also the optical polarization degree can be 
higher, $\sim 30\%$. Here we find that the X-ray/\gm-ray polarization in the leptonic scenario is $\sim 15\%$, while the hadronic 
polarization is $\gtrsim 30\%$. \cite{Zhang16} have shown that the hadronic high-energy polarization signatures are similar between the 
quiescent state and the active state, and we expect the same for the leptonic model. Therefore, we recommend that the next
generation polarimeters should focus on the active state for best high-energy polarization detection. We notice that the hadronic polarization 
degree is consistently higher (or comparable) than the optical counterpart. This is easy to understand, because the optical 
polarization is generally contaminated by an unpolarized thermal component in FSRQs, while the X-ray and $\gamma$-ray polarization is 
mostly due to synchrotron.

To summarize, the current generation polarimeters are likely to detect X-ray and $\gamma$-ray polarization during active states of FSRQs 
for both leptonic and hadronic models, and in particular, when the optical polarization degree is high. Considering the fact that FSRQs peaks 
between MeV and GeV, a $\gamma$-ray polarimeter may be easier to detect polarization. Three polarization features of FSRQs can distinguish 
the leptonic and hadronic models: 1. the hadronic model shows a systematic higher polarization degree than the leptonic; 2. in the hadronic 
model, the X-ray and $\gamma$-ray polarization degrees are similar, while in the leptonic model, the X-ray polarization is higher than $\gamma$-ray; 
3. the hadronic polarization is generally higher than the optical counterpart, while the leptonic polarization is only half of that.

\subsection{A Detection Beyond 50 GeV and the \gm-ray Absorption}
In the leptonic emission framework, the origin of the \gm-ray radiation in FSRQs is believed to be due to IC 
scattering of BLR photons by the jet electrons. However, the same BLR radiation field can also absorb \gm-rays 
via the \gm\gm~pair production process \citep[e.g.,][]{2003APh....18..377D,2006ApJ...653.1089L}. This is aligned 
with the fact that only a handful of FSRQs are detected in the VHE band\footnote{http://tevcat.uchicago.edu/}. 
Knowledge of the BLR absorption enables us to constrain the location of the \gm-ray emission region by 
requiring that the optical depth for \gm-ray absorption should be small ($\tau_{\gamma\gamma}<1$) at 
$R_{\rm diss}$. Therefore, it is of great interest to study the effect of the BLR absorption on the 
\gm-ray spectra of 2FHL FSRQs.

\begin{figure*}
\hbox{
\includegraphics[width=6cm]{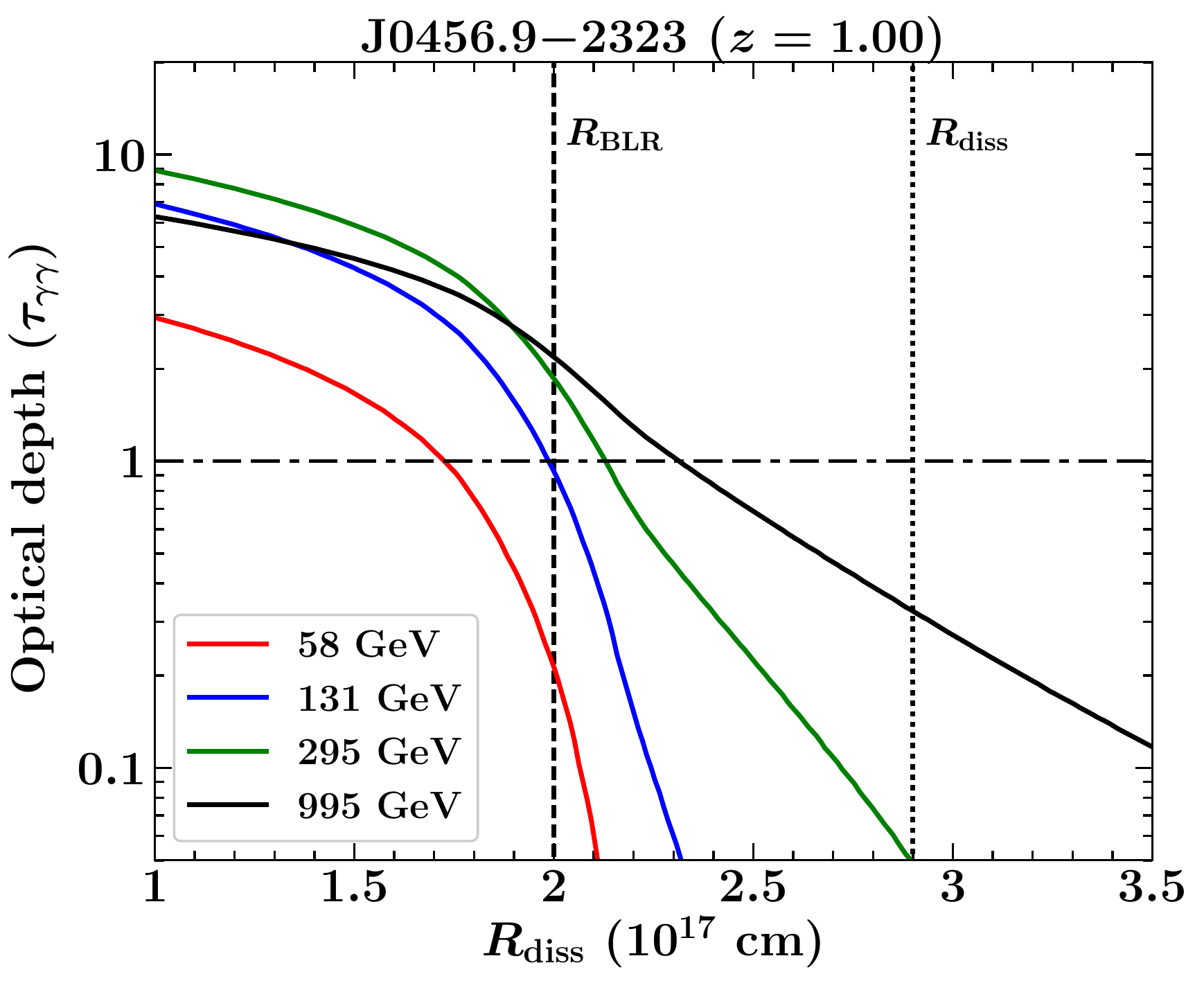}
\includegraphics[width=6cm]{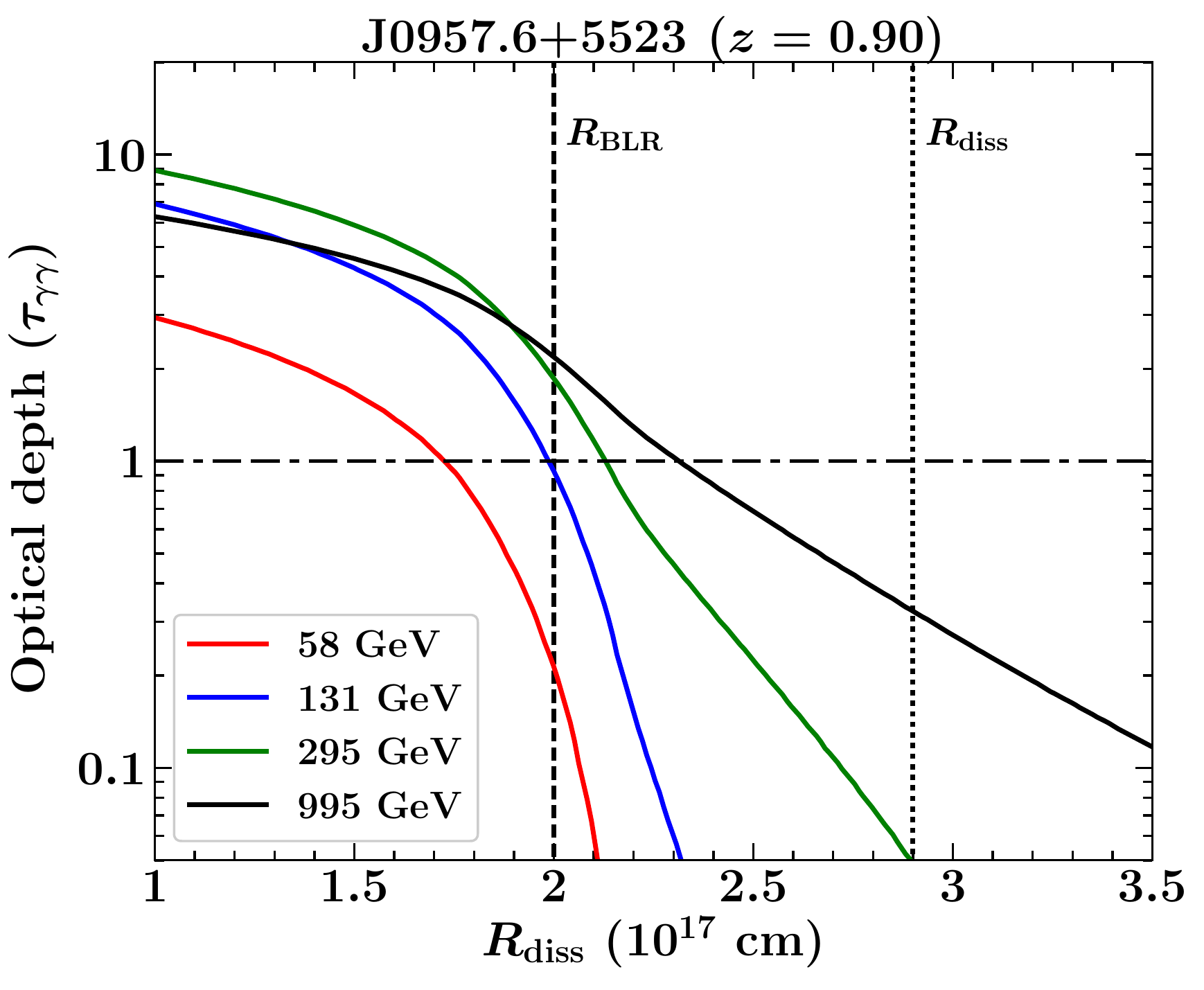}
\includegraphics[width=6cm]{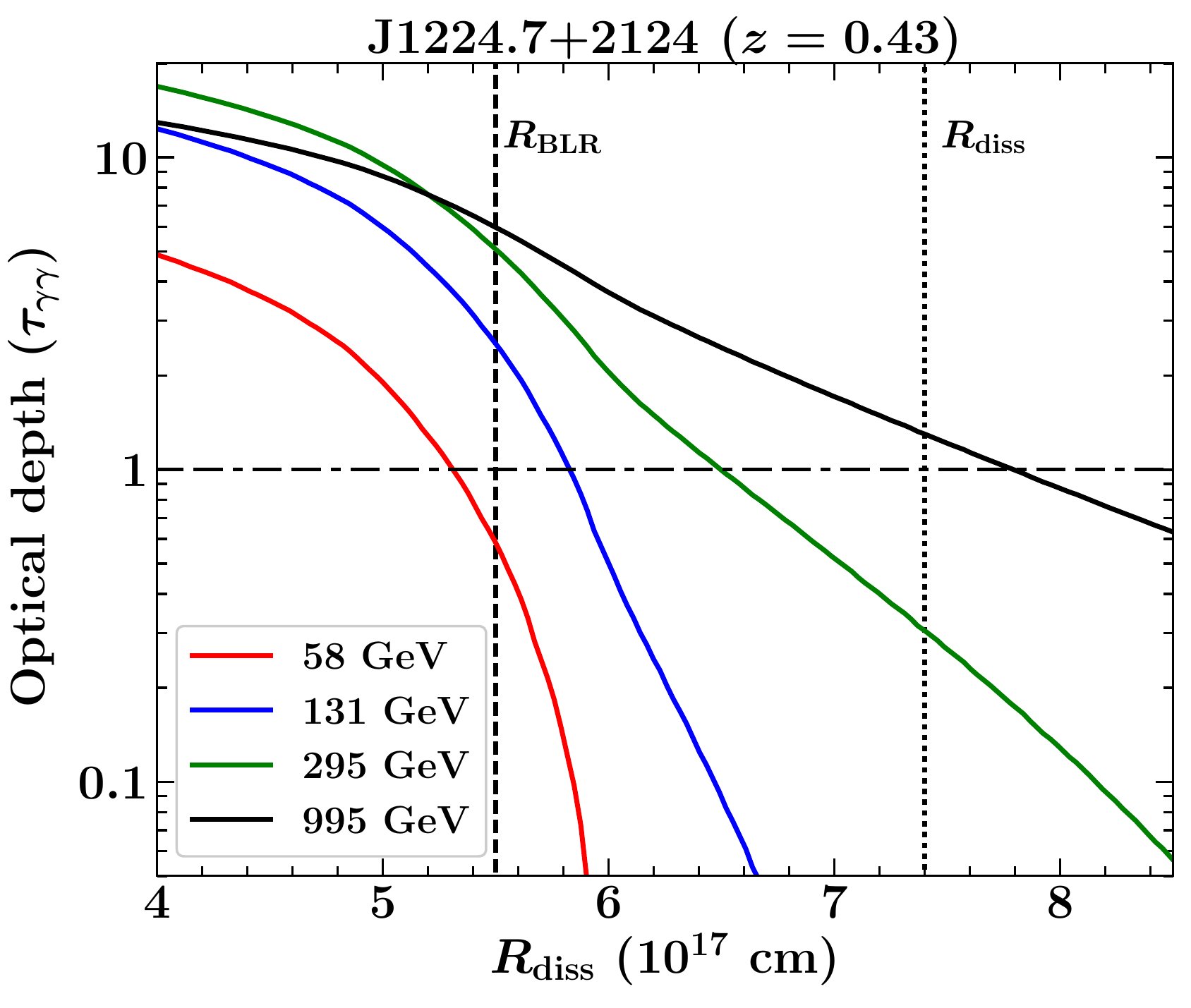}
}
\hbox{
\includegraphics[width=6cm]{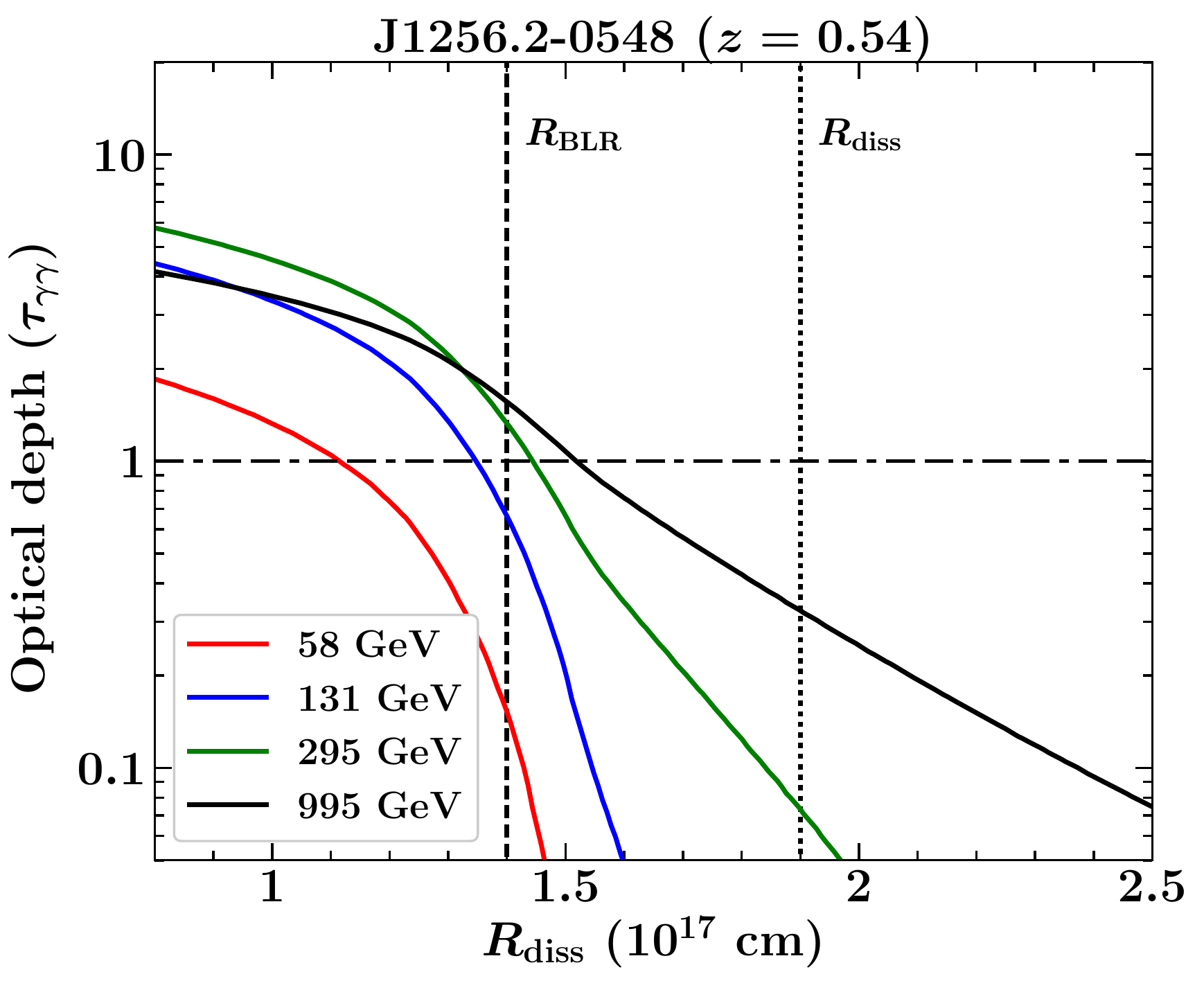}
\includegraphics[width=6cm]{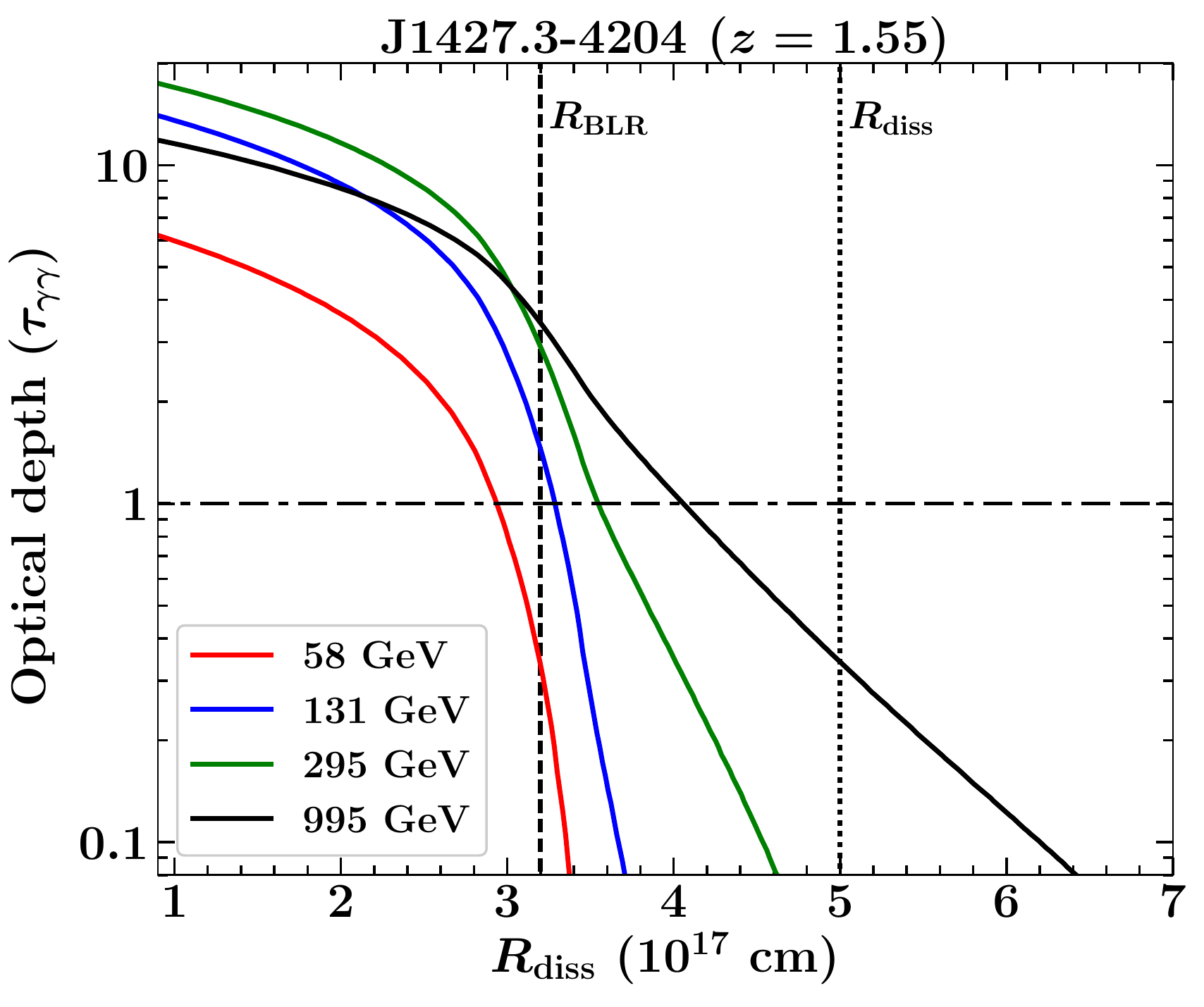}
\includegraphics[width=6cm]{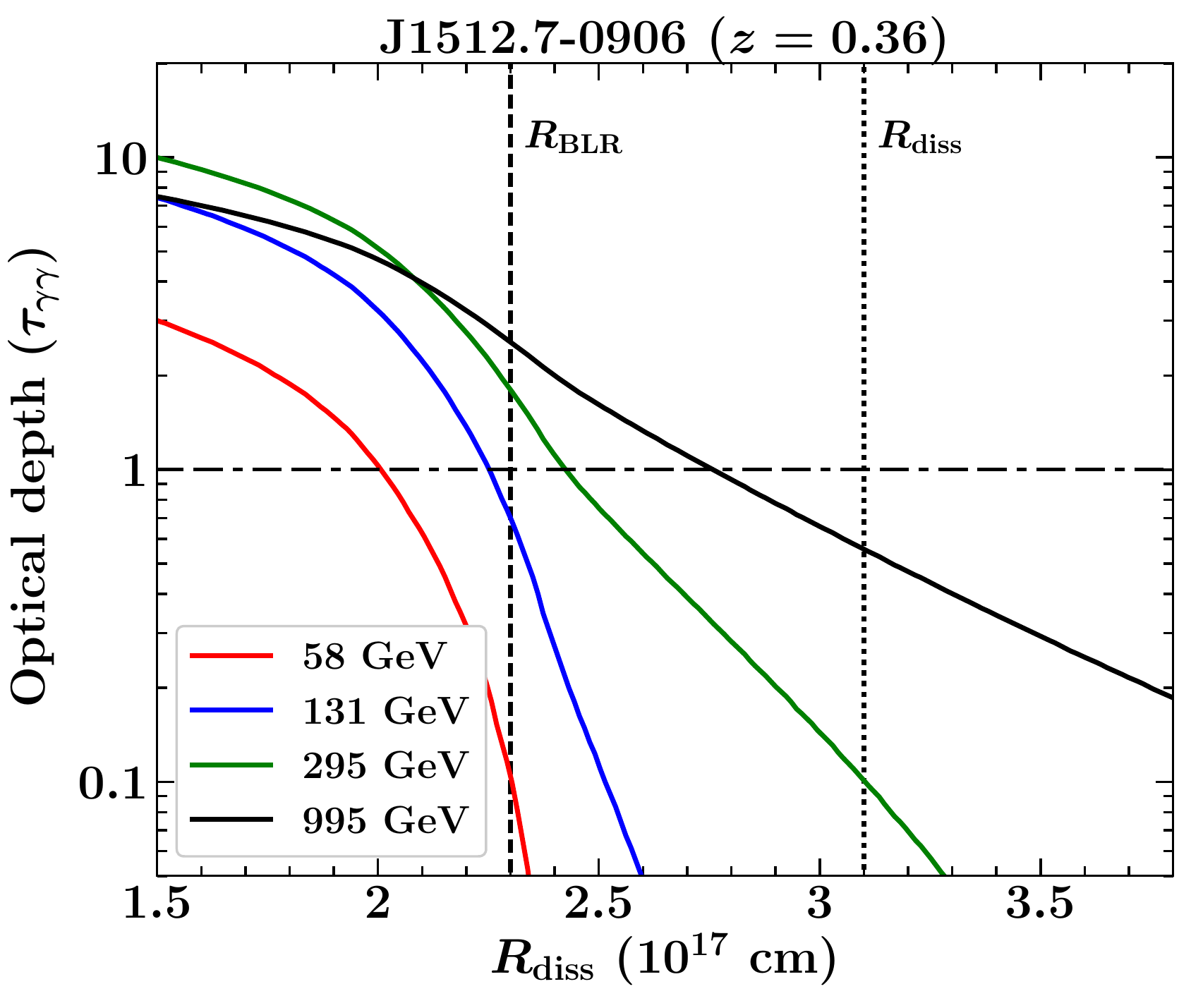}
}
\hbox{\hspace{3.0cm}
\includegraphics[width=6cm]{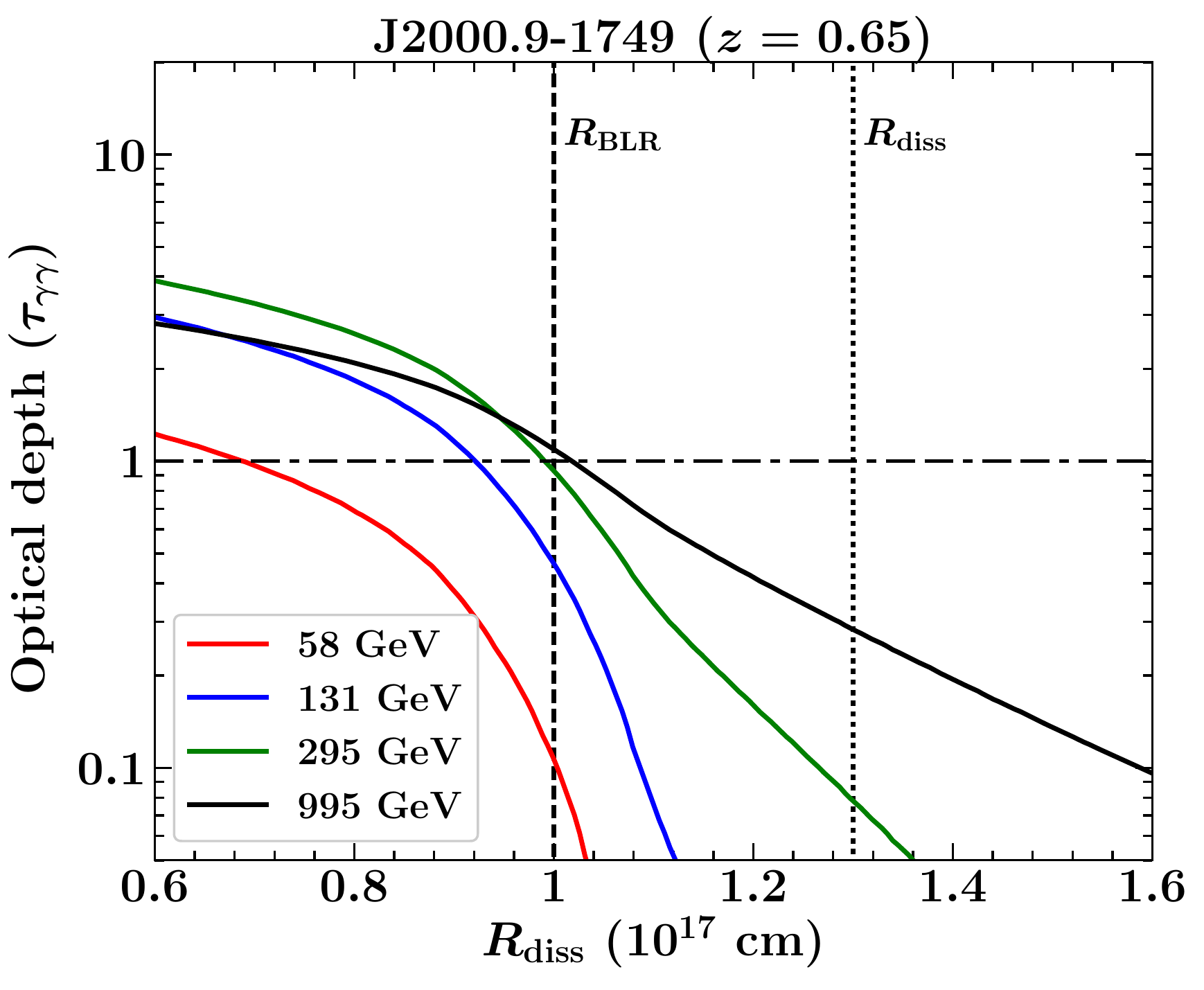}
\includegraphics[width=6cm]{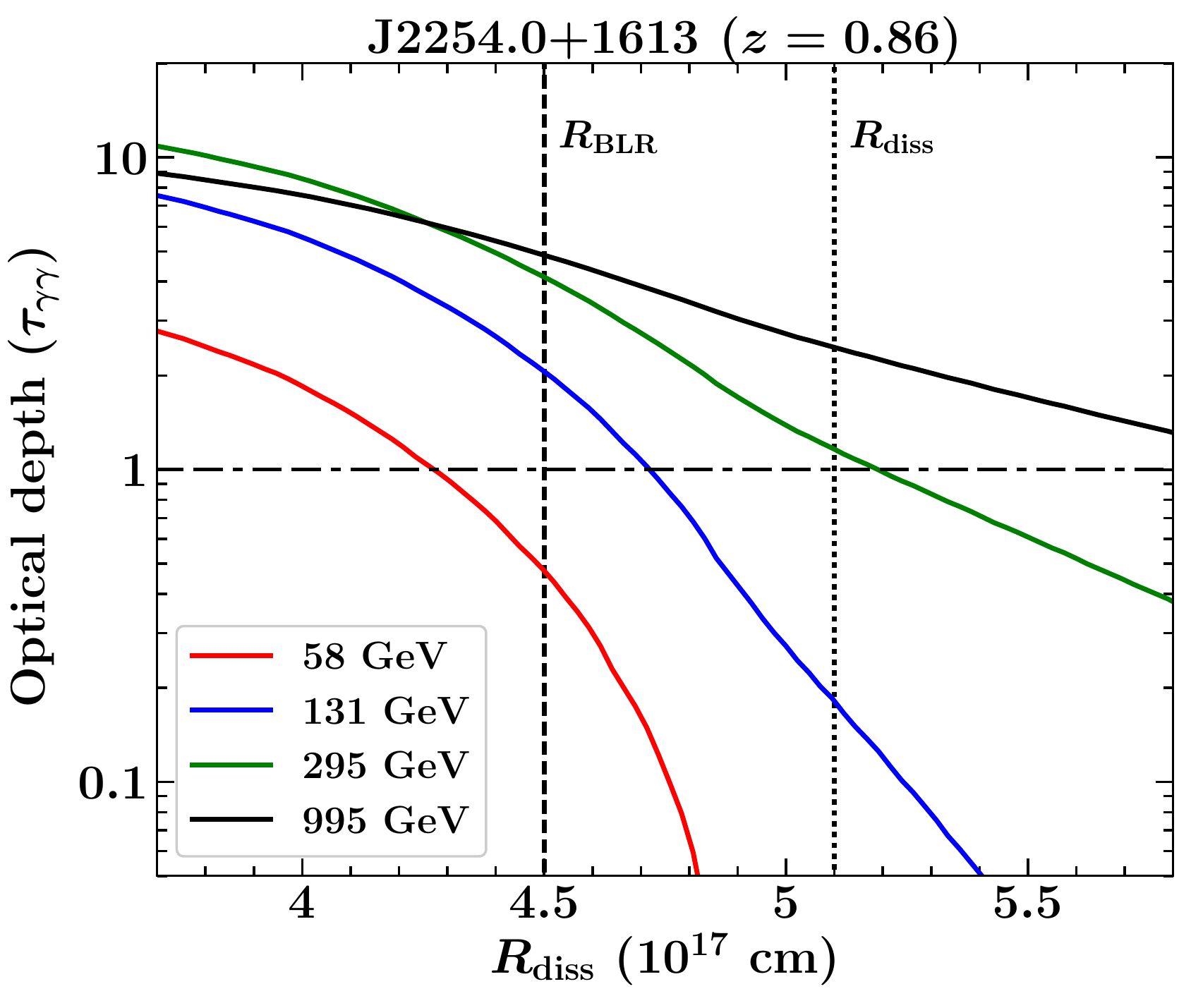}
}
\caption{The variation of $\gamma\gamma$ absorption optical depth ($\tau_{\gamma\gamma}$) as a function 
of the location of the emission region ($R_{\rm diss}$) along the jet. Vertical black dashed and dotted lines correspond 
to the inner radius of the BLR and the location of the emission region, respectively, as inferred from the leptonic SED 
modeling. Horizontal black dash-dash-dot line represent $\tau_{\gamma\gamma}=1$. Various color lines denote the variation of the 
optical depths derived for \gm-ray photons of different energies, as labelled.}\label{fig_gamma_opacity} 
\end{figure*}

Recently, \citet[][]{2016ApJ...821..102B} have developed a novel approach to quantify the \gm\gm~opacity due 
to the BLR radiation field which primarly depends only on the BLR luminosity and energy density. These parameters 
can be constrained from the observations: either from the emission line luminosities or from the modeling of the 
big blue bump (see Section \ref{sec5} for details). Therefore, we adopt the methodology described in 
\citet[][]{2016ApJ...821..102B} to derive $\tau_{\gamma\gamma}$ as a function of $R_{\rm diss}$. The results 
are shown in Figure \ref{fig_gamma_opacity}. In this figure, various color lines correspond to 
$\tau_{\gamma\gamma}$ as a function of $R_{\rm diss}$ derived for \gm-ray photons of different energies, as labelled. 
The location of the \gm-ray emitting region is also shown with the vertical dotted line. As can be seen, $\tau_{\gamma\gamma}$ 
is very small at $R_{\rm diss}$, even for a \gm-ray photon of $\sim$300 GeV energy (except for J2254.0+1613 where 
$\tau_{\gamma\gamma}\gtrsim1$). Interestingly, according to our calculation, the BLR is transparent to $\sim$50 GeV photon 
even at its inner boundary (shown with the vertical dashed line). This implies that the \gm-ray emission region can be located close 
to the BLR where the BLR radiative energy density is dense enough to act as a primary reservoir of seed photon for EC process, 
however, sufficiently transparent to high energy (50$-$100 GeV) \gm-ray photons.

\subsection{Prospects for VHE Emission and EBL Studies}
A statistically significant detection above 50 GeV by \fermi-LAT makes blazars viable 
candidates for observations in the VHE band with ground based Cherenkov telescopes 
such as H.E.S.S., MAGIC and VERITAS. As of now, only seven FSRQs are known as VHE emitters and 
this work includes 3 of them, i.e., J1224.7+2124 \citep{2011ApJ...730L...8A}, J1256.2$-$0548 
\citep{2008Sci...320.1752M}, and J1512.7$-$0906 \citep{2013A&A...554A.107H}. 

\begin{figure*}
\hbox{
\includegraphics[width=6.1cm]{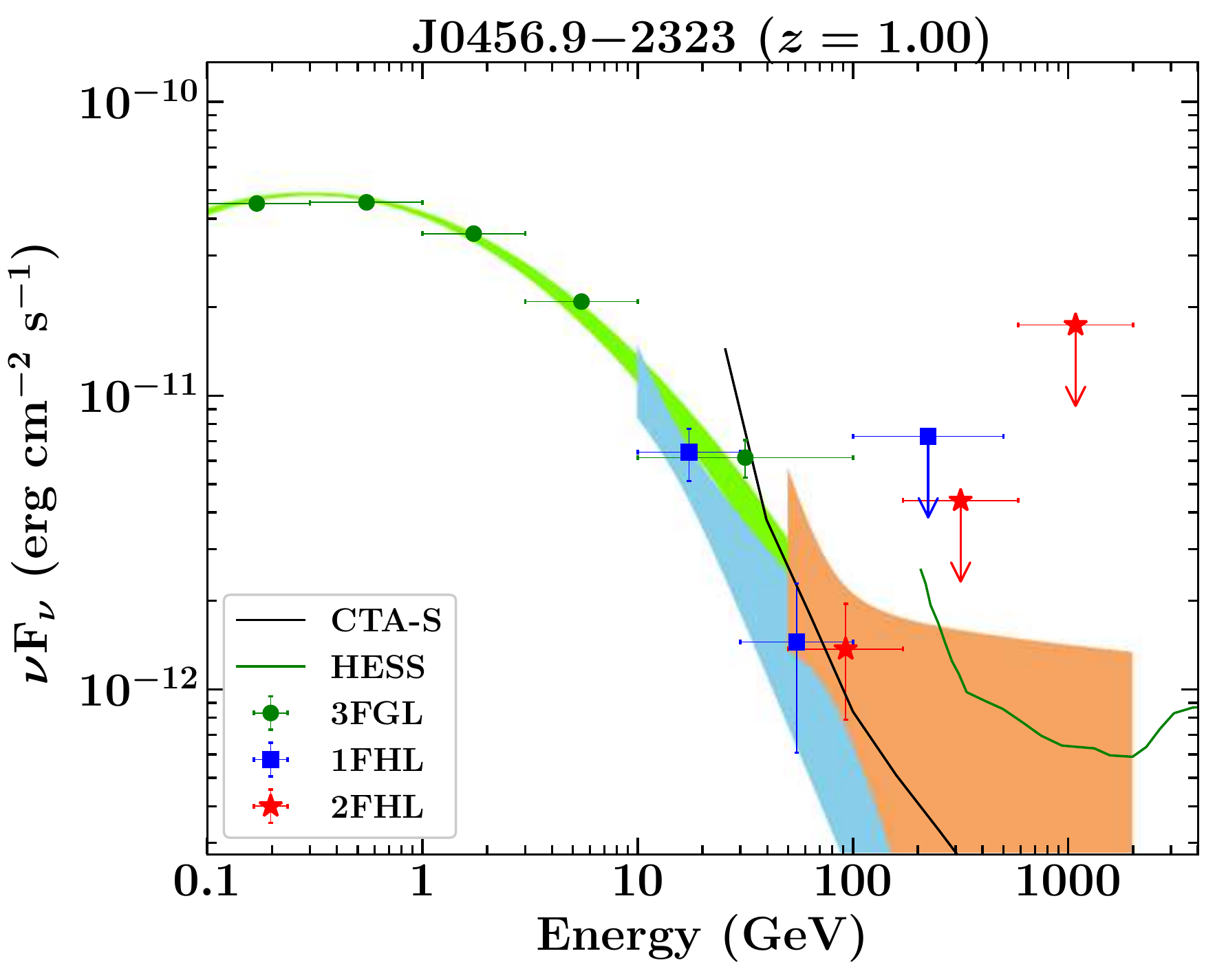}
\includegraphics[width=6.1cm]{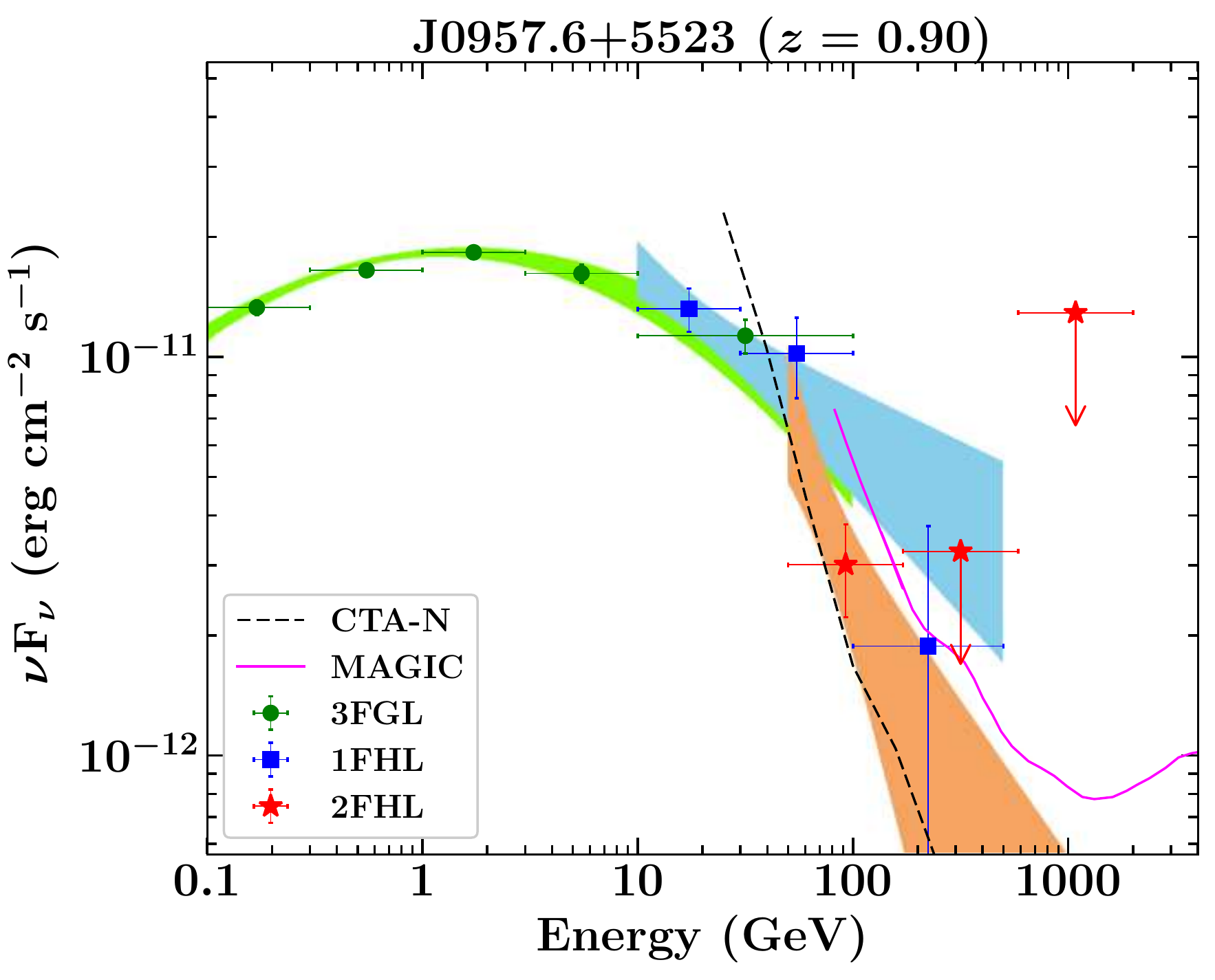}
\includegraphics[width=6.1cm]{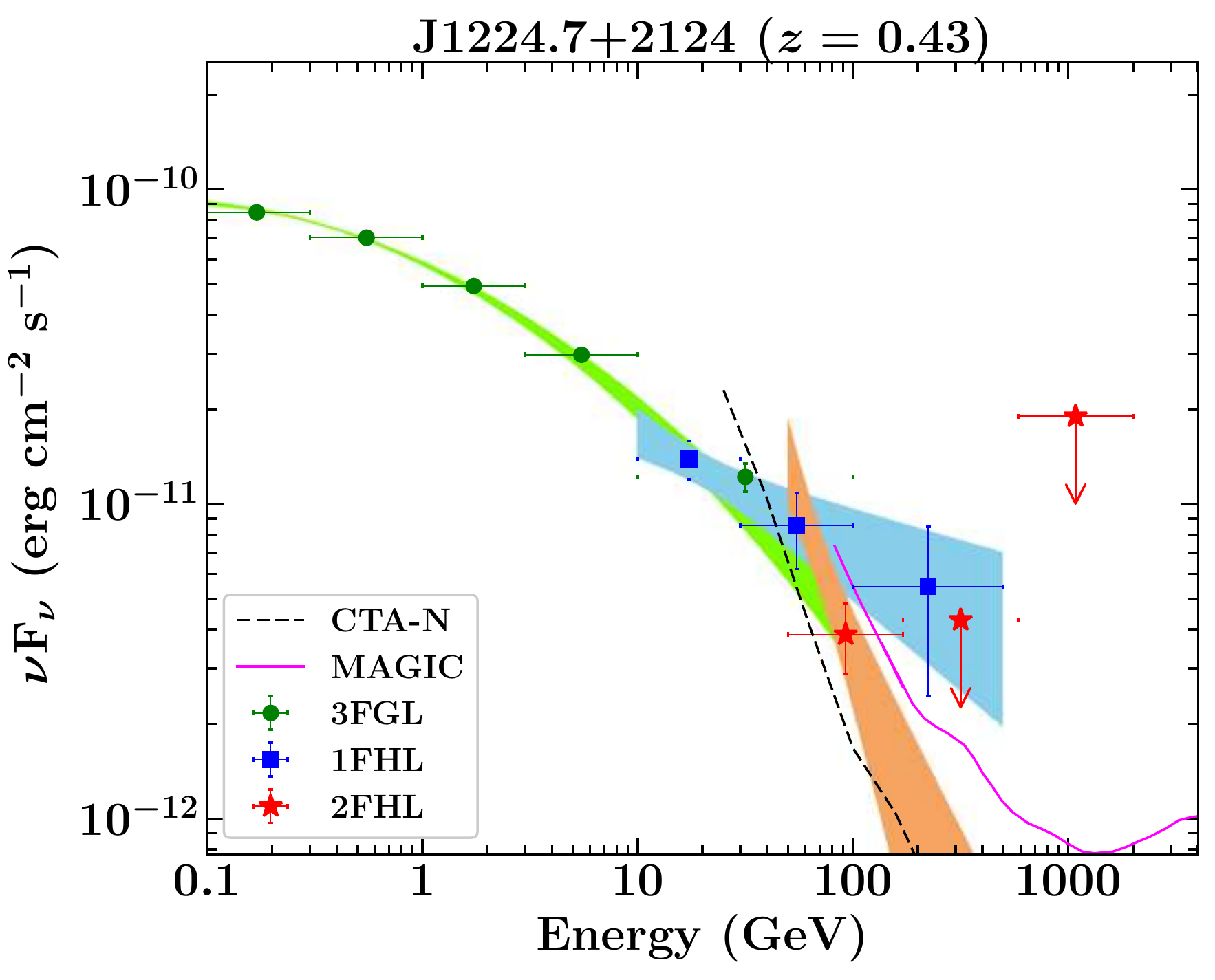}
}
\hbox{
\includegraphics[width=6.1cm]{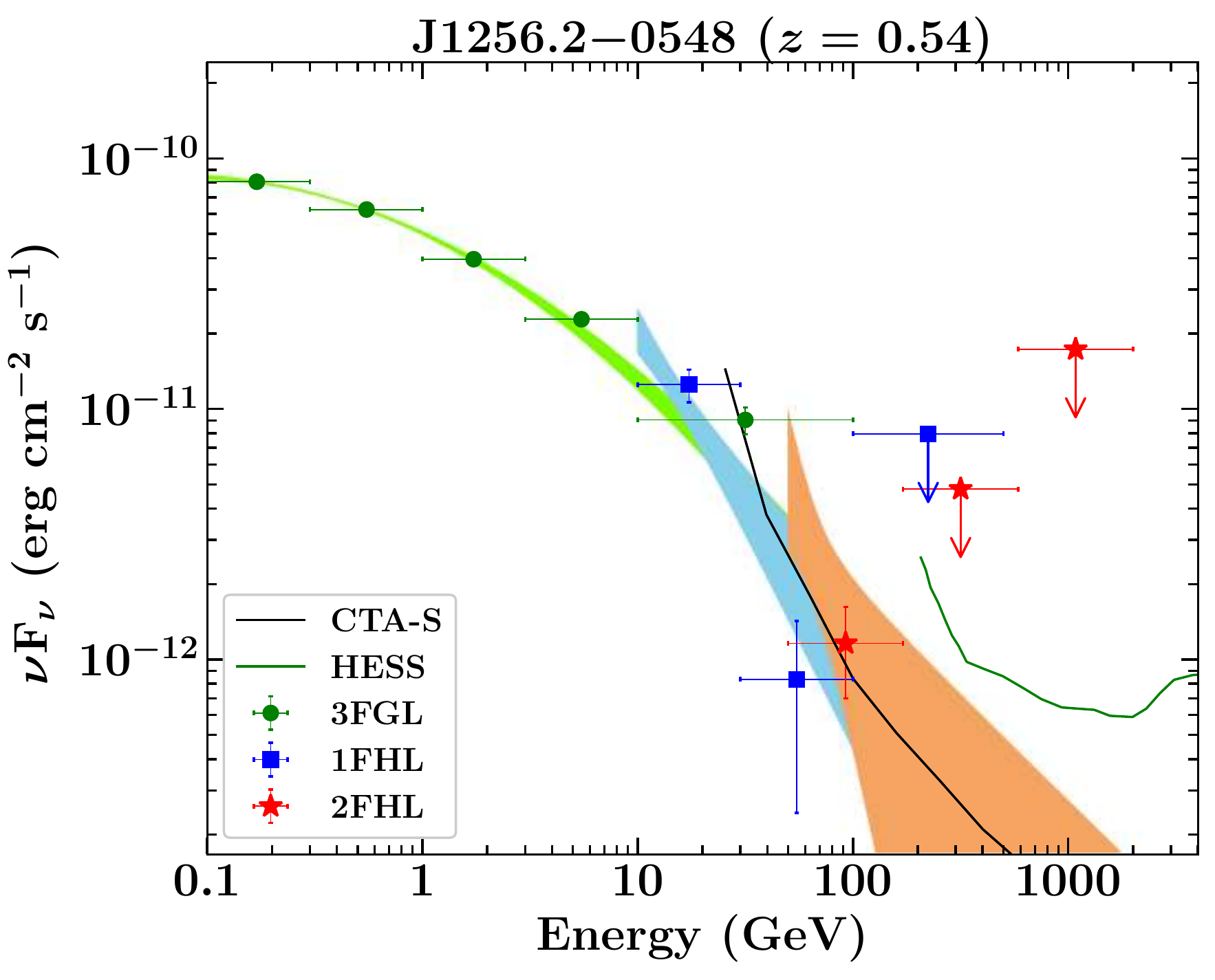}
\includegraphics[width=6.1cm]{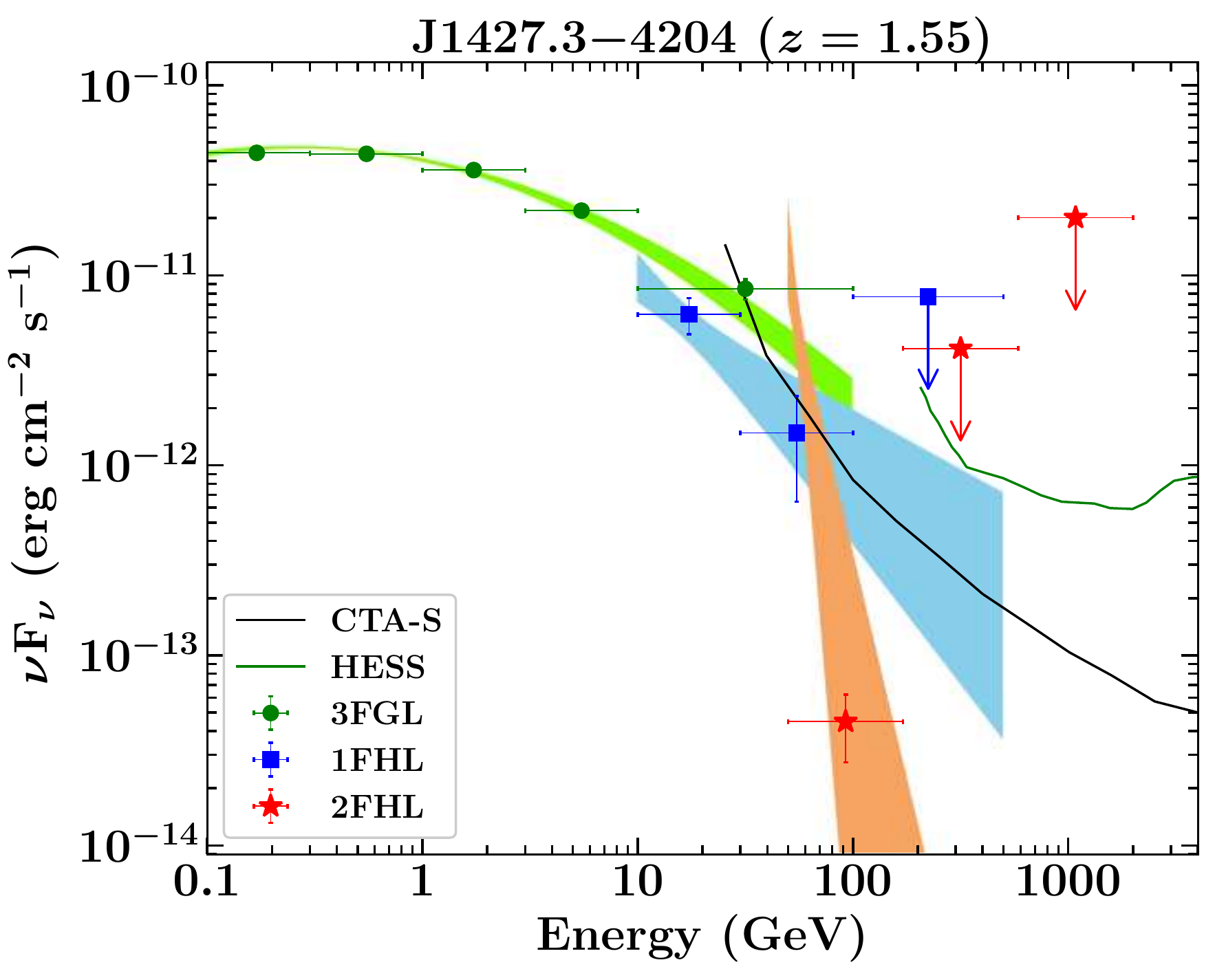}
\includegraphics[width=6.1cm]{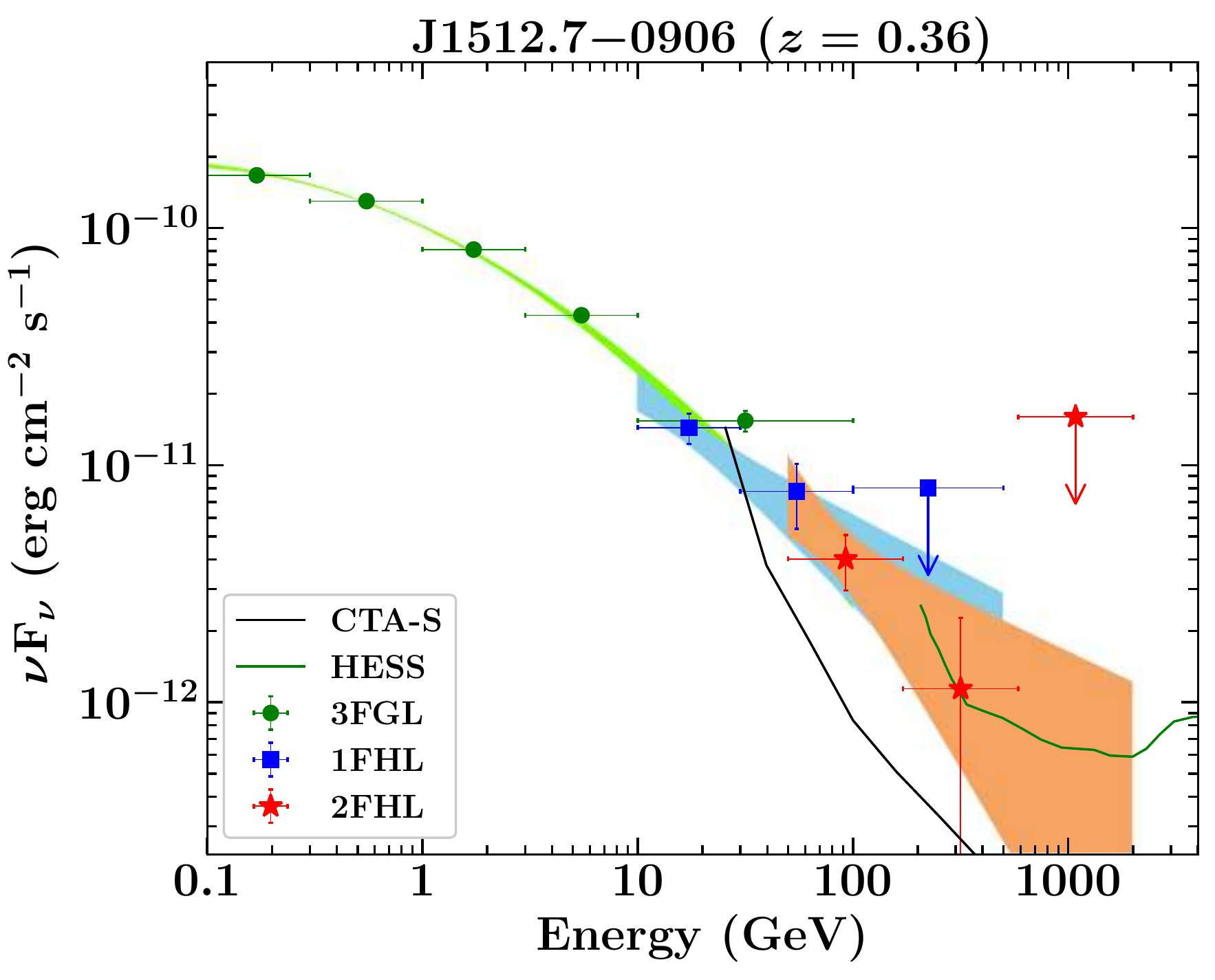}
}
\hbox{\hspace{3.0cm}
\includegraphics[width=6.1cm]{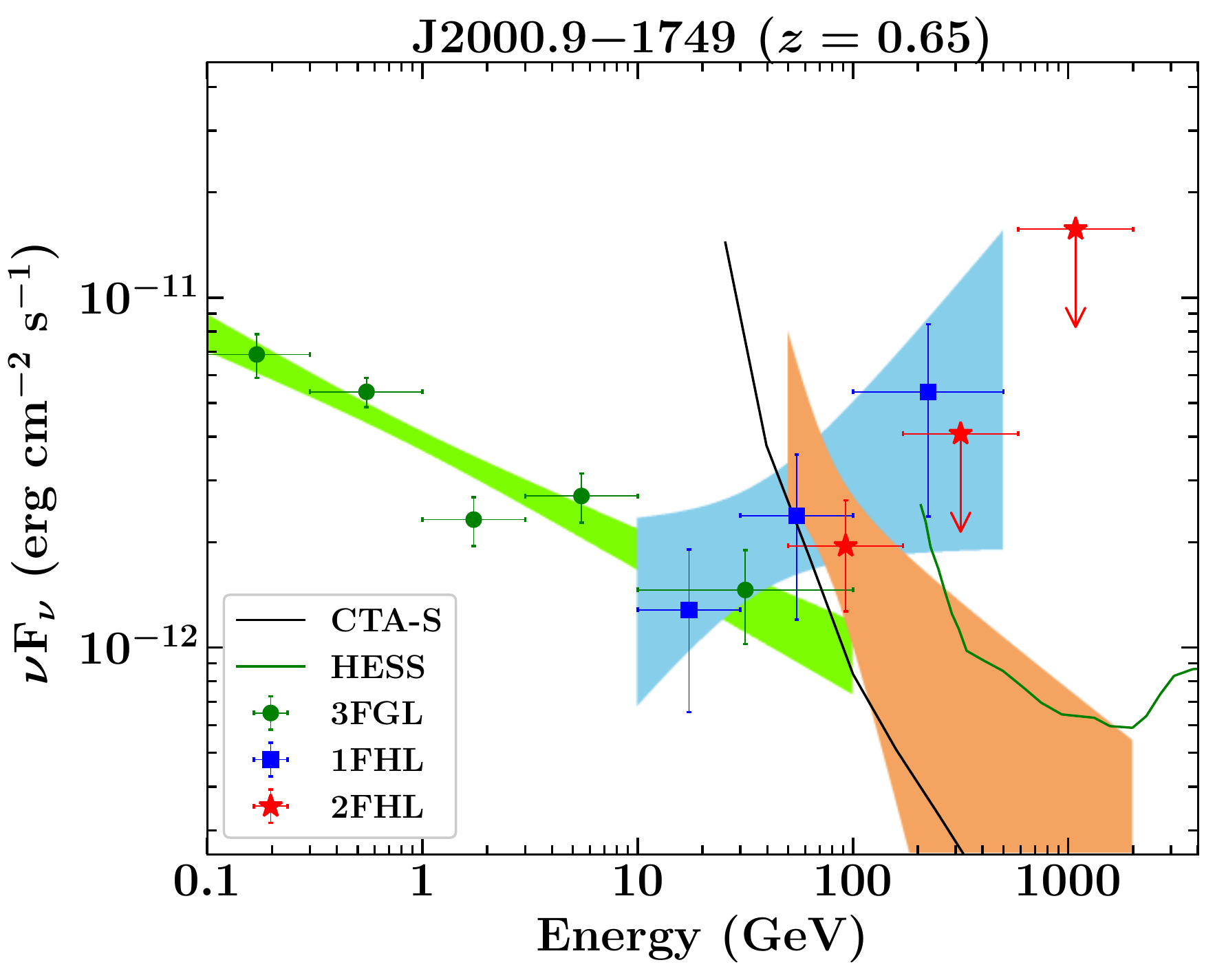}
\includegraphics[width=6.1cm]{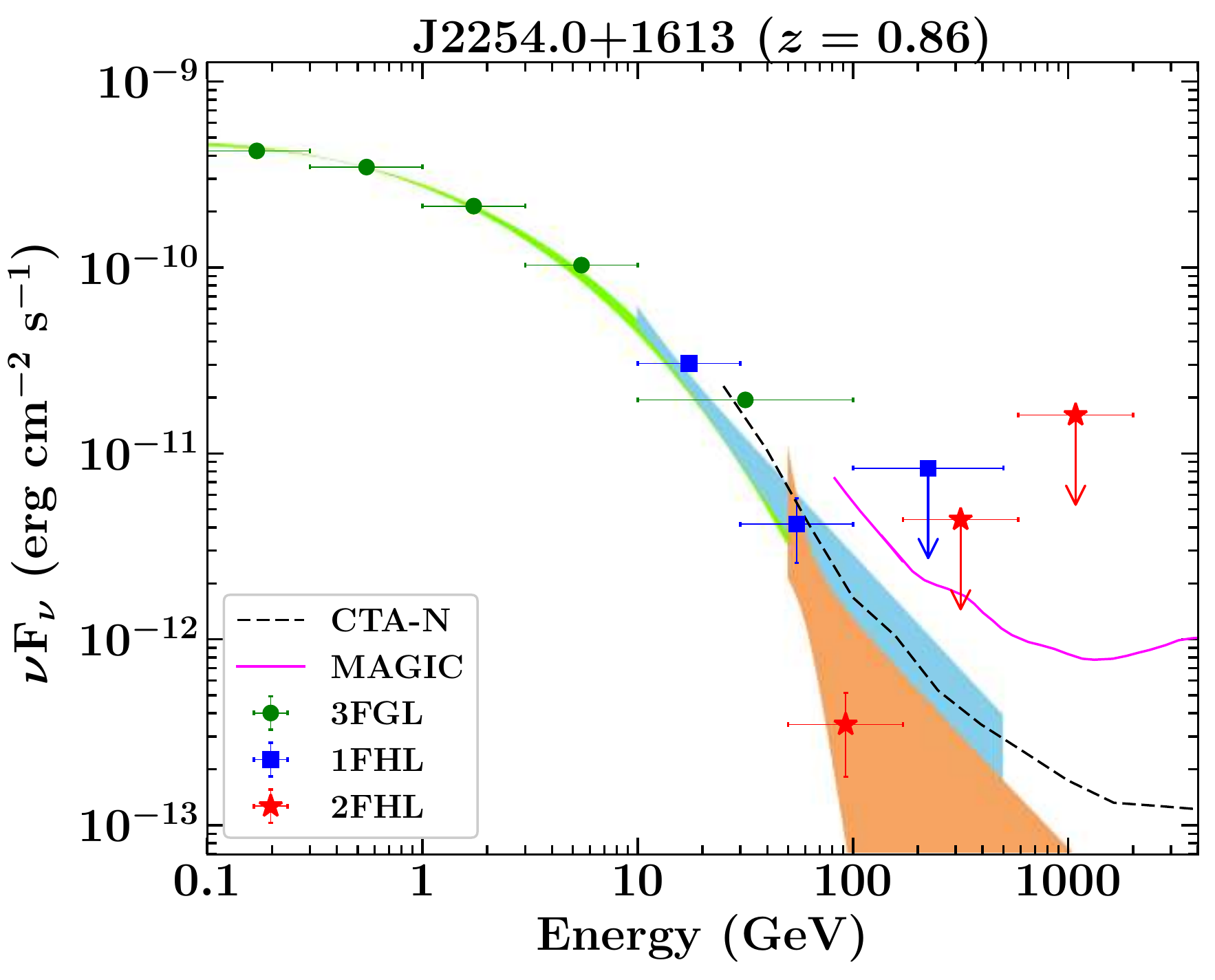}
}
\caption{Gamma-ray SEDs of FSRQs studied in this work. Spectral data from the 3FGL, 1FHL, and 2FHL catalogs
are represented by green circles, blue squares, and reds stars, respectively. Associated 1$\sigma$ uncertainties are shown with 
shaded butterfly regions. We also show the sensitivity limits for 50 hours of integration with the MAGIC and HESS telescopes 
\citep[pink and green solid lines,][]{2016APh....72...76A,2015ICRC...34..980H}, 
and the upcoming CTA-North (black dashed line) and CTA-South (black solid line). Note that we plot CTA-North and MAGIC sensitivities
for northern hemisphere objects, whereas, for southern hemisphere sources, we show the sensitivity curves for CTA-South and HESS
observatories. \label{fig_gamma_sed}} 
\end{figure*}

The 2FHL spectral shapes of the FSRQs provides us a clue about their VHE 
detection possibility due to the broad energy coverage of the 2FHL catalog (up to 2 TeV). 
In Figure \ref{fig_gamma_sed}, where the 100~MeV$-$2~TeV \gm-ray SEDs are shown, 
we over plot the sensitivity limits of the MAGIC and HESS telescopes (pink and green solid lines, respectively)
 and the future CTA-North and CTA-South
observatories\footnote{The sensitivity limits for all instruments are extracted from the public 
CTA page: https://www.cta-observatory.org/science/cta-performance/\#1472563157332-1ef9e83d-426c. 
Note that not all of the sources are visible from all of the facilities. Therefore, for positive declination sources, we show sensitivity plots
of CTA-North and MAGIC, whereas, HESS and CTA-South sensitivity curves are used for southern hemisphere objects.} 
(black dashed and solid lines, respectively) for an integration 
time of 50 hours and a given zenith angle \citep[][]{2016APh....72...76A,2015ICRC...34..980H}. 
By comparing the 2FHL spectral points with CTA sensitivity curves, we expect that CTA will possibly be able to 
detect all of them except J1427.3$-$4204 and J2254.0+1613. These two objects have the two softest 2FHL 
spectra among all the FSRQs and there is only marginal overlap between the CTA sensitivity 
curve and their 2FHL bow-tie plot. For both of them, the 2FHL spectral points lie well below the CTA sensitivity plots. 
In fact, the 3FGL spectrum of J2254.0+1613 is modeled 
as a power law with an exponential cutoff \citep[][]{2015ApJS..218...23A} indicating 
the presence of a sharp decline in the flux above 50 GeV. \citet[][]{2011ApJ...738..148M} 
predicted J0957.6+5523 as a plausible candidate for VHE detection due to its hard 
0.1$-$300~GeV spectrum and the lack of significant \gm-ray flux variability, however, 
a dedicated observing campaign of 35 hours with MAGIC and 45 hours from VERITAS only resulted in flux upper 
limits \citep[][]{2014MNRAS.440..530A,2013arXiv1303.1103F}. 
A comparison with the MAGIC sensitivity curve in the \gm-ray spectrum of J0957.6+5523 
(Figure \ref{fig_gamma_sed}) suggests that the source would be difficult to detect even in 50 hours 
of integration. Furthermore, as of now, all the FSRQs are detected in the VHE band during 
flaring activity periods. However, the unprecedented sensitivity of CTA will allow us 
to observe these FSRQs (and many more) even during their average low activity states.

\begin{figure*}
\includegraphics[width=\linewidth]{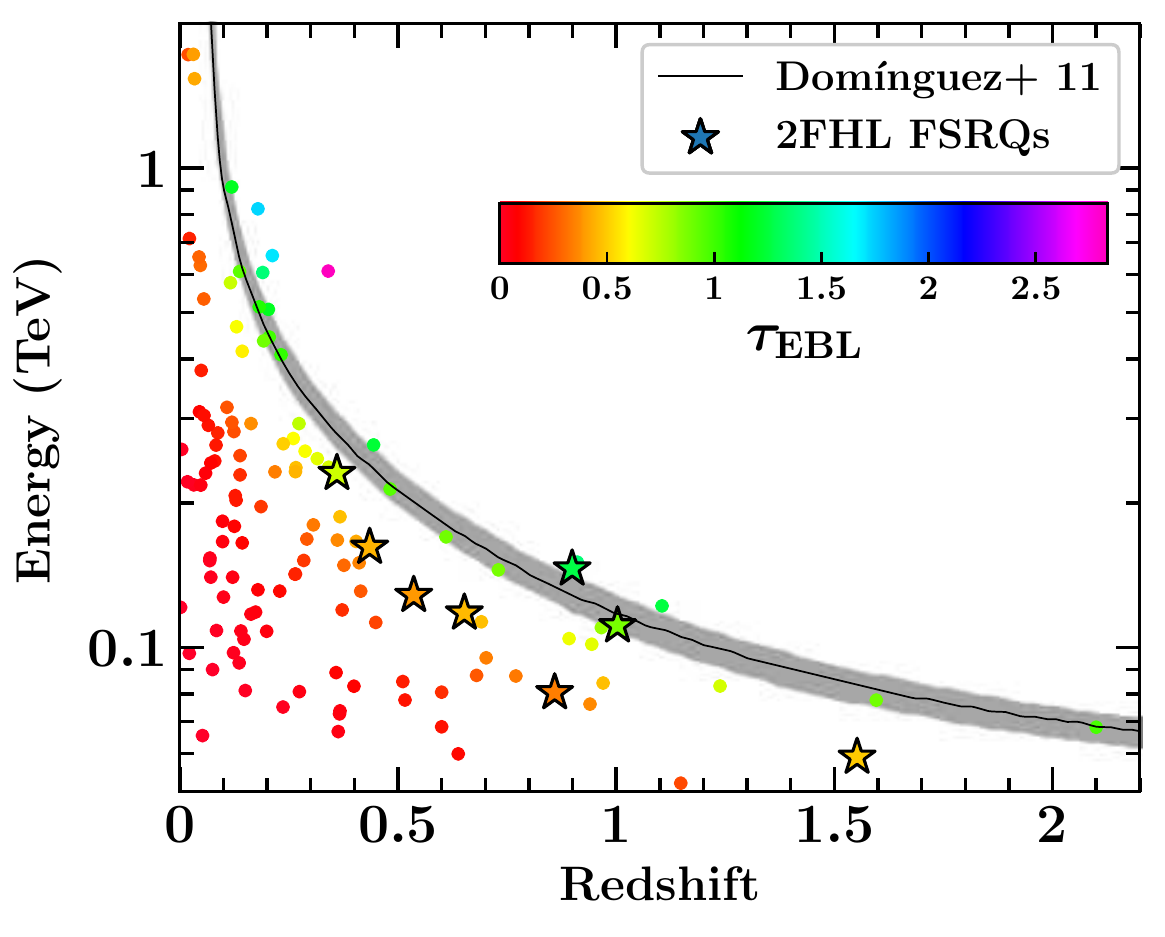}
\caption{A plot of the energy of the highest energy photons detected from 2FHL sources versus redshift, as reported 
in \citet[][]{2016ApJS..222....5A}. The 2FHL FSRQs are shown with stars. The color scheme represents the optical 
depth ($\tau_{\rm EBL}$) for a given energy and redshift following the EBL attenuation model of \citet[][]{2011MNRAS.410.2556D}, 
as shown in the colorbar. Solid black line denotes the cosmic \gm-ray horizon with 1$\sigma$ uncertainties (shaded area) adopting the 
same EBL model. As can be seen, on average, the FSRQs follow the opacity pattern expected from the EBL model.}\label{fig_ebl} 
\end{figure*}

In general, EBL studies using blazars are more prone towards BL Lac objects \citep[][]{2015ApJ...813L..34D}. This is mainly 
due to the hard \gm-ray spectra of these sources on which the EBL imprint can easily be 
observed \citep[e.g.,][]{2012Sci...338.1190A}. FSRQs, on the other hand, are rarely 
detected above 10$-$20~GeV due to their soft \gm-ray spectrum. Furthermore, $\gamma\gamma$
absorption on the BLR radiation field may be difficult to disentangle from EBL absorption
effects. Therefore, EBL studies with FSRQs are generally more difficult than with BL Lac 
sources. In this regard, 2FHL FSRQs can be used to probe the theories of the redshift 
dependence of EBL evolution as they are observed above 50 GeV and also they are located 
at high redshifts ($z>0.5$). Furthermore, based on the \gm\gm~opacity estimation for the BLR absorption, 
we found $\tau_{\gamma\gamma}<1$ at the location of the \gm-ray emitting regions, thus indicating a negligible 
BLR absorption effect on the \gm-ray spectra of 2FHL FSRQs (Figure \ref{fig_gamma_opacity}). In Figure \ref{fig_ebl}, 
we plot the energy of the highest 
energy photons (HEP) detected from 2FHL sources as a function of their redshifts; 2FHL 
FSRQs are marked with stars. We use the EBL attenuation model of \citet[][]{2011MNRAS.410.2556D} 
to derive the EBL optical depth ($\tau_{\rm EBL}$) for all the sources (see the color 
scheme in Figure \ref{fig_ebl}) and show the cosmic \gm-ray horizon with 1$\sigma$ 
uncertainty, as derived by this model \citep[see also,][]{2013ApJ...770...77D}. We do not see any major deviation from the 
predicted horizon and they are located in the more transparent region. However, there are a couple of 
noteworthy observations. As can be seen, the optical depth towards the FSRQ J0957.6+5523 
($z=0.9$ and HEP = 145 GeV) matches the \gm-ray horizon within the 1$\sigma$ 
uncertainty of the latter. 
Moreover, at a redshift of 
1.55, J1427.3$-$4204 is the most distant FSRQ in the 2FHL catalog and has the softest 
2FHL spectrum in our sample. It lies well below the $\tau_{\rm EBL}=1$ line in Figure 
\ref{fig_ebl}, which is consistent with its observed \gm-ray spectral behavior. 

\section{Summary}{\label{sec8}}
We have performed a broadband analysis of eight FSRQs present in the 2FHL catalog. Our findings are summarized below.
\begin{enumerate}
\item Both leptonic and hadronic emission models reasonably explain the broadband SEDs and are marginally 
consistent with 2FHL spectra.
\item The location of the \gm-ray emission is found to be at the outer edge of the BLR and it is consistent with our 
quantitative estimate of the \gm\gm~opacity for the \gm-ray absorption with the BLR radiation field.
\item According to our analysis, leptonic emission models predicts a significantly lower degree of high energy polarization 
compared to the hadronic ones.
\item In the hadronic scenario, the degrees of both X-ray and \gm-ray polarization are expected to be similar, but the 
X-ray polarization is predicted to be higher than \gm-rays if blazar jets are powered by leptonic emission mechanisms.
\item It is likely that the X-ray polarimeters (e.g., IXPE) may detect a significant degree of polarization from FSRQs during 
their flaring activity states. If so, it will provide supportive evidence for the hadronic origin of the observed radiation.
\item A majority of the hard \gm-ray spectrum FSRQs would be detectable with the upcoming TeV facility CTA, though 
J2254.0+1613 may remain below the detection threshold, especially during the non-flaring states, due to a strong cutoff in its \gm-ray spectra.
\end{enumerate}

\acknowledgments
We are grateful to an anonymous referee for constructive criticism.
The \textit{Fermi} LAT Collaboration acknowledges generous ongoing support
from a number of agencies and institutes that have supported both the
development and the operation of the LAT as well as scientific data analysis.
These include the National Aeronautics and Space Administration and the
Department of Energy in the United States, the Commissariat \`a l'Energie Atomique
and the Centre National de la Recherche Scientifique / Institut National de Physique
Nucl\'eaire et de Physique des Particules in France, the Agenzia Spaziale Italiana
and the Istituto Nazionale di Fisica Nucleare in Italy, the Ministry of Education,
Culture, Sports, Science and Technology (MEXT), High Energy Accelerator Research
Organization (KEK) and Japan Aerospace Exploration Agency (JAXA) in Japan, and
the K.~A.~Wallenberg Foundation, the Swedish Research Council and the
Swedish National Space Board in Sweden. Additional support for science analysis 
during the operations phase is gratefully acknowledged from the Istituto Nazionale di 
Astrofisica in Italy and the Centre National d'\'Etudes Spatiales in France.
This research has made use of data obtained through the High Energy Astrophysics 
Science Archive Research Center Online Service, provided by the NASA/Goddard Space 
Flight Center.  Part of this work is based on archival data, software or online services provided by the
ASI Data Center (ASDC). This research has made use of the XRT Data Analysis Software
(XRTDAS). This work made use of data supplied by the UK Swift Science Data Centre at the University of Leicester.
AD thanks the support of the Juan de la Cierva program from the Spanish MEC. The work of M.B. is supported by 
the South African Research Chair Initiative (SARChI) of the Department of Science and Technology and the 
National Research Foundation\footnote{Any opinion, finding, and conclusion or recommendation expressed in this material is that of the authors and the NRF does not accept any liability in this regard.} of South Africa. HZ acknowledges support from Fermi Guest Investigator program Cycle 10, grant number 80NSSC17K0753.

\software{XSPEC \citep{1996ASPC..101...17A}, Swift-XRT data product generator \citep{2009MNRAS.397.1177E}}.

\bibliographystyle{aasjournal}
\bibliography{Master}

\begin{thebibliography}{}
\expandafter\ifx\csname natexlab\endcsname\relax\def\natexlab#1{#1}\fi

\bibitem[{{Abdo} {et~al.}(2010){Abdo}, {Ackermann}, {Agudo}, {Ajello}, {Aller},
  {Aller}, {Angelakis}, {Arkharov}, {Axelsson}, {Bach}, \&
  et~al.}]{2010ApJ...716...30A}
{Abdo}, A.~A., {Ackermann}, M., {Agudo}, I., {et~al.} 2010, \apj, 716, 30

\bibitem[{{Abdo} {et~al.}(2011){Abdo}, {Ackermann}, {Ajello}, {Baldini},
  {Ballet}, {Barbiellini}, {Bastieri}, {Bechtol}, {Bellazzini}, {Berenji}, \&
  et~al.}]{2011ApJ...736..131A}
{Abdo}, A.~A., {Ackermann}, M., {Ajello}, M., {et~al.} 2011, \apj, 736, 131

\bibitem[{{Acero} {et~al.}(2015){Acero}, {Ackermann}, {Ajello}, {Albert},
  {Atwood}, {Axelsson}, {Baldini}, {Ballet}, {Barbiellini}, {Bastieri},
  {Belfiore}, {Bellazzini}, {Bissaldi}, {Blandford}, {Bloom}, {Bogart},
  {Bonino}, {Bottacini}, {Bregeon}, {Britto}, {Bruel}, {Buehler}, {Burnett},
  {Buson}, {Caliandro}, {Cameron}, {Caputo}, {Caragiulo}, {Caraveo},
  {Casandjian}, {Cavazzuti}, {Charles}, {Chaves}, {Chekhtman}, {Cheung},
  {Chiang}, {Chiaro}, {Ciprini}, {Claus}, {Cohen-Tanugi}, {Cominsky}, {Conrad},
  {Cutini}, {D'Ammando}, {de Angelis}, {DeKlotz}, {de Palma}, {Desiante},
  {Digel}, {Di Venere}, {Drell}, {Dubois}, {Dumora}, {Favuzzi}, {Fegan},
  {Ferrara}, {Finke}, {Franckowiak}, {Fukazawa}, {Funk}, {Fusco}, {Gargano},
  {Gasparrini}, {Giebels}, {Giglietto}, {Giommi}, {Giordano}, {Giroletti},
  {Glanzman}, {Godfrey}, {Grenier}, {Grondin}, {Grove}, {Guillemot}, {Guiriec},
  {Hadasch}, {Harding}, {Hays}, {Hewitt}, {Hill}, {Horan}, {Iafrate}, {Jogler},
  {J{\'o}hannesson}, {Johnson}, {Johnson}, {Johnson}, {Johnson}, {Kamae},
  {Kataoka}, {Katsuta}, {Kuss}, {La Mura}, {Landriu}, {Larsson}, {Latronico},
  {Lemoine-Goumard}, {Li}, {Li}, {Longo}, {Loparco}, {Lott}, {Lovellette},
  {Lubrano}, {Madejski}, {Massaro}, {Mayer}, {Mazziotta}, {McEnery},
  {Michelson}, {Mirabal}, {Mizuno}, {Moiseev}, {Mongelli}, {Monzani},
  {Morselli}, {Moskalenko}, {Murgia}, {Nuss}, {Ohno}, {Ohsugi}, {Omodei},
  {Orienti}, {Orlando}, {Ormes}, {Paneque}, {Panetta}, {Perkins},
  {Pesce-Rollins}, {Piron}, {Pivato}, {Porter}, {Racusin}, {Rando}, {Razzano},
  {Razzaque}, {Reimer}, {Reimer}, {Reposeur}, {Rochester}, {Romani},
  {Salvetti}, {S{\'a}nchez-Conde}, {Saz Parkinson}, {Schulz}, {Siskind},
  {Smith}, {Spada}, {Spandre}, {Spinelli}, {Stephens}, {Strong}, {Suson},
  {Takahashi}, {Takahashi}, {Tanaka}, {Thayer}, {Thayer}, {Thompson},
  {Tibaldo}, {Tibolla}, {Torres}, {Torresi}, {Tosti}, {Troja}, {Van Klaveren},
  {Vianello}, {Winer}, {Wood}, {Wood}, {Zimmer}, \& {Fermi-LAT
  Collaboration}}]{2015ApJS..218...23A}
{Acero}, F., {Ackermann}, M., {Ajello}, M., {et~al.} 2015, \apjs, 218, 23

\bibitem[{{Ackermann} {et~al.}(2012){Ackermann}, {Ajello}, {Allafort},
  {Schady}, {Baldini}, {Ballet}, {Barbiellini}, {Bastieri}, {Bellazzini},
  {Blandford}, {Bloom}, {Borgland}, {Bottacini}, {Bouvier}, {Bregeon},
  {Brigida}, {Bruel}, {Buehler}, {Buson}, {Caliandro}, {Cameron}, {Caraveo},
  {Cavazzuti}, {Cecchi}, {Charles}, {Chaves}, {Chekhtman}, {Cheung}, {Chiang},
  {Chiaro}, {Ciprini}, {Claus}, {Cohen-Tanugi}, {Conrad}, {Cutini},
  {D'Ammando}, {de Palma}, {Dermer}, {Digel}, {do Couto e Silva},
  {Dom{\'{\i}}nguez}, {Drell}, {Drlica-Wagner}, {Favuzzi}, {Fegan}, {Focke},
  {Franckowiak}, {Fukazawa}, {Funk}, {Fusco}, {Gargano}, {Gasparrini},
  {Gehrels}, {Germani}, {Giglietto}, {Giordano}, {Giroletti}, {Glanzman},
  {Godfrey}, {Grenier}, {Grove}, {Guiriec}, {Gustafsson}, {Hadasch},
  {Hayashida}, {Hays}, {Jackson}, {Jogler}, {Kataoka}, {Kn{\"o}dlseder},
  {Kuss}, {Lande}, {Larsson}, {Latronico}, {Longo}, {Loparco}, {Lovellette},
  {Lubrano}, {Mazziotta}, {McEnery}, {Mehault}, {Michelson}, {Mizuno}, {Monte},
  {Monzani}, {Morselli}, {Moskalenko}, {Murgia}, {Tramacere}, {Nuss},
  {Greiner}, {Ohno}, {Ohsugi}, {Omodei}, {Orienti}, {Orlando}, {Ormes},
  {Paneque}, {Perkins}, {Pesce-Rollins}, {Piron}, {Pivato}, {Porter},
  {Rain{\`o}}, {Rando}, {Razzano}, {Razzaque}, {Reimer}, {Reimer}, {Reyes},
  {Ritz}, {Rau}, {Romoli}, {Roth}, {S{\'a}nchez-Conde}, {Sanchez}, {Scargle},
  {Sgr{\`o}}, {Siskind}, {Spandre}, {Spinelli}, {Stawarz}, {Suson},
  {Takahashi}, {Tanaka}, {Thayer}, {Thompson}, {Tibaldo}, {Tinivella},
  {Torres}, {Tosti}, {Troja}, {Usher}, {Vandenbroucke}, {Vasileiou},
  {Vianello}, {Vitale}, {Waite}, {Winer}, {Wood}, \&
  {Wood}}]{2012Sci...338.1190A}
{Ackermann}, M., {Ajello}, M., {Allafort}, A., {et~al.} 2012, Science, 338,
  1190

\bibitem[{{Ackermann} {et~al.}(2013){Ackermann}, {Ajello}, {Allafort},
  {Atwood}, {Baldini}, {Ballet}, {Barbiellini}, {Bastieri}, {Bechtol},
  {Belfiore}, {Bellazzini}, {Bernieri}, {Bissaldi}, {Bloom}, {Bonamente},
  {Brandt}, {Bregeon}, {Brigida}, {Bruel}, {Buehler}, {Burnett}, {Buson},
  {Caliandro}, {Cameron}, {Campana}, {Caraveo}, {Casandjian}, {Cavazzuti},
  {Cecchi}, {Charles}, {Chaves}, {Chekhtman}, {Cheung}, {Chiang}, {Chiaro},
  {Ciprini}, {Claus}, {Cohen-Tanugi}, {Cominsky}, {Conrad}, {Cutini},
  {D'Ammando}, {de Angelis}, {de Palma}, {Dermer}, {Desiante}, {Digel}, {Di
  Venere}, {Drell}, {Drlica-Wagner}, {Favuzzi}, {Fegan}, {Ferrara}, {Focke},
  {Fortin}, {Franckowiak}, {Funk}, {Fusco}, {Gargano}, {Gasparrini}, {Gehrels},
  {Germani}, {Giglietto}, {Giommi}, {Giordano}, {Giroletti}, {Godfrey},
  {Gomez-Vargas}, {Grenier}, {Guiriec}, {Hadasch}, {Hanabata}, {Harding},
  {Hayashida}, {Hays}, {Hewitt}, {Hill}, {Horan}, {Hughes}, {Jogler},
  {J{\'o}hannesson}, {Johnson}, {Johnson}, {Johnson}, {Kamae}, {Kataoka},
  {Kawano}, {Kn{\"o}dlseder}, {Kuss}, {Lande}, {Larsson}, {Latronico},
  {Lemoine-Goumard}, {Longo}, {Loparco}, {Lott}, {Lovellette}, {Lubrano},
  {Massaro}, {Mayer}, {Mazziotta}, {McEnery}, {Mehault}, {Michelson}, {Mizuno},
  {Moiseev}, {Monzani}, {Morselli}, {Moskalenko}, {Murgia}, {Nemmen}, {Nuss},
  {Ohsugi}, {Okumura}, {Orienti}, {Ormes}, {Paneque}, {Perkins},
  {Pesce-Rollins}, {Piron}, {Pivato}, {Porter}, {Rain{\`o}}, {Razzano},
  {Reimer}, {Reimer}, {Reposeur}, {Ritz}, {Romani}, {Roth}, {Saz Parkinson},
  {Schulz}, {Sgr{\`o}}, {Siskind}, {Smith}, {Spandre}, {Spinelli}, {Stawarz},
  {Strong}, {Suson}, {Takahashi}, {Thayer}, {Thayer}, {Thompson}, {Tibaldo},
  {Tinivella}, {Torres}, {Tosti}, {Troja}, {Uchiyama}, {Usher},
  {Vandenbroucke}, {Vasileiou}, {Vianello}, {Vitale}, {Werner}, {Winer},
  {Wood}, \& {Wood}}]{2013ApJS..209...34A}
---. 2013, \apjs, 209, 34

\bibitem[{{Ackermann} {et~al.}(2015){Ackermann}, {Ajello}, {Atwood}, {Baldini},
  {Ballet}, {Barbiellini}, {Bastieri}, {Becerra Gonzalez}, {Bellazzini},
  {Bissaldi}, {Blandford}, {Bloom}, {Bonino}, {Bottacini}, {Brandt}, {Bregeon},
  {Britto}, {Bruel}, {Buehler}, {Buson}, {Caliandro}, {Cameron}, {Caragiulo},
  {Caraveo}, {Carpenter}, {Casandjian}, {Cavazzuti}, {Cecchi}, {Charles},
  {Chekhtman}, {Cheung}, {Chiang}, {Chiaro}, {Ciprini}, {Claus},
  {Cohen-Tanugi}, {Cominsky}, {Conrad}, {Cutini}, {D'Abrusco}, {D'Ammando}, {de
  Angelis}, {Desiante}, {Digel}, {Di Venere}, {Drell}, {Favuzzi}, {Fegan},
  {Ferrara}, {Finke}, {Focke}, {Franckowiak}, {Fuhrmann}, {Fukazawa},
  {Furniss}, {Fusco}, {Gargano}, {Gasparrini}, {Giglietto}, {Giommi},
  {Giordano}, {Giroletti}, {Glanzman}, {Godfrey}, {Grenier}, {Grove},
  {Guiriec}, {Hewitt}, {Hill}, {Horan}, {Itoh}, {J{\'o}hannesson}, {Johnson},
  {Johnson}, {Kataoka}, {Kawano}, {Krauss}, {Kuss}, {La Mura}, {Larsson},
  {Latronico}, {Leto}, {Li}, {Li}, {Longo}, {Loparco}, {Lott}, {Lovellette},
  {Lubrano}, {Madejski}, {Mayer}, {Mazziotta}, {McEnery}, {Michelson},
  {Mizuno}, {Moiseev}, {Monzani}, {Morselli}, {Moskalenko}, {Murgia}, {Nuss},
  {Ohno}, {Ohsugi}, {Ojha}, {Omodei}, {Orienti}, {Orlando}, {Paggi}, {Paneque},
  {Perkins}, {Pesce-Rollins}, {Piron}, {Pivato}, {Porter}, {Rain{\`o}},
  {Rando}, {Razzano}, {Razzaque}, {Reimer}, {Reimer}, {Romani}, {Salvetti},
  {Schaal}, {Schinzel}, {Schulz}, {Sgr{\`o}}, {Siskind}, {Sokolovsky}, {Spada},
  {Spandre}, {Spinelli}, {Stawarz}, {Suson}, {Takahashi}, {Takahashi},
  {Tanaka}, {Thayer}, {Thayer}, {Tibaldo}, {Torres}, {Torresi}, {Tosti},
  {Troja}, {Uchiyama}, {Vianello}, {Winer}, {Wood}, \&
  {Zimmer}}]{2015ApJ...810...14A}
{Ackermann}, M., {Ajello}, M., {Atwood}, W.~B., {et~al.} 2015, \apj, 810, 14

\bibitem[{{Ackermann} {et~al.}(2016){Ackermann}, {Ajello}, {Atwood}, {Baldini},
  {Ballet}, {Barbiellini}, {Bastieri}, {Becerra Gonzalez}, {Bellazzini},
  {Bissaldi}, {Blandford}, {Bloom}, {Bonino}, {Bottacini}, {Brandt}, {Bregeon},
  {Bruel}, {Buehler}, {Buson}, {Caliandro}, {Cameron}, {Caputo}, {Caragiulo},
  {Caraveo}, {Cavazzuti}, {Cecchi}, {Charles}, {Chekhtman}, {Cheung}, {Chiang},
  {Chiaro}, {Ciprini}, {Cohen}, {Cohen-Tanugi}, {Cominsky}, {Conrad}, {Cuoco},
  {Cutini}, {D'Ammando}, {de Angelis}, {de Palma}, {Desiante}, {Di Mauro}, {Di
  Venere}, {Dom{\'{\i}}nguez}, {Drell}, {Favuzzi}, {Fegan}, {Ferrara}, {Focke},
  {Fortin}, {Franckowiak}, {Fukazawa}, {Funk}, {Furniss}, {Fusco}, {Gargano},
  {Gasparrini}, {Giglietto}, {Giommi}, {Giordano}, {Giroletti}, {Glanzman},
  {Godfrey}, {Grenier}, {Grondin}, {Guillemot}, {Guiriec}, {Harding}, {Hays},
  {Hewitt}, {Hill}, {Horan}, {Iafrate}, {Hartmann}, {Jogler},
  {J{\'o}hannesson}, {Johnson}, {Kamae}, {Kataoka}, {Kn{\"o}dlseder}, {Kuss},
  {La Mura}, {Larsson}, {Latronico}, {Lemoine-Goumard}, {Li}, {Li}, {Longo},
  {Loparco}, {Lott}, {Lovellette}, {Lubrano}, {Madejski}, {Maldera},
  {Manfreda}, {Mayer}, {Mazziotta}, {Michelson}, {Mirabal}, {Mitthumsiri},
  {Mizuno}, {Moiseev}, {Monzani}, {Morselli}, {Moskalenko}, {Murgia}, {Nuss},
  {Ohsugi}, {Omodei}, {Orienti}, {Orlando}, {Ormes}, {Paneque}, {Perkins},
  {Pesce-Rollins}, {Petrosian}, {Piron}, {Pivato}, {Porter}, {Rain{\`o}},
  {Rando}, {Razzano}, {Razzaque}, {Reimer}, {Reimer}, {Reposeur}, {Romani},
  {S{\'a}nchez-Conde}, {Saz Parkinson}, {Schmid}, {Schulz}, {Sgr{\`o}},
  {Siskind}, {Spada}, {Spandre}, {Spinelli}, {Suson}, {Tajima}, {Takahashi},
  {Takahashi}, {Takahashi}, {Thayer}, {Thompson}, {Tibaldo}, {Torres}, {Tosti},
  {Troja}, {Vianello}, {Wood}, {Wood}, {Yassine}, {Zaharijas}, \&
  {Zimmer}}]{2016ApJS..222....5A}
---. 2016, \apjs, 222, 5

\bibitem[{{Aharonian}(2000)}]{Aharonian00}
{Aharonian}, F.~A. 2000, \na, 5, 377

\bibitem[{{Ajello} {et~al.}(2017){Ajello}, {Atwood}, {Baldini}, {Ballet},
  {Barbiellini}, {Bastieri}, {Bellazzini}, {Bissaldi}, {Blandford}, {Bloom},
  {Bonino}, {Bregeon}, {Britto}, {Bruel}, {Buehler}, {Buson}, {Cameron},
  {Caputo}, {Caragiulo}, {Caraveo}, {Cavazzuti}, {Cecchi}, {Charles},
  {Chekhtman}, {Cheung}, {Chiaro}, {Ciprini}, {Cohen}, {Costantin}, {Costanza},
  {Cuoco}, {Cutini}, {D'Ammando}, {de Palma}, {Desiante}, {Digel}, {Di Lalla},
  {Di Mauro}, {Di Venere}, {Dom{\'{\i}}nguez}, {Drell}, {Dumora}, {Favuzzi},
  {Fegan}, {Ferrara}, {Fortin}, {Franckowiak}, {Fukazawa}, {Funk}, {Fusco},
  {Gargano}, {Gasparrini}, {Giglietto}, {Giommi}, {Giordano}, {Giroletti},
  {Glanzman}, {Green}, {Grenier}, {Grondin}, {Grove}, {Guillemot}, {Guiriec},
  {Harding}, {Hays}, {Hewitt}, {Horan}, {J{\'o}hannesson}, {Kensei}, {Kuss},
  {La Mura}, {Larsson}, {Latronico}, {Lemoine-Goumard}, {Li}, {Longo},
  {Loparco}, {Lott}, {Lubrano}, {Magill}, {Maldera}, {Manfreda}, {Mazziotta},
  {McEnery}, {Meyer}, {Michelson}, {Mirabal}, {Mitthumsiri}, {Mizuno},
  {Moiseev}, {Monzani}, {Morselli}, {Moskalenko}, {Negro}, {Nuss}, {Ohsugi},
  {Omodei}, {Orienti}, {Orlando}, {Palatiello}, {Paliya}, {Paneque}, {Perkins},
  {Persic}, {Pesce-Rollins}, {Piron}, {Porter}, {Principe}, {Rain{\`o}},
  {Rando}, {Razzano}, {Razzaque}, {Reimer}, {Reimer}, {Reposeur}, {Saz
  Parkinson}, {Sgr{\`o}}, {Simone}, {Siskind}, {Spada}, {Spandre}, {Spinelli},
  {Stawarz}, {Suson}, {Takahashi}, {Tak}, {Thayer}, {Thayer}, {Thompson},
  {Torres}, {Torresi}, {Troja}, {Vianello}, {Wood}, \&
  {Wood}}]{2017ApJS..232...18A}
{Ajello}, M., {Atwood}, W.~B., {Baldini}, L., {et~al.} 2017, \apjs, 232, 18

\bibitem[{{Aleksi{\'c}} {et~al.}(2011){Aleksi{\'c}}, {Antonelli}, {Antoranz},
  {Backes}, {Barrio}, {Bastieri}, {Becerra Gonz{\'a}lez}, {Bednarek},
  {Berdyugin}, {Berger}, {Bernardini}, {Biland}, {Blanch}, {Bock}, {Boller},
  {Bonnoli}, {Borla Tridon}, {Braun}, {Bretz}, {Ca{\~n}ellas}, {Carmona},
  {Carosi}, {Colin}, {Colombo}, {Contreras}, {Cortina}, {Cossio}, {Covino},
  {Dazzi}, {De Angelis}, {De Cea del Pozo}, {De Lotto}, {Delgado Mendez},
  {Diago Ortega}, {Doert}, {Dom{\'{\i}}nguez}, {Dominis Prester}, {Dorner},
  {Doro}, {Elsaesser}, {Ferenc}, {Fonseca}, {Font}, {Fruck}, {Garc{\'{\i}}a
  L{\'o}pez}, {Garczarczyk}, {Garrido}, {Giavitto}, {Godinovi{\'c}}, {Hadasch},
  {H{\"a}fner}, {Herrero}, {Hildebrand}, {H{\"o}hne-M{\"o}nch}, {Hose},
  {Hrupec}, {Huber}, {Jogler}, {Klepser}, {Kr{\"a}henb{\"u}hl}, {Krause}, {La
  Barbera}, {Lelas}, {Leonardo}, {Lindfors}, {Lombardi}, {L{\'o}pez}, {Lorenz},
  {Makariev}, {Maneva}, {Mankuzhiyil}, {Mannheim}, {Maraschi}, {Mariotti},
  {Mart{\'{\i}}nez}, {Mazin}, {Meucci}, {Miranda}, {Mirzoyan}, {Miyamoto},
  {Mold{\'o}n}, {Moralejo}, {Nieto}, {Nilsson}, {Orito}, {Oya}, {Paneque},
  {Paoletti}, {Pardo}, {Paredes}, {Partini}, {Pasanen}, {Pauss},
  {Perez-Torres}, {Persic}, {Peruzzo}, {Pilia}, {Pochon}, {Prada}, {Prada
  Moroni}, {Prandini}, {Puljak}, {Reichardt}, {Reinthal}, {Rhode}, {Rib{\'o}},
  {Rico}, {R{\"u}gamer}, {Saggion}, {Saito}, {Saito}, {Salvati}, {Satalecka},
  {Scalzotto}, {Scapin}, {Schultz}, {Schweizer}, {Shayduk}, {Shore},
  {Sillanp{\"a}{\"a}}, {Sitarek}, {Sobczynska}, {Spanier}, {Spiro}, {Stamerra},
  {Steinke}, {Storz}, {Strah}, {Suri{\'c}}, {Takalo}, {Tavecchio}, {Temnikov},
  {Terzi{\'c}}, {Tescaro}, {Teshima}, {Thom}, {Tibolla}, {Torres}, {Treves},
  {Vankov}, {Vogler}, {Wagner}, {Weitzel}, {Zabalza}, {Zandanel}, {Zanin},
  {MAGIC Collaboration}, {Tanaka}, {Wood}, \& {Buson}}]{2011ApJ...730L...8A}
{Aleksi{\'c}}, J., {Antonelli}, L.~A., {Antoranz}, P., {et~al.} 2011, \apjl,
  730, L8

\bibitem[{{Aleksi{\'c}} {et~al.}(2014){Aleksi{\'c}}, {Ansoldi}, {Antonelli},
  {Antoranz}, {Babic}, {Bangale}, {de Almeida}, {Barrio}, {Gonz{\'a}lez},
  {Bednarek}, {Berger}, {Bernardini}, {Biland}, {Blanch}, {Bock}, {Bonnefoy},
  {Bonnoli}, {Borracci}, {Bretz}, {Carmona}, {Carosi}, {Fidalgo}, {Colin},
  {Colombo}, {Contreras}, {Cortina}, {Covino}, {Da Vela}, {Dazzi}, {De
  Angelis}, {De Caneva}, {De Lotto}, {Mendez}, {Doert}, {Dom{\'{\i}}nguez},
  {Prester}, {Dorner}, {Doro}, {Einecke}, {Eisenacher}, {Elsaesser}, {Farina},
  {Ferenc}, {Fonseca}, {Font}, {Frantzen}, {Fruck}, {L{\'o}pez}, {Garczarczyk},
  {Terrats}, {Gaug}, {Giavitto}, {Godinovi{\'c}}, {Mu{\~n}oz}, {Gozzini},
  {Hadasch}, {Hayashida}, {Herrero}, {Hildebrand}, {Hose}, {Hrupec}, {Idec},
  {Kadenius}, {Kellermann}, {Knoetig}, {Kodani}, {Konno}, {Krause}, {Kubo},
  {Kushida}, {Barbera}, {Lelas}, {Lewandowska}, {Lindfors}, {Lombardi},
  {L{\'o}pez}, {L{\'o}pez-Coto}, {L{\'o}pez-Oramas}, {Lorenz}, {Lozano},
  {Makariev}, {Mallot}, {Maneva}, {Mankuzhiyil}, {Mannheim}, {Maraschi},
  {Marcote}, {Mariotti}, {Mart{\'{\i}}nez}, {Mazin}, {Menzel}, {Meucci},
  {Miranda}, {Mirzoyan}, {Moralejo}, {Munar-Adrover}, {Nakajima},
  {Niedzwiecki}, {Nilsson}, {Nishijima}, {Nowak}, {Orito}, {Overkemping},
  {Paiano}, {Palatiello}, {Paneque}, {Paoletti}, {Paredes}, {Paredes-Fortuny},
  {Partini}, {Persic}, {Prada}, {Moroni}, {Prandini}, {Preziuso}, {Puljak},
  {Reinthal}, {Rhode}, {Rib{\'o}}, {Rico}, {Garcia}, {R{\"u}gamer}, {Saggion},
  {Saito}, {Saito}, {Salvati}, {Satalecka}, {Scalzotto}, {Scapin}, {Schultz},
  {Schweizer}, {Shore}, {Sillanp{\"a}{\"a}}, {Sitarek}, {Snidaric},
  {Sobczynska}, {Spanier}, {Stamatescu}, {Stamerra}, {Steinbring}, {Storz},
  {Sun}, {Suri{\'c}}, {Takalo}, {Takami}, {Tavecchio}, {Temnikov},
  {Terzi{\'c}}, {Tescaro}, {Teshima}, {Thaele}, {Tibolla}, {Torres}, {Toyama},
  {Treves}, {Vogler}, {Wagner}, {Zandanel}, \& {Zanin}}]{2014MNRAS.440..530A}
{Aleksi{\'c}}, J., {Ansoldi}, S., {Antonelli}, L.~A., {et~al.} 2014, \mnras,
  440, 530

\bibitem[{{Aleksi{\'c}} {et~al.}(2016){Aleksi{\'c}}, {Ansoldi}, {Antonelli},
  {Antoranz}, {Babic}, {Bangale}, {Barcel{\'o}}, {Barrio}, {Becerra
  Gonz{\'a}lez}, {Bednarek}, {Bernardini}, {Biasuzzi}, {Biland}, {Bitossi},
  {Blanch}, {Bonnefoy}, {Bonnoli}, {Borracci}, {Bretz}, {Carmona}, {Carosi},
  {Cecchi}, {Colin}, {Colombo}, {Contreras}, {Corti}, {Cortina}, {Covino}, {Da
  Vela}, {Dazzi}, {De Angelis}, {De Caneva}, {De Lotto}, {de O{\~n}a Wilhelmi},
  {Delgado Mendez}, {Dettlaff}, {Dominis Prester}, {Dorner}, {Doro}, {Einecke},
  {Eisenacher}, {Elsaesser}, {Fidalgo}, {Fink}, {Fonseca}, {Font}, {Frantzen},
  {Fruck}, {Galindo}, {Garc{\'{\i}}a L{\'o}pez}, {Garczarczyk}, {Garrido
  Terrats}, {Gaug}, {Giavitto}, {Godinovi{\'c}}, {Gonz{\'a}lez Mu{\~n}oz},
  {Gozzini}, {Haberer}, {Hadasch}, {Hanabata}, {Hayashida}, {Herrera},
  {Hildebrand}, {Hose}, {Hrupec}, {Idec}, {Illa}, {Kadenius}, {Kellermann},
  {Knoetig}, {Kodani}, {Konno}, {Krause}, {Kubo}, {Kushida}, {La Barbera},
  {Lelas}, {Lemus}, {Lewandowska}, {Lindfors}, {Lombardi}, {Longo},
  {L{\'o}pez}, {L{\'o}pez-Coto}, {L{\'o}pez-Oramas}, {Lorca}, {Lorenz},
  {Lozano}, {Makariev}, {Mallot}, {Maneva}, {Mankuzhiyil}, {Mannheim},
  {Maraschi}, {Marcote}, {Mariotti}, {Mart{\'{\i}}nez}, {Mazin}, {Menzel},
  {Miranda}, {Mirzoyan}, {Moralejo}, {Munar-Adrover}, {Nakajima}, {Negrello},
  {Neustroev}, {Niedzwiecki}, {Nilsson}, {Nishijima}, {Noda}, {Orito},
  {Overkemping}, {Paiano}, {Palatiello}, {Paneque}, {Paoletti}, {Paredes},
  {Paredes-Fortuny}, {Persic}, {Poutanen}, {Prada Moroni}, {Prandini},
  {Puljak}, {Reinthal}, {Rhode}, {Rib{\'o}}, {Rico}, {Rodriguez Garcia},
  {R{\"u}gamer}, {Saito}, {Saito}, {Satalecka}, {Scalzotto}, {Scapin},
  {Schultz}, {Schlammer}, {Schmidl}, {Schweizer}, {Shore}, {Sillanp{\"a}{\"a}},
  {Sitarek}, {Snidaric}, {Sobczynska}, {Spanier}, {Stamerra}, {Steinbring},
  {Storz}, {Strzys}, {Takalo}, {Takami}, {Tavecchio}, {Tejedor}, {Temnikov},
  {Terzi{\'c}}, {Tescaro}, {Teshima}, {Thaele}, {Tibolla}, {Torres}, {Toyama},
  {Treves}, {Vogler}, {Wetteskind}, {Will}, \& {Zanin}}]{2016APh....72...76A}
---. 2016, Astroparticle Physics, 72, 76

\bibitem[{{Arnaud}(1996)}]{1996ASPC..101...17A}
{Arnaud}, K.~A. 1996, in Astronomical Society of the Pacific Conference Series,
  Vol. 101, Astronomical Data Analysis Software and Systems V, ed. G.~H.
  {Jacoby} \& J.~{Barnes}, 17

\bibitem[{{Atwood} {et~al.}(2013){Atwood}, {Albert}, {Baldini}, {Tinivella},
  {Bregeon}, {Pesce-Rollins}, {Sgr{\`o}}, {Bruel}, {Charles}, {Drlica-Wagner},
  {Franckowiak}, {Jogler}, {Rochester}, {Usher}, {Wood}, {Cohen-Tanugi}, \&
  {S.~Zimmer for the Fermi-LAT Collaboration}}]{2013arXiv1303.3514A}
{Atwood}, W., {Albert}, A., {Baldini}, L., {et~al.} 2013, ArXiv e-prints,
  arXiv:1303.3514

\bibitem[{{Barnacka} {et~al.}(2016){Barnacka}, {Geller}, {Dell'Antonio}, \&
  {Zitrin}}]{2016ApJ...821...58B}
{Barnacka}, A., {Geller}, M.~J., {Dell'Antonio}, I.~P., \& {Zitrin}, A. 2016,
  \apj, 821, 58

\bibitem[{{Begelman} \& {Sikora}(1987)}]{1987ApJ...322..650B}
{Begelman}, M.~C., \& {Sikora}, M. 1987, \apj, 322, 650

\bibitem[{{Bonnoli} {et~al.}(2011){Bonnoli}, {Ghisellini}, {Foschini},
  {Tavecchio}, \& {Ghirlanda}}]{2011MNRAS.410..368B}
{Bonnoli}, G., {Ghisellini}, G., {Foschini}, L., {Tavecchio}, F., \&
  {Ghirlanda}, G. 2011, \mnras, 410, 368

\bibitem[{{Bonometto} \& {Saggion}(1973)}]{Bonometto73}
{Bonometto}, S., \& {Saggion}, A. 1973, \aap, 23, 9

\bibitem[{{B{\"o}ttcher} \& {Els}(2016)}]{2016ApJ...821..102B}
{B{\"o}ttcher}, M., \& {Els}, P. 2016, \apj, 821, 102

\bibitem[{{B{\"o}ttcher} {et~al.}(2013){B{\"o}ttcher}, {Reimer}, {Sweeney}, \&
  {Prakash}}]{Boettcher13}
{B{\"o}ttcher}, M., {Reimer}, A., {Sweeney}, K., \& {Prakash}, A. 2013, \apj,
  768, 54

\bibitem[{{Breeveld} {et~al.}(2011){Breeveld}, {Landsman}, {Holland}, {Roming},
  {Kuin}, \& {Page}}]{2011AIPC.1358..373B}
{Breeveld}, A.~A., {Landsman}, W., {Holland}, S.~T., {et~al.} 2011, in American
  Institute of Physics Conference Series, Vol. 1358, American Institute of
  Physics Conference Series, ed. J.~E. {McEnery}, J.~L. {Racusin}, \&
  N.~{Gehrels}, 373--376

\bibitem[{{Burrows} {et~al.}(2005){Burrows}, {Hill}, {Nousek}, {Kennea},
  {Wells}, {Osborne}, {Abbey}, {Beardmore}, {Mukerjee}, {Short}, {Chincarini},
  {Campana}, {Citterio}, {Moretti}, {Pagani}, {Tagliaferri}, {Giommi},
  {Capalbi}, {Tamburelli}, {Angelini}, {Cusumano}, {Br{\"a}uninger}, {Burkert},
  \& {Hartner}}]{2005SSRv..120..165B}
{Burrows}, D.~N., {Hill}, J.~E., {Nousek}, J.~A., {et~al.} 2005, \ssr, 120, 165

\bibitem[{{Celotti} \& {Ghisellini}(2008)}]{2008MNRAS.385..283C}
{Celotti}, A., \& {Ghisellini}, G. 2008, \mnras, 385, 283

\bibitem[{{Celotti} {et~al.}(1997){Celotti}, {Padovani}, \&
  {Ghisellini}}]{1997MNRAS.286..415C}
{Celotti}, A., {Padovani}, P., \& {Ghisellini}, G. 1997, \mnras, 286, 415

\bibitem[{{Cerruti} {et~al.}(2013){Cerruti}, {Dermer}, {Lott}, {Boisson}, \&
  {Zech}}]{2013ApJ...771L...4C}
{Cerruti}, M., {Dermer}, C.~D., {Lott}, B., {Boisson}, C., \& {Zech}, A. 2013,
  \apjl, 771, L4

\bibitem[{{Dai} {et~al.}(2007){Dai}, {Xie}, {Zhou}, {Li}, {Chen}, \&
  {Ma}}]{2007AJ....133.2187D}
{Dai}, H., {Xie}, G.~Z., {Zhou}, S.~B., {et~al.} 2007, \aj, 133, 2187

\bibitem[{{Dermer}(1995)}]{1995ApJ...446L..63D}
{Dermer}, C.~D. 1995, \apjl, 446, L63

\bibitem[{{Dermer} {et~al.}(2014){Dermer}, {Cerruti}, {Lott}, {Boisson}, \&
  {Zech}}]{2014ApJ...782...82D}
{Dermer}, C.~D., {Cerruti}, M., {Lott}, B., {Boisson}, C., \& {Zech}, A. 2014,
  \apj, 782, 82

\bibitem[{{Dom{\'{\i}}nguez} \& {Ajello}(2015)}]{2015ApJ...813L..34D}
{Dom{\'{\i}}nguez}, A., \& {Ajello}, M. 2015, \apjl, 813, L34

\bibitem[{{Dom{\'{\i}}nguez} {et~al.}(2013){Dom{\'{\i}}nguez}, {Finke},
  {Prada}, {Primack}, {Kitaura}, {Siana}, \& {Paneque}}]{2013ApJ...770...77D}
{Dom{\'{\i}}nguez}, A., {Finke}, J.~D., {Prada}, F., {et~al.} 2013, \apj, 770,
  77

\bibitem[{{Dom{\'{\i}}nguez} {et~al.}(2011){Dom{\'{\i}}nguez}, {Primack},
  {Rosario}, {Prada}, {Gilmore}, {Faber}, {Koo}, {Somerville},
  {P{\'e}rez-Torres}, {P{\'e}rez-Gonz{\'a}lez}, {Huang}, {Davis},
  {Guhathakurta}, {Barmby}, {Conselice}, {Lozano}, {Newman}, \&
  {Cooper}}]{2011MNRAS.410.2556D}
{Dom{\'{\i}}nguez}, A., {Primack}, J.~R., {Rosario}, D.~J., {et~al.} 2011,
  \mnras, 410, 2556

\bibitem[{{Donea} \& {Protheroe}(2003)}]{2003APh....18..377D}
{Donea}, A.-C., \& {Protheroe}, R.~J. 2003, Astroparticle Physics, 18, 377

\bibitem[{{Evans} {et~al.}(2009){Evans}, {Beardmore}, {Page}, {Osborne},
  {O'Brien}, {Willingale}, {Starling}, {Burrows}, {Godet}, {Vetere}, {Racusin},
  {Goad}, {Wiersema}, {Angelini}, {Capalbi}, {Chincarini}, {Gehrels}, {Kennea},
  {Margutti}, {Morris}, {Mountford}, {Pagani}, {Perri}, {Romano}, \&
  {Tanvir}}]{2009MNRAS.397.1177E}
{Evans}, P.~A., {Beardmore}, A.~P., {Page}, K.~L., {et~al.} 2009, \mnras, 397,
  1177

\bibitem[{{Falomo} {et~al.}(2017){Falomo}, {Treves}, {Scarpa}, {Paiano}, \&
  {Landoni}}]{2017MNRAS.470.2814F}
{Falomo}, R., {Treves}, A., {Scarpa}, R., {Paiano}, S., \& {Landoni}, M. 2017,
  \mnras, 470, 2814

\bibitem[{{Francis} {et~al.}(1991){Francis}, {Hewett}, {Foltz}, {Chaffee},
  {Weymann}, \& {Morris}}]{1991ApJ...373..465F}
{Francis}, P.~J., {Hewett}, P.~C., {Foltz}, C.~B., {et~al.} 1991, \apj, 373,
  465

\bibitem[{{Frank} {et~al.}(2002){Frank}, {King}, \&
  {Raine}}]{2002apa..book.....F}
{Frank}, J., {King}, A., \& {Raine}, D.~J. 2002, {Accretion Power in
  Astrophysics, by Juhan Frank and Andrew King and Derek Raine, pp.~398.~ISBN
  0521620538.~Cambridge, UK: Cambridge University Press, February 2002}

\bibitem[{{Furniss} \& {McConville}(2013)}]{2013arXiv1303.1103F}
{Furniss}, A., \& {McConville}, W. 2013, ArXiv e-prints, arXiv:1303.1103

\bibitem[{{Ghisellini} \& {Tavecchio}(2009)}]{2009MNRAS.397..985G}
{Ghisellini}, G., \& {Tavecchio}, F. 2009, \mnras, 397, 985

\bibitem[{{Ghisellini} {et~al.}(2014){Ghisellini}, {Tavecchio}, {Maraschi},
  {Celotti}, \& {Sbarrato}}]{2014Natur.515..376G}
{Ghisellini}, G., {Tavecchio}, F., {Maraschi}, L., {Celotti}, A., \&
  {Sbarrato}, T. 2014, \nat, 515, 376

\bibitem[{{Guo} {et~al.}(2016){Guo}, {Li}, {Li}, {Daughton}, {Zhang},
  {Lloyd-Ronning}, {Liu}, {Zhang}, \& {Deng}}]{Guo16}
{Guo}, F., {Li}, X., {Li}, H., {et~al.} 2016, \apjl, 818, L9

\bibitem[{{Hauser} \& {Dwek}(2001)}]{2001ARA&A..39..249H}
{Hauser}, M.~G., \& {Dwek}, E. 2001, \araa, 39, 249

\bibitem[{{Hervet} {et~al.}(2016){Hervet}, {Boisson}, \&
  {Sol}}]{2016A&A...592A..22H}
{Hervet}, O., {Boisson}, C., \& {Sol}, H. 2016, \aap, 592, A22

\bibitem[{{H.E.S.S.~Collaboration} {et~al.}(2013){H.E.S.S.~Collaboration},
  {Abramowski}, {Acero}, {Aharonian}, {Akhperjanian}, {Anton}, {Balenderan},
  {Balzer}, {Barnacka}, {Becherini}, {Becker Tjus}, {Behera}, {Bernl{\"o}hr},
  {Birsin}, {Biteau}, {Bochow}, {Boisson}, {Bolmont}, {Bordas}, {Brucker},
  {Brun}, {Brun}, {Bulik}, {Carrigan}, {Casanova}, {Cerruti}, {Chadwick},
  {Chaves}, {Cheesebrough}, {Colafrancesco}, {Cologna}, {Conrad}, {Couturier},
  {Dalton}, {Daniel}, {Davids}, {Degrange}, {Deil}, {deWilt}, {Dickinson},
  {Djannati-Ata{\"i}}, {Domainko}, {O'C.~Drury}, {Dubus}, {Dutson}, {Dyks},
  {Dyrda}, {Egberts}, {Eger}, {Espigat}, {Fallon}, {Farnier}, {Fegan},
  {Feinstein}, {Fernandes}, {Fernandez}, {Fiasson}, {Fontaine}, {F{\"o}rster},
  {F{\"u}{\ss}ling}, {Gajdus}, {Gallant}, {Garrigoux}, {Gast}, {Giebels},
  {Glicenstein}, {Gl{\"u}ck}, {G{\"o}ring}, {Grondin}, {Grudzi{\'n}ska},
  {H{\"a}ffner}, {Hague}, {Hahn}, {Hampf}, {Harris}, {Hauser}, {Heinz},
  {Heinzelmann}, {Henri}, {Hermann}, {Hillert}, {Hinton}, {Hofmann},
  {Hofverberg}, {Holler}, {Horns}, {Jacholkowska}, {Jahn}, {Jamrozy}, {Jung},
  {Kastendieck}, {Katarzy{\'n}ski}, {Katz}, {Kaufmann}, {Kh{\'e}lifi},
  {Klepser}, {Klochkov}, {Klu{\'z}niak}, {Kneiske}, {Kolitzus}, {Komin},
  {Kosack}, {Kossakowski}, {Krayzel}, {Kr{\"u}ger}, {Laffon}, {Lamanna},
  {Lefaucheur}, {Lemoine-Goumard}, {Lenain}, {Lennarz}, {Lohse}, {Lopatin},
  {Lu}, {Marandon}, {Marcowith}, {Masbou}, {Maurin}, {Maxted}, {Mayer},
  {McComb}, {Medina}, {M{\'e}hault}, {Menzler}, {Moderski}, {Mohamed},
  {Moulin}, {Naumann}, {Naumann-Godo}, {de Naurois}, {Nedbal}, {Nguyen},
  {Niemiec}, {Nolan}, {Ohm}, {de O{\~n}a Wilhelmi}, {Opitz}, {Ostrowski},
  {Oya}, {Panter}, {Parsons}, {Paz Arribas}, {Pekeur}, {Pelletier}, {Perez},
  {Petrucci}, {Peyaud}, {Pita}, {P{\"u}hlhofer}, {Punch}, {Quirrenbach},
  {Raab}, {Raue}, {Reimer}, {Reimer}, {Renaud}, {de los Reyes}, {Rieger},
  {Ripken}, {Rob}, {Rosier-Lees}, {Rowell}, {Rudak}, {Rulten}, {Sahakian},
  {Sanchez}, {Santangelo}, {Schlickeiser}, {Schulz}, {Schwanke}, {Schwarzburg},
  {Schwemmer}, {Sheidaei}, {Skilton}, {Sol}, {Spengler}, {Stawarz},
  {Steenkamp}, {Stegmann}, {Stinzing}, {Stycz}, {Sushch}, {Szostek},
  {Tavernet}, {Terrier}, {Tluczykont}, {Trichard}, {Valerius}, {van Eldik},
  {Vasileiadis}, {Venter}, {Viana}, {Vincent}, {V{\"o}lk}, {Volpe}, {Vorobiov},
  {Vorster}, {Wagner}, {Ward}, {White}, {Wierzcholska}, {Wouters}, {Zacharias},
  {Zajczyk}, {Zdziarski}, {Zech}, \& {Zechlin}}]{2013A&A...554A.107H}
{H.E.S.S.~Collaboration}, {Abramowski}, A., {Acero}, F., {et~al.} 2013, \aap,
  554, A107

\bibitem[{{Holler} {et~al.}(2015){Holler}, {de Naurois}, {Zaborov}, {Balzer},
  \& {Chalm{\'e}-Calvet}}]{2015ICRC...34..980H}
{Holler}, M., {de Naurois}, M., {Zaborov}, D., {Balzer}, A., \&
  {Chalm{\'e}-Calvet}, R. 2015, in International Cosmic Ray Conference,
  Vol.~34, 34th International Cosmic Ray Conference (ICRC2015), 980

\bibitem[{{Janiak} {et~al.}(2015){Janiak}, {Sikora}, \&
  {Moderski}}]{2015MNRAS.449..431J}
{Janiak}, M., {Sikora}, M., \& {Moderski}, R. 2015, \mnras, 449, 431

\bibitem[{{Jorstad} {et~al.}(2005){Jorstad}, {Marscher}, {Lister}, {Stirling},
  {Cawthorne}, {Gear}, {G{\'o}mez}, {Stevens}, {Smith}, {Forster}, \&
  {Robson}}]{2005AJ....130.1418J}
{Jorstad}, S.~G., {Marscher}, A.~P., {Lister}, M.~L., {et~al.} 2005, \aj, 130,
  1418

\bibitem[{{Joshi} {et~al.}(2014){Joshi}, {Marscher}, \&
  {B{\"o}ttcher}}]{2014ApJ...785..132J}
{Joshi}, M., {Marscher}, A.~P., \& {B{\"o}ttcher}, M. 2014, \apj, 785, 132

\bibitem[{{Kalberla} {et~al.}(2005){Kalberla}, {Burton}, {Hartmann}, {Arnal},
  {Bajaja}, {Morras}, \& {P{\"o}ppel}}]{2005A&A...440..775K}
{Kalberla}, P.~M.~W., {Burton}, W.~B., {Hartmann}, D., {et~al.} 2005, \aap,
  440, 775

\bibitem[{{Kapanadze} {et~al.}(2017){Kapanadze}, {Dorner}, {Romano},
  {Vercellone}, {Kapanadze}, \& {Tabagari}}]{2017ApJ...848..103K}
{Kapanadze}, B., {Dorner}, D., {Romano}, P., {et~al.} 2017, \apj, 848, 103

\bibitem[{{Lister} {et~al.}(2013){Lister}, {Aller}, {Aller}, {Homan},
  {Kellermann}, {Kovalev}, {Pushkarev}, {Richards}, {Ros}, \&
  {Savolainen}}]{2013AJ....146..120L}
{Lister}, M.~L., {Aller}, M.~F., {Aller}, H.~D., {et~al.} 2013, \aj, 146, 120

\bibitem[{{Liu} \& {Bai}(2006)}]{2006ApJ...653.1089L}
{Liu}, H.~T., \& {Bai}, J.~M. 2006, \apj, 653, 1089

\bibitem[{{Madejski} {et~al.}(2016){Madejski}, {Nalewajko}, {Madsen}, {Chiang},
  {Balokovi{\'c}}, {Paneque}, {Furniss}, {Hayashida}, {Urry}, {Sikora},
  {Ajello}, {Blandford}, {Harrison}, {Sanchez}, {Giebels}, {Stern},
  {Alexander}, {Barret}, {Boggs}, {Christensen}, {Craig}, {Forster}, {Giommi},
  {Grefenstette}, {Hailey}, {Hornstrup}, {Kitaguchi}, {Koglin}, {Mao},
  {Miyasaka}, {Mori}, {Perri}, {Pivovaroff}, {Puccetti}, {Rana}, {Westergaard},
  {Zhang}, \& {Zoglauer}}]{2016ApJ...831..142M}
{Madejski}, G.~M., {Nalewajko}, K., {Madsen}, K.~K., {et~al.} 2016, \apj, 831,
  142

\bibitem[{{MAGIC Collaboration} {et~al.}(2008){MAGIC Collaboration}, {Albert},
  {Aliu}, {Anderhub}, {Antonelli}, {Antoranz}, {Backes}, {Baixeras}, {Barrio},
  {Bartko}, {Bastieri}, {Becker}, {Bednarek}, {Berger}, {Bernardini},
  {Bigongiari}, {Biland}, {Bock}, {Bonnoli}, {Bordas}, {Bosch-Ramon}, {Bretz},
  {Britvitch}, {Camara}, {Carmona}, {Chilingarian}, {Commichau}, {Contreras},
  {Cortina}, {Costado}, {Covino}, {Curtef}, {Dazzi}, {De Angelis}, {Cea del
  Pozo}, {de los Reyes}, {De Lotto}, {De Maria}, {De Sabata}, {Mendez},
  {Dominguez}, {Dorner}, {Doro}, {Errando}, {Fagiolini}, {Ferenc},
  {Fern{\'a}ndez}, {Firpo}, {Fonseca}, {Font}, {Galante}, {Garc{\'{\i}}a
  L{\'o}pez}, {Garczarczyk}, {Gaug}, {Goebel}, {Hayashida}, {Herrero},
  {H{\"o}hne}, {Hose}, {Hsu}, {Huber}, {Jogler}, {Kneiske}, {Kranich}, {La
  Barbera}, {Laille}, {Leonardo}, {Lindfors}, {Lombardi}, {Longo}, {L{\'o}pez},
  {Lorenz}, {Majumdar}, {Maneva}, {Mankuzhiyil}, {Mannheim}, {Maraschi},
  {Mariotti}, {Mart{\'{\i}}nez}, {Mazin}, {Meucci}, {Meyer}, {Miranda},
  {Mirzoyan}, {Mizobuchi}, {Moles}, {Moralejo}, {Nieto}, {Nilsson}, {Ninkovic},
  {Otte}, {Oya}, {Panniello}, {Paoletti}, {Paredes}, {Pasanen}, {Pascoli},
  {Pauss}, {Pegna}, {Perez-Torres}, {Persic}, {Peruzzo}, {Piccioli}, {Prada},
  {Prandini}, {Puchades}, {Raymers}, {Rhode}, {Rib{\'o}}, {Rico}, {Rissi},
  {Robert}, {R{\"u}gamer}, {Saggion}, {Saito}, {Salvati}, {Sanchez-Conde},
  {Sartori}, {Satalecka}, {Scalzotto}, {Scapin}, {Schmitt}, {Schweizer},
  {Shayduk}, {Shinozaki}, {Shore}, {Sidro}, {Sierpowska-Bartosik},
  {Sillanp{\"a}{\"a}}, {Sobczynska}, {Spanier}, {Stamerra}, {Stark}, {Takalo},
  {Tavecchio}, {Temnikov}, {Tescaro}, {Teshima}, {Tluczykont}, {Torres},
  {Turini}, {Vankov}, {Venturini}, {Vitale}, {Wagner}, {Wittek}, {Zabalza},
  {Zandanel}, {Zanin}, \& {Zapatero}}]{2008Sci...320.1752M}
{MAGIC Collaboration}, {Albert}, J., {Aliu}, E., {et~al.} 2008, Science, 320,
  1752

\bibitem[{{Mannheim} \& {Biermann}(1992)}]{Mannheim92}
{Mannheim}, K., \& {Biermann}, P.~L. 1992, \aap, 253, L21

\bibitem[{{Marscher}(2014)}]{Marscher14}
{Marscher}, A.~P. 2014, \apj, 780, 87

\bibitem[{{Marscher} \& {Gear}(1985)}]{1985ApJ...298..114M}
{Marscher}, A.~P., \& {Gear}, W.~K. 1985, \apj, 298, 114

\bibitem[{{McConville} {et~al.}(2011){McConville}, {Ostorero}, {Moderski},
  {Stawarz}, {Cheung}, {Ajello}, {Bouvier}, {Bregeon}, {Donato}, {Finke},
  {Furniss}, {McEnery}, {Monzani}, {Orienti}, {Reyes}, {Rossetti}, \&
  {Williams}}]{2011ApJ...738..148M}
{McConville}, W., {Ostorero}, L., {Moderski}, R., {et~al.} 2011, \apj, 738, 148

\bibitem[{{M{\"u}cke} \& {Protheroe}(2001)}]{Mucke01}
{M{\"u}cke}, A., \& {Protheroe}, R.~J. 2001, Astroparticle Physics, 15, 121

\bibitem[{{Oke} {et~al.}(1984){Oke}, {Shields}, \&
  {Korycansky}}]{1984ApJ...277...64O}
{Oke}, J.~B., {Shields}, G.~A., \& {Korycansky}, D.~G. 1984, \apj, 277, 64

\bibitem[{{Paliya} {et~al.}(2017){Paliya}, {Marcotulli}, {Ajello}, {Joshi},
  {Sahayanathan}, {Rao}, \& {Hartmann}}]{2017ApJ...851...33P}
{Paliya}, V.~S., {Marcotulli}, L., {Ajello}, M., {et~al.} 2017, \apj, 851, 33

\bibitem[{{Paliya} {et~al.}(2015){Paliya}, {Sahayanathan}, \&
  {Stalin}}]{2015ApJ...803...15P}
{Paliya}, V.~S., {Sahayanathan}, S., \& {Stalin}, C.~S. 2015, \apj, 803, 15

\bibitem[{{Pavlidou} {et~al.}(2014){Pavlidou}, {Angelakis}, {Myserlis},
  {Blinov}, {King}, {Papadakis}, {Tassis}, {Hovatta}, {Pazderska},
  {Paleologou}, {Balokovi{\'c}}, {Feiler}, {Fuhrmann}, {Khodade}, {Kus},
  {Kylafis}, {Modi}, {Panopoulou}, {Papamastorakis}, {Pazderski}, {Pearson},
  {Rajarshi}, {Ramaprakash}, {Readhead}, {Reig}, \&
  {Zensus}}]{2014MNRAS.442.1693P}
{Pavlidou}, V., {Angelakis}, E., {Myserlis}, I., {et~al.} 2014, \mnras, 442,
  1693

\bibitem[{{Pian} {et~al.}(1999){Pian}, {Urry}, {Maraschi}, {Madejski},
  {McHardy}, {Koratkar}, {Treves}, {Chiappetti}, {Grandi}, {Hartman}, {Kubo},
  {Leach}, {Pesce}, {Imhoff}, {Thompson}, \& {Wehrle}}]{1999ApJ...521..112P}
{Pian}, E., {Urry}, C.~M., {Maraschi}, L., {et~al.} 1999, \apj, 521, 112

\bibitem[{{Pjanka} {et~al.}(2017){Pjanka}, {Zdziarski}, \&
  {Sikora}}]{2017MNRAS.465.3506P}
{Pjanka}, P., {Zdziarski}, A.~A., \& {Sikora}, M. 2017, \mnras, 465, 3506

\bibitem[{{Planck Collaboration} {et~al.}(2016){Planck Collaboration}, {Ade},
  {Aghanim}, {Arnaud}, {Ashdown}, {Aumont}, {Baccigalupi}, {Banday},
  {Barreiro}, {Bartlett}, \& et~al.}]{2016A&A...594A..13P}
{Planck Collaboration}, {Ade}, P.~A.~R., {Aghanim}, N., {et~al.} 2016, \aap,
  594, A13

\bibitem[{{Poutanen} \& {Stern}(2010)}]{2010ApJ...717L.118P}
{Poutanen}, J., \& {Stern}, B. 2010, \apjl, 717, L118

\bibitem[{{Roming} {et~al.}(2005){Roming}, {Kennedy}, {Mason}, {Nousek}, {Ahr},
  {Bingham}, {Broos}, {Carter}, {Hancock}, {Huckle}, {Hunsberger}, {Kawakami},
  {Killough}, {Koch}, {McLelland}, {Smith}, {Smith}, {Soto}, {Boyd},
  {Breeveld}, {Holland}, {Ivanushkina}, {Pryzby}, {Still}, \&
  {Stock}}]{2005SSRv..120...95R}
{Roming}, P.~W.~A., {Kennedy}, T.~E., {Mason}, K.~O., {et~al.} 2005, \ssr, 120,
  95

\bibitem[{{Rybicki} \& {Lightman}(1985)}]{Rybicki85}
{Rybicki}, G.~B., \& {Lightman}, A.~P. 1985, {Radiative processes in
  astrophysics.}

\bibitem[{{Sbarrato} {et~al.}(2012){Sbarrato}, {Ghisellini}, {Maraschi}, \&
  {Colpi}}]{2012MNRAS.421.1764S}
{Sbarrato}, T., {Ghisellini}, G., {Maraschi}, L., \& {Colpi}, M. 2012, \mnras,
  421, 1764

\bibitem[{{Schlafly} \& {Finkbeiner}(2011)}]{2011ApJ...737..103S}
{Schlafly}, E.~F., \& {Finkbeiner}, D.~P. 2011, \apj, 737, 103

\bibitem[{{Shakura} \& {Sunyaev}(1973)}]{1973A&A....24..337S}
{Shakura}, N.~I., \& {Sunyaev}, R.~A. 1973, \aap, 24, 337

\bibitem[{{Shaw} {et~al.}(2012){Shaw}, {Romani}, {Cotter}, {Healey},
  {Michelson}, {Readhead}, {Richards}, {Max-Moerbeck}, {King}, \&
  {Potter}}]{2012ApJ...748...49S}
{Shaw}, M.~S., {Romani}, R.~W., {Cotter}, G., {et~al.} 2012, \apj, 748, 49

\bibitem[{{Sikora} {et~al.}(2016){Sikora}, {Rutkowski}, \&
  {Begelman}}]{2016MNRAS.457.1352S}
{Sikora}, M., {Rutkowski}, M., \& {Begelman}, M.~C. 2016, \mnras, 457, 1352

\bibitem[{{Sironi} \& {Spitkovsky}(2011)}]{Sironi11}
{Sironi}, L., \& {Spitkovsky}, A. 2011, \apj, 726, 75

\bibitem[{{Smith} {et~al.}(2009){Smith}, {Montiel}, {Rightley}, {Turner},
  {Schmidt}, \& {Jannuzi}}]{2009arXiv0912.3621S}
{Smith}, P.~S., {Montiel}, E., {Rightley}, S., {et~al.} 2009, arXiv:0912.3621,
  arXiv:0912.3621

\bibitem[{{Stickel} {et~al.}(1989){Stickel}, {Fried}, \&
  {Kuehr}}]{1989A&AS...80..103S}
{Stickel}, M., {Fried}, J.~W., \& {Kuehr}, H. 1989, \aaps, 80, 103

\bibitem[{{Stickel} \& {Kuehr}(1993)}]{1993A&AS..100..395S}
{Stickel}, M., \& {Kuehr}, H. 1993, \aaps, 100, 395

\bibitem[{{Tavecchio} {et~al.}(2011){Tavecchio}, {Becerra-Gonzalez},
  {Ghisellini}, {Stamerra}, {Bonnoli}, {Foschini}, \&
  {Maraschi}}]{2011A&A...534A..86T}
{Tavecchio}, F., {Becerra-Gonzalez}, J., {Ghisellini}, G., {et~al.} 2011, \aap,
  534, A86

\bibitem[{{Urry} \& {Padovani}(1995)}]{1995PASP..107..803U}
{Urry}, C.~M., \& {Padovani}, P. 1995, \pasp, 107, 803

\bibitem[{{Weisskopf} {et~al.}(2016){Weisskopf}, {Ramsey}, {O'Dell}, {Tennant},
  {Elsner}, {Soffitta}, {Bellazzini}, {Costa}, {Kolodziejczak}, {Kaspi},
  {Muleri}, {Marshall}, {Matt}, \& {Romani}}]{2016SPIE.9905E..17W}
{Weisskopf}, M.~C., {Ramsey}, B., {O'Dell}, S., {et~al.} 2016, in \procspie,
  Vol. 9905, Space Telescopes and Instrumentation 2016: Ultraviolet to Gamma
  Ray, 990517

\bibitem[{{Woo} \& {Urry}(2002)}]{2002ApJ...579..530W}
{Woo}, J.-H., \& {Urry}, C.~M. 2002, \apj, 579, 530

\bibitem[{{Zdziarski} \& {B{\"o}ttcher}(2015)}]{Zdziarski15}
{Zdziarski}, A.~A., \& {B{\"o}ttcher}, M. 2015, \mnras, 450, L21

\bibitem[{{Zhang} \& B{\"o}ttcher(2013)}]{Zhang13}
{Zhang}, H., \& B{\"o}ttcher, M. 2013, \apj, 774, 18

\bibitem[{{Zhang} {et~al.}(2015){Zhang}, {Chen}, {B{\"o}ttcher}, {Guo}, \&
  {Li}}]{Zhang15}
{Zhang}, H., {Chen}, X., {B{\"o}ttcher}, M., {Guo}, F., \& {Li}, H. 2015, \apj,
  804, 58

\bibitem[{{Zhang} {et~al.}(2016){Zhang}, {Diltz}, \& {B{\"o}ttcher}}]{Zhang16}
{Zhang}, H., {Diltz}, C., \& {B{\"o}ttcher}, M. 2016, \apj, 829, 69

\end{thebibliography}

\begin{table*}
\caption{Basic 2FHL properties of the $\gamma$-ray blazars studied in this work. Column information are as follows: 
(1) 2FHL name; (2) other name; (3) redshift; (4) test statistic; (5) photon flux (50GeV$-$2 TeV energy range, in 
units of 10$^{-11}$ \phflux);  (6) spectral index (50GeV$-$2 TeV energy range); (7) number of predicted photons; and (8) 3FGL association. All the information are taken from \citet[][]{2016ApJS..222....5A}. \label{tab:basic_info}
}
\begin{center}
\begin{tabular}{lccccccc}
\hline
2FHL name  & Other name & $z$ & TS & $F_{0.05-2~{\rm TeV}}$ & $\Gamma_{0.05-2~{\rm TeV}}$ &  $N_{\rm pred}$ & 3FGL  \\
(1) & (2) & (3)  & (4) & (5) &  (6)  & (7) & (8) \\ 

\hline
J0456.9$-$2323  &  PKS 0454$-$234  &  1.00 &  30.7  & 1.53$\pm$0.78 & 3.23$\pm$1.16  &  4.1  & J0457.0$-$2324 \\
J0957.6+5523	&  4C +55.17       &  0.90 &  120.3 & 3.59$\pm$1.03 & 3.49$\pm$0.72  & 12.9  & J0957.6+5523    \\
J1224.7+2124    &  4C +21.35       &  0.43 &  108.0 & 5.44$\pm$1.29 & 4.06$\pm$0.74  &  15.0 & J1224.9+2122     \\
J1256.2$-$0548  &  3C 279          &  0.54 &  47.4  & 1.87$\pm$0.87 & 4.44$\pm$1.61  &  4.9  & J1256.1$-$0547  \\
J1427.3$-$4204	&  PKS B1424$-$418 &  1.55 &  41.8  & 1.65$\pm$0.74 & 11.30$\pm$4.60 &  5.0  & J1427.9$-$4206  \\
J1512.7$-$0906  &  PKS 1510$-$08   &  0.36 &  124.0 & 4.59$\pm$1.28 & 2.99$\pm$0.57  &  13.1 & J1512.8$-$0906  \\
J2000.9$-$1749  &  PKS 1958$-$179  &  0.65 &  45.9  & 2.30$\pm$0.92 & 3.46$\pm$0.98  &  6.7  & J2001.0$-$1750  \\ 
J2254.0+1613    &  3C 454.3        &  0.86 &  28.5  & 1.13$\pm$0.66 & 6.26$\pm$3.06  &  3.0  & J2254.0+1608     \\
\hline
\end{tabular}
\end{center}
\end{table*}

\begin{table*}[h!]
\begin{center}
\caption{Summary of the X-ray analysis.\label{tab:x-ray}}
\begin{tabular}{cccccccc}
\tableline\tableline
 Name                  & \#  & Exp.    &      $N_{\rm H}$             & $\Gamma_{\rm X}$           & $\beta_{\rm X}$                   & $F_{X}$                                             & Stat.\\
                           &   & (ksec) &(10$^{20}$ cm$^{-2}$)  &        &                                              & (10$^{-12}$ erg cm$^{-2}$ s$^{-1}$) & $\chi^2$/dof \\ 
\tableline
 J0456.9$-$2323 & 16   & 54.1    & 2.84 &1.55$^{+0.06}_{-0.07}$   &                                              & 1.24$^{+0.07}_{-0.07}$                      & 48.42/58 \\
 J0957.6+5523    & 8     & 27.3    & 0.93 & 1.72$^{+0.12}_{-0.11}$   &                                              & 0.70$^{+0.08}_{-0.08}$                      & 24.01/18 \\
 J1224.7+2124    & 97   & 182.6  & 2.01 & 1.85$^{+0.02}_{-0.02}$   &  $-$0.48$^{+0.04}_{-0.05}$  & 5.93$^{+0.08}_{-0.12}$                      & 397.06/399 \\
 J1256.2$-$0547 & 244 & 422.3  & 2.05 & 1.49$^{+0.01}_{-0.01}$   &  0.15$^{+0.02}_{-0.02}$       & 15.40$^{+0.10}_{-0.10}$                    & 772.77/675 \\
 J1427.3$-$4204 & 53   & 182.4  & 7.63 & 1.15$^{+0.04}_{-0.04}$   &  0.51$^{+0.07}_{-0.06}$       & 3.49$^{+0.09}_{-0.09}$                      & 376.83/345 \\
 J1512.7$-$0906 & 168 & 318.6  & 6.89 & 1.38$^{+0.02}_{-0.02}$   &  $-$0.07$^{+0.03}_{-0.03}$  & 11.00$^{+0.11}_{-0.15}$                     & 656.58/638 \\
 J2000.9$-$1749 & 3     & 9.4      & 6.93 & 1.65$^{+0.18}_{-0.18}$   &                                              & 1.67$^{+0.24}_{-0.19}$                       & 8.26/11 \\
 J2254.0+1613    & 153 & 303.4  & 6.63 & 1.21$^{+0.01}_{-0.01}$   &  0.31$^{+0.02}_{-0.02}$       & 42.65$^{+0.30}_{-0.30}$                     & 968.02/731 \\
 \tableline
\end{tabular}
\end{center}
\tablecomments{Second column represent the total number of XRT observations for each source. The spectral parameter $\Gamma_{\rm X}$ represents the best-fit photon index for a power-law model or slope at the pivot energy (fixed to 1 keV) for the log-parabola model. On the other hand, $\beta_{\rm X}$ is the curvature term for the log-parabola model. $F_{X}$ is the integrated X-ray flux in the energy range of 0.3$-$10 keV.}
\end{table*}

\begin{table*}
{\small
\begin{center}
\caption{List of the Parameters used/derived in the leptonic SED modeling of 2FHL FSRQs. Col.[1]: 2FHL Name; 
Col.[2]: viewing angle, in degrees; Col.[3]: central black hole mass, in log scale; Col.[4]: accretion disk luminosity 
(\lum), in log scale; Col.[5] and [6]: broken power law spectral indices; Col.[7]: magnetic field, in Gauss; 
Col.[8]: particle energy density, in erg cm$^{-3}$; Col.[9]: bulk Lorentz factor; Col.[10]: break Lorentz factor; 
Col.[11]: maximum Lorentz factor; Col.[12]: distance of the emission region from the black hole, in parsec; 
Col.[13]: Size of the BLR, in parsec; Col.[14]: size of the emission region, in parsec; Col.[15], [16], and [17]: jet powers in magnetic field, 
electrons, and protons, respectively. The characteristic temperature of the dusty torus is taken as 500 K and 
we assume $\gamma_{\rm min}$ as unity, for all the sources.}\label{tab:leptonic_sed_param}
\begin{tabular}{lcccccccccccccccc}
\tableline
\tableline
2FHL Name & $\theta_{\rm v}$ & $M_{\rm BH}$ & $L_{\rm disk}$ & $s1$ & $s2$ & $B$ & $U_{\rm e}$ 
& $\Gamma$ & $\gamma_{\rm b}$ & $\gamma_{\rm max}$ & $R_{\rm diss}$ & $R_{\rm BLR}$ & $R_{\rm blob}$ & $P_{\rm m}$ 
 & $P_{\rm e}$ & $P_{\rm p}$\\
~[1] & [2] & [3] & [4] & [5] & [6] & [7] & [8] & [9] & [10] & [11] & [12] & [13] & [14] & [15] & [16] & [17] \\
\tableline
J0456.9$-$2323  & 2 & 9.0 & 45.6 & 1.6 & 3.9 & 0.8 & 0.01 & 18 & 833 & 5000 & 0.10 & 0.06 & 0.010 & 44.8 & 44.6 & 46.5 \\
J0957.6+5523     & 3 & 8.4 & 45.6 & 2.0 & 3.4 & 0.6 & 0.02 & 14 & 1762& 7000& 0.10 & 0.06 & 0.010 & 44.4 & 44.6 & 47.0 \\
J1224.7+2124     & 3 & 8.8 & 46.5 & 1.8 & 3.7 & 0.5 & 0.01 & 13 & 226 & 8000& 0.24 & 0.18 & 0.024 & 44.9  & 44.4 & 46.8 \\
J1256.2$-$0548  & 3 & 8.5 & 45.3 & 1.8 & 3.9 & 1.7 & 0.10 & 14 & 352 & 5000 & 0.06 & 0.05 & 0.006 & 44.9 & 44.8 & 47.1 \\
J1427.3$-$4204  & 3 & 9.0 & 46.0 & 1.6 & 3.7 & 0.8 & 0.02 & 15 & 852 & 7000 & 0.16 & 0.10 & 0.016 & 45.1 & 45.0 & 46.9 \\
J1512.7$-$0906  & 3 & 8.8 & 45.7 & 1.9 & 4.0 & 1.2 & 0.02 & 18 & 206 &10000 & 0.10& 0.08  & 0.010 & 45.2 & 44.8 & 46.8 \\
J2000.9$-$1749  & 3 & 9.0 & 45.2 & 1.8 & 3.6 & 1.3 & 0.05 & 16 & 272 & 8000 & 0.06 & 0.04 & 0.006 & 44.7  & 44.6 & 46.9 \\
J2254.0+1613     & 3 & 9.0 & 46.3 & 2.1 & 3.8 & 3.0 & 0.03 & 18 & 215 & 3500& 0.16 & 0.14 & 0.016 & 46.4  & 45.4 & 48.0 \\
\tableline
\end{tabular}
\end{center}
}
\end{table*}

\begin{table*}
{\small
\begin{center}
\caption{List of the Parameters used/derived in the hadronic SED modeling of 2FHL FSRQs. Col.[1]: 2FHL Name; 
Col.[2]: viewing angle, in degrees; Col.[3]: magnetic field, in Gauss;
Col.[4]: bulk Lorentz factor; Col.[5]: power law spectral indices for primary electrons; Col.[6] and [7]: minimal and maximal Lorentz factors for primary electrons; Col.[8] and [9]: power law spectral indices for primary protons before and after the spectral break;
Col.[10] and [11]: spectral break and maximal Lorentz factors for primary protons; Col.[12]: size of the emission region, in parsec; Col.[13], [14], and [15]: jet powers in magnetic field, 
electrons, and protons, respectively.
We assume $\gamma_{\rm 1,p}$ as unity for all the sources.}\label{tab:hadronic_sed_param}
\begin{tabular}{lcccccccccccccc}
\tableline
\tableline
2FHL Name & $\theta_{\rm v}$ & $B$ & $\Gamma$ & $s_e$ & $\gamma_{\rm 1,e}$ & $\gamma_{\rm 2,e}$ & $s_{\rm 1,p}$ & $s_{\rm 2,p}$ & $\gamma_{\rm b,p}$ & $\gamma_{\rm 2,p}$ & $R_{\rm blob}$ 
& $P_{\rm m}$ & $P_{\rm e}$ & $P_{\rm p}$\\
~[1] & [2] & [3] & [4] & [5] & [6] & [7] & [8] & [9] & [10] & [11] & [12] & [13] & [14] & [15] \\
\tableline
J0456.9$-$2323  & 2 & 100 & 18  & 2.5 & 80 & 700 & 1.9 & 3.5 & $10^8$ & $8\times 10^8$ & $3.2\times 10^{-4}$ & 46.1 & 43.2 & 47.9 \\
J0957.6+5523    & 3 & 100 & 14  & 2.0 & 80 & 700 & 1.9 & 3.0 & $2\times 10^8$ & $10^9$ & $3.2\times 10^{-4}$ & 45.9 & 42.8 & 47.3 \\
J1224.7+2124    & 3 & 30  & 13  & 2.6 & 30 & 900 & 2.0 & 3.8 & $1.2\times 10^8$ & $10^9$ & $2.3\times 10^{-4}$ & 44.4 & 43.3 & 49.0 \\
J1256.2$-$0548  & 3 & 50  & 14  & 2.8 & 90 & 1000& 2.0 & 3.8 & $10^8$ & $10^9$  & $4.8\times 10^{-4}$ & 45.6 & 43.6 & 48.5 \\
J1427.3$-$4204  & 3 & 100 & 11  & 2.5 & 70 & 1000& 2.0 & 2.4 & $4\times 10^8$ & $5\times 10^9$ & $1.3\times 10^{-3}$ & 46.9 & 43.5 & 48.3 \\
J1512.7$-$0906  & 3 & 50  & 18  & 2.2 & 20 & 200 & 2.0 & 3.7 & $7\times 10^7$ & $5\times 10^9$ & $2.6\times 10^{-3}$ & 47.3 & 43.0 & 48.0 \\
J2000.9$-$1749  & 3 & 50  & 14  & 2.5 & 50 & 2000& 2.0 & 3.7 & $8\times 10^7$ & $10^9$ & $2.9\times 10^{-4}$ & 45.2 & 43.5 & 48.0 \\
J2254.0+1613    & 3 & 50  & 18  & 2.6 & 50 & 1000& 2.0 & 3.5 & $10^8$ & $10^9$ & $3.2\times 10^{-3}$ & 47.5 & 44.2 & 49.2 \\
\tableline
\end{tabular}
\end{center}
}
\end{table*}

\begin{table*}
\caption{High energy polarization at 1 keV and 1 MeV derived for 2FHL FSRQs. Column information are as follows: 
(1) 2FHL name; (2) average optical polarization taken from Steward or RoboPol observatories
\citep[marked with $^*$ and $^{\dagger}$, respectively;][]{2009arXiv0912.3621S,2014MNRAS.442.1693P}; (3) depolarization factor $Z_{\rm m}$, as described in Section \ref{xpol}; (4) and (5) degree of polarization predicted at 1 keV from leptonic and hadronic modeling, respectively; and (6) and (7) degree of polarization predicted at 1 MeV from leptonic and hadronic modeling, respectively. Note that we derive the polarization at 1 keV and 1 MeV both for the average and elevated activity states appropriately correcting for partially ordered magnetic field. The high activity state optical polarizations are collected from the Steward observatory database. \label{tab:pol}
}
\begin{center}
\begin{tabular}{ccccccc}
\hline
2FHL name  & optical Pol. & $Z_{\rm m}$ & lep. Pol. & had. Pol. & lep. Pol. & had. Pol.\\
 & (\%) &   & (1 keV, \%) & (1 keV, \%) & (1 MeV, \%) & (1 MeV, \%)\\ 
(1) & (2) & (3)  & (4) & (5) & (6) & (7)\\ 

\hline
& & average activity state & & \\
J0456.9$-$2323  &  9.9$^*$          & 0.16 & 4.2 & 9.5  & 1.0 & 10.1 \\
J0957.6+5523	&  5.7$^{\dagger}$  & 0.10 & 1.6 & 6.6  & 1.0 & 7.2  \\
J1224.7+2124    &  5.4$^*$          & 0.99 & 1.8 & 50.8 & 0.0 & 55.3 \\
J1256.2$-$0548  &  15.0$^*$         & 0.23 & 8.7 & 14.4 & 1.6 & 16.4 \\
J1427.3$-$4204	&  ---              & ---  & --- & ---  & --- & ---  \\
J1512.7$-$0906  &  3.8$^*$          & 0.14 & 2.5 & 9.7  & 0.0 & 9.7  \\
J2000.9$-$1749  &  13.0$^{\dagger}$ & 0.25 & 6.0 & 13.4 & 0.5 & 15.7 \\ 
J2254.0+1613    &  5.8$^*$          & 0.09 & 1.9 & 6.2  & 0.0 & 6.9  \\
\hline
& & elevated activity state & & \\
J0456.9$-$2323  &  35.3$^*$ & 0.55 & 15.3 & 34.7 & 3.6  & 38.3 \\
J0957.6+5523	&  ---      & ---  & ---  & ---  & ---  & ---  \\
J1224.7+2124    &  29.1$^*$ & 0.99 & 1.8  & 50.8 & 0.0  & 55.3 \\
J1256.2$-$0548  &  34.5$^*$ & 0.53 & 20.1 & 33.1 & 3.7  & 37.9 \\
J1427.3$-$4204	&  ---      & ---  & ---  & ---  & ---  & ---  \\
J1512.7$-$0906  &  25.8$^*$ & 0.96 & 17.1 & 66.4 & 0.1  & 66.6 \\
J2000.9$-$1749  &  ---      & ---  & ---  & ---  & ---  & ---  \\ 
J2254.0+1613    &  25.0$^*$ & 0.41 & 8.7  & 28.0 & 0.2  & 28.4 \\
\hline
\end{tabular}
\end{center}
\end{table*}

\end{document}